\documentclass[a4paper,12pt]{article}
\usepackage[utf8]{inputenc}
\usepackage[T1]{fontenc}
\usepackage{lmodern}
\usepackage[margin=2.5cm]{geometry}
\usepackage{amsmath,amssymb,bm}
\usepackage{enumitem}
\usepackage{xurl}
\usepackage{booktabs}
\usepackage[colorlinks=true,linkcolor=blue,citecolor=blue,urlcolor=blue]{hyperref}
\setlist{nosep,leftmargin=2.4em}
\title{SPICE-Q and Large-Scale Quantum Chip Production}
\author{\parbox{0.95\textwidth}{\centering
\normalsize Cai\textsuperscript{\dag},
Ling Qiao\textsuperscript{1,2,\dag} ,
Bin Yang\textsuperscript{1,2,*},
Fumin Luo\textsuperscript{1,2},
Chang Liu\textsuperscript{1,2},
\\
WeiGui Guo\textsuperscript{1,2},
GuoRong Zhang\textsuperscript{1,2},
XueFei Liu\textsuperscript{1,2},
Qinglang Guo\textsuperscript{1,2,*},
and Bin Wu\textsuperscript{*}
\\[0.6em]
\small \textsuperscript{1}Yangtze Delta Industrial Innovation Center of Quantum Science and Technology, Suzhou, China, 215100\\
\small \textsuperscript{2}China Academy of Electronics and Information Technology, No. 11 Shuangyuan Road,\\
\small Shijingshan District, Beijing, China, 100041\\[0.4em]
\small qiaoling@tgqs.net, 1763098000@qq.com\\
\small \textsuperscript{\dag}These authors contributed equally to this work.\\
\small \textsuperscript{*}Corresponding authors: Bin Yang, Qinglang Guo, and Bin Wu.\\
\small Correspondence to: gql1993@mail.ustc.edu.cn
}}
\date{June 16, 2026}

\begin{document}
\maketitle
\tableofcontents
\newpage
\section{Abstract}

\begin{quote}
Literature index: The historical analogy with SPICE is based mainly on the SPICE report by Nagel and Pederson, Nagel's SPICE2 report, the MNA method, reviews of SPICE numerical methods, the BSIM device model, and experience from CMOS VLSI design [1][2][3][4][24][30]; the requirements for NISQ/FTQC and surface-code scaling refer to the work of Preskill, Fowler, Gambetta, and others [5][6][32]; the background on superconducting qubits, transmons, cQED, microwave networks, material loss, and 3D interconnects is based mainly on the literature by Krantz, Kjaergaard, Blais, Koch, Rosenberg, Vahidpour, Dunsworth, Place, Pozar, and others [7][8][9][10][11][12][16][17][18][19][21][22]; the background on EPR parameter extraction, SQuADDS, SQcircuit, scqubits, QuTiP, DTCO, and manufacturability is based mainly on the literature by Minev, Shanto, Rajabzadeh, Groszkowski, Johansson, Verjauw, Van Damme, and others [25][26][27][28][29][33][34].
\end{quote}

This paper discusses a proposed SPICE-Q framework for superconducting quantum chips. Its goal is not to replace existing tools such as HFSS, Qiskit Metal, pyEPR, SQcircuit, SQuADDS, scqubits, or QuTiP, but to organize layout generation, electromagnetic simulation, energy-participation-ratio/circuit quantization, Hamiltonian solving, open-system noise analysis, cryogenic testing, and manufacturing-data feedback into a unified and traceable data chain [26][27][28][33][34]. The framework draws on the role played by classical SPICE in the integrated-circuit industry: through standardized device models, MNA/numerical solvers, and process-parameter interfaces, it shifts design from empirical trial and error toward a reproducible, model-driven workflow [1][2][3][4][24][30].

For superconducting quantum chips, the key mapping chain can be summarized as follows: process/PDK constraints $\rightarrow$ layout geometry $\rightarrow$ electromagnetic modes and equivalent parameters $\rightarrow$ EPR/circuit-quantization parameters $\rightarrow$ effective Hamiltonian $\rightarrow$ metrics for frequency, coupling, anharmonicity, decoherence, readout, and yield. This chain must simultaneously handle factors such as Josephson-junction discreteness, transmon frequency allocation, readout resonators and Purcell constraints, coupler crosstalk, microwave routing, 3D interconnects, material/interface loss, and package modes [7][8][9][12][15][16][17][18][19][21]. On the manufacturing side, the model is calibrated using critical-current distributions, resonant-frequency deviations, cryogenic $T_1/T_2$ statistics, test structures, and cross-wafer process data [19][25][29].

This paper further positions SPICE-Q as a design-technology co-optimization (DTCO) framework rather than a single simulator or an already mature commercial quantum EDA platform. Its core value lies in providing unified model interfaces, statistical parameter models, model cards, version governance, and a closed manufacturing-feedback loop, so that discontinuities among layout, electromagnetics, quantum dynamics, noise models, and process statistics can be managed in an engineering workflow. Recent results from SQuADDS, EPR/pyEPR, SQcircuit, scqubits, and 300 mm CMOS-compatible qubit fabrication show that superconducting quantum-chip design is moving from empirical single-device optimization toward an engineering flow that is verifiable, reusable, and statistically manufacturable [25][26][27][28][29][33].

Therefore, the central argument of this paper is as follows: for larger-scale and fault-tolerantly scalable superconducting quantum processors, hardware design cannot rely only on isolated device optimization or one-off EM simulation. Instead, it requires a continuous model chain from device physics, layout, electromagnetic fields, Hamiltonians, noise, and manufacturing statistics to system-level yield. As the unified representation and DTCO vehicle for this chain, SPICE-Q can provide a methodological foundation for future quantum chips to evolve from laboratory prototypes into manufacturable, testable, and iteratively improvable engineering systems [5][6][14][15][25][29][32].

\section{The Emergence of SPICE and Large-Scale Classical Circuits and Its Implications for Quantum Chips}

\subsection{Introduction}

\begin{quote}
Literature index: The discussion in this subsection on the origin of SPICE, the unified circuit-simulation framework, and the EDA paradigm shift mainly refers to Nagel [1], the MNA method [2], and the SPICE Book [3].
\end{quote}

SPICE (Simulation Program with Integrated Circuit Emphasis) is one of the core simulation tools that enabled the industrialization of modern large-scale semiconductors. Its emergence transformed integrated-circuit design from a "physical-experiment-driven mode" dependent on experience and manual trial and error into a "computer-aided engineering design mode" centered on numerical simulation. It fundamentally changed the chip-design workflow and substantially raised the upper bound of design complexity (Nagel, 1975).

The central contribution of SPICE is that it provided, for the first time, a unified circuit-level simulation framework, allowing circuits with thousands or even millions of device-level elements to be accurately predicted and verified before fabrication. In this framework, a circuit is abstracted as a set of nonlinear equations based on nodal analysis:

\[
\mathbf{G}(\mathbf{x}) \mathbf{x} = \mathbf{b}
\]

Here, $\mathbf{x}$ denotes the state vector of node voltages and branch currents, $\mathbf{G}$ denotes the admittance matrix formed by device models such as MOSFETs, BJTs, resistors, and capacitors, and $\mathbf{b}$ denotes external excitation sources. Through the Newton-Raphson method, SPICE can solve the steady-state and transient responses of strongly nonlinear circuits.

The introduction of SPICE directly promoted the development of the large-scale semiconductor industry, enabling transistor counts to expand rapidly from several tens in the early period to tens of billions in modern systems. Its significance lies not only in improved computational capability, but also in establishing a closed-loop engineering paradigm of "design-simulation-verification", in which complex-system design no longer depends on point-by-point experimental validation but instead on repeatable numerical prediction models.

Against this background, SPICE is not merely a simulation tool; it has also become a foundational pillar of modern electronic design automation (EDA). Without SPICE, today's highly modular and reusable IC design flow could not have emerged, nor could the fabless model have appeared and scaled (Nagel \textbackslash{}\& Pederson, 1973).

For quantum chips, the current design flow remains in an early stage analogous to the period before the emergence of SPICE: simulation tools are fragmented, models are not unified, and a closed loop for manufacturing feedback is lacking. Therefore, the success of SPICE in classical integrated circuits offers a key lesson for quantum computing: a unified, multiphysics "quantum SPICE" framework must be constructed to support scalable design methods from the device level to the system level.

This historical analogy also forms the theoretical motivation for SPICE-Q: just as SPICE unified the design of electronic circuits, SPICE-Q aims to unify the complex relationships among electromagnetics, quantum dynamics, and manufacturing processes in quantum hardware, thereby laying an engineering foundation for large-scale fault-tolerant quantum computers.

\subsection{The Components of SPICE}

\begin{quote}
Literature index: Device models/parameter extraction refer to the BSIM and process-measurement literature [4], while numerical solvers and MNA/nonlinear circuit equations refer to Nagel, Ho-Ruehli-Brennan, and Vladimirescu [1][2][3].
\end{quote}

The core structure of SPICE (Simulation Program with Integrated Circuit Emphasis) can be rigorously divided into two complementary but tightly coupled components: \textbf{device model data (device data / model parameters)} and \textbf{numerical simulator (numerical simulator / solver)}. This two-layer structure constitutes the basic abstraction framework of modern circuit simulation and electronic design automation (EDA) (Nagel, 1975; Vladimirescu, 1994).

\textbf{(1) Device data layer: physical model parameters based on process and measurement}

The first component of SPICE is device data, whose essence is a parameterized abstraction of real semiconductor processes. These data are usually obtained from physical measurements after wafer fabrication (wafer-level characterization), including I--V characteristic curves across temperatures and bias conditions, parasitic capacitance, resistance, and noise spectral density.

For typical CMOS or BJT processes, device models are usually given in parameterized form, such as the BSIM or Gummel--Poon models. Their parameter set can be written as:

\[
\mathcal{M}_{device} = \{ V_{th}, \mu, C_{ox}, R_s, R_d, \gamma, \lambda, \ldots \}
\]

These parameters are not theoretical constants; rather, they are empirical models obtained through process statistics, wafer test structures, and batch-data fitting. Model families such as BSIM are typical implementations of this parameter-extraction logic in MOS devices (Sheu et al., 1987). Thus, the device layer of SPICE is essentially a "process-driven statistical physical model", whose accuracy directly depends on the consistency of the manufacturing process and the completeness of measurement data.

In modern advanced process nodes, this data layer is usually provided by a Process Design Kit (PDK) and updated through continuous manufacturing feedback, thereby forming a closed-loop calibration mechanism for process variation.

\textbf{(2) Simulator layer: a circuit-solution engine based on numerical analysis}

The second core component of SPICE is the numerical simulator, whose function is to transform circuit topology into a solvable mathematical system. Specifically, the circuit is first converted into the form of Modified Nodal Analysis (MNA):

\[
\mathbf{F}(\mathbf{x}, t) = \mathbf{0}
\]

where $\mathbf{x}$ denotes the state vector of node voltages and branch currents. For nonlinear devices, this system of equations is usually solved by Newton-Raphson iteration:

\[
\mathbf{J}(\mathbf{x}^{(k)}) \Delta \mathbf{x}^{(k)} = -\mathbf{F}(\mathbf{x}^{(k)})
\]

where $\mathbf{J}$ is the Jacobian matrix, representing the local linearized derivative structure of the circuit system. Through time-discretization methods such as implicit trapezoidal integration, SPICE can be extended to transient analysis, frequency-domain analysis, noise analysis, and other simulation modes (Vladimirescu, 1994).

Therefore, the SPICE simulator is essentially a "nonlinear multiscale numerical solver", and its performance determines the computational scalability of the entire EDA flow.

\textbf{(3) Coupling mechanism and engineering significance of the two-layer structure}

The key innovation of SPICE lies in the strict decoupling of the "device data layer" and the "numerical solution layer", while coupling them through a unified device model interface. This structure allows:

\begin{itemize}
\item different device models to be replaced without changing the solver, such as BSIM4 $\rightarrow$ BSIM-CMG;
\item numerical algorithms to be upgraded without changing the device models, such as improving Newton convergence or sparse-matrix solvers;
\item only parameter sets to be updated during process evolution while the simulation framework remains stable.
\end{itemize}

This "model-solver separation" architecture made SPICE an extensible industrial-grade simulation platform and directly supported the formation of modern standard cell libraries and Process Design Kits (PDKs) (Nagel \textbackslash{}\& Pederson, 1973).

\textbf{(4) Direct implications for quantum-chip simulation frameworks}

In the quantum-chip setting, this two-layer structure has a direct analogical significance: device data correspond to Josephson-junction (JJ) parameters, microwave loss spectra, and decoherence statistical distributions, whereas the simulator corresponds to a coupled system of electromagnetic field solvers and Hamiltonian-dynamics evolution engines. Therefore, the core idea of SPICE-Q is precisely to extend this "two-layer structure" to multiphysics quantum systems, thereby enabling engineering-oriented modeling of manufacturable quantum hardware.

\subsection{The Transition of SPICE from Physical Design to Engineering Design}

\begin{quote}
Literature index: The discussion in this subsection on the transition from direct solution of physical equations to parameterized engineering abstraction mainly refers to SPICE2, the SPICE Book, and the BSIM device-model literature [1][3][4].
\end{quote}

SPICE gradually evolved from an early computational tool that "directly drove design verification with physical equations" into a core EDA infrastructure that "drives system design with engineering abstractions and standardized models". The essence of this transition is a shift from direct numerical solution of continuous physical systems toward engineering management of uncertainty, statistical behavior, and scalability in complex manufacturing systems (Nagel, 1975; Vladimirescu, 1994).

\textbf{(1) Limitations of the physical-design paradigm}

In the early stage of SPICE, circuit design relied mainly on "physics-based design", namely system solution directly based on device-physics equations such as drift-diffusion equations, Poisson equations, and semiconductor carrier-transport models. However, as the scale of integrated circuits grew exponentially, this method gradually exposed three core problems:

\begin{itemize}
\item \textbf{Computational complexity is not scalable:} the scale of the nonlinear equation system grows exponentially with the number of transistors, causing solution cost to rise rapidly;
\item \textbf{Process fluctuations are difficult to control:} physical parameters such as threshold voltage $V_{th}$ and mobility $\mu$ show significant statistical fluctuations at the nanometer scale;
\item \textbf{Cross-layer design discontinuity:} a unified abstraction interface is lacking between device physics and system-level performance.
\end{itemize}

Thus, a purely physics-driven design method cannot satisfy the engineering requirements of large-scale integrated circuits.

\textbf{(2) Introduction of the engineering-design paradigm}

To solve the above problems, the SPICE system gradually introduced "engineering abstraction layers". The core idea is to replace direct physical solution with parameterized models and to describe process uncertainty through statistical modeling.

Within this framework, device behavior is no longer calculated explicitly from physical equations, but represented by empirical models:

\[
I_D = f(V_{GS}, V_{DS}; \theta_{process})
\]

where $\theta_{process}$ denotes a parameter set extracted from process statistics. Its sources include wafer test structures, Monte Carlo sampling, and batch-data analysis (Sheu et al., 1987).

This method transforms "uncontrollable microscopic physical complexity" into a "manageable statistical parameter space", thereby substantially reducing design complexity.

\textbf{(3) Core mechanisms of the engineering solution}

The key to SPICE's engineering transformation lies in the establishment of three mechanisms:

\textbf{1. Model Hierarchization}\textbackslash{}\textbackslash{} Through model families such as BSIM and EKV, device behavior is divided into different accuracy levels, achieving a tradeoff between computational efficiency and physical accuracy.

\textbf{2. Process Parameterization}\textbackslash{}\textbackslash{} Random variables in the manufacturing process are explicitly incorporated into the model, for example:

\[
V_{th} = V_{th,0} + \Delta V_{random} + \Delta V_{systematic}
\]

thereby enabling predictable modeling of process deviations.

\textbf{3. Design-Simulation Loop}\textbackslash{}\textbackslash{} Through the combination of PDKs and SPICE simulation, design outputs can be fed back into the process-optimization procedure, forming a closed-loop optimization system.

\textbf{(4) Paradigm migration from physical control to engineering control}

The core contribution of SPICE is that it realized a transition from "physical controllability" to "engineering predictability". In the physical-design paradigm, system performance depends on precise control of microscopic physical processes; in the engineering-design paradigm, system performance depends on the stability of statistical models and the manageability of parameter spaces.

The result of this transition is that:

\begin{itemize}
\item chip design no longer depends on physical debugging of individual devices;
\item process fluctuations are incorporated into the design space rather than treated as errors;
\item system-level performance can be predicted directly through model optimization.
\end{itemize}

\textbf{(5) Implications for quantum-chip SPICE-Q}

For quantum chips, the field is still in a "physical-design-dominated stage", relying on electromagnetic simulation and accurate Hamiltonian modeling but lacking mature engineering abstraction layers. Therefore, the historical evolution of SPICE toward engineering design shows that, for quantum computing to scale, similar engineering abstraction mechanisms must be introduced. Fluctuations in Josephson junctions, microwave loss, and decoherence processes must be parameterized, thereby constructing the statistical engineering-model foundation required by SPICE-Q.

\subsection{SPICE in the 1970s}

\begin{quote}
Literature index: The development of SPICE in the 1969-1970s, transistor-level simulation, and its numerical core mainly refer to Nagel's SPICE2 technical report [1], the MNA paper [2], and the SPICE Book [3].
\end{quote}

SPICE (Simulation Program with Integrated Circuit Emphasis) was born at the University of California, Berkeley between 1969 and 1972. It was developed by Laurence W. Nagel under the supervision of Donald O. Pederson, and its initial goal was to solve the problem that integrated-circuit design at the time "lacked a unified circuit-level simulation tool" (Nagel \textbackslash{}\& Pederson, 1973; Nagel, 1975). During this period, integrated circuits were still in an early stage of development. Design relied mainly on manual calculation and experimental verification, and the complexity of transistor-level circuits had already begun to exceed the tractable range of traditional analytical methods.

\textbf{(1) Basic goals and functional definition of SPICE}

The core role of SPICE in the 1970s was to provide a unified transistor-level circuit simulation environment, allowing designers to predict circuit behavior before fabrication. Its basic task can be formalized as follows: given a circuit topology and device models, solve the nonlinear dynamic response of the system.

The circuit is transformed into a system of differential-algebraic equations (DAEs) of the following form:

\[
\mathbf{F}(\mathbf{x}(t), \dot{\mathbf{x}}(t), t) = 0
\]

where $\mathbf{x}(t)$ denotes state variables consisting of circuit node voltages and branch currents. SPICE discretizes the continuous-time problem through numerical integration methods such as the implicit trapezoidal rule, thereby enabling transient analysis.

\textbf{(2) Significance of transistor-level simulation}

In the 1970s, one of the most important breakthroughs of SPICE was transistor-level simulation, especially accurate modeling of BJTs (bipolar junction transistors) and early MOSFETs. This capability enabled it to simulate the dynamic behavior of basic logic circuits, amplifiers, and analog circuits.

Taking the BJT as an example, its core current relationship can be expressed as:

\[
I_C = I_S \left( e^{\frac{V_{BE}}{V_T}} - 1 \right)
\]

This nonlinear exponential relationship is solved in SPICE by numerical iteration rather than analytical calculation, allowing complex coupling effects in large-scale interconnect structures to be handled (Nagel \textbackslash{}\& Pederson, 1973).

The significance of transistor-level simulation is that it embedded "device physical behavior" directly into "circuit system behavior" for the first time, enabling designers to observe at the system level how device nonidealities, such as saturation effects and parasitic capacitance, affect overall performance.

\textbf{(3) The role of SPICE in early circuit design}

In the 1970s, SPICE was mainly used for the following three types of tasks:

\begin{itemize}
\item \textbf{Amplifier design and verification:} analyzing gain, bandwidth, and nonlinear distortion;
\item \textbf{Digital logic circuit analysis:} simulating the delay and power consumption of basic gate circuits such as AND, OR, and NAND;
\item \textbf{Parasitic-effect modeling:} including capacitive coupling, interconnect resistance, and noise effects.
\end{itemize}

These capabilities enabled engineers to identify design defects before chip fabrication, thereby substantially reducing the cost of trial and error.

\textbf{(4) Numerical solution methods and computational framework}

The core numerical methods adopted by SPICE in the 1970s include:

\textbf{1. Modified Nodal Analysis (MNA)}\textbackslash{}\textbackslash{} Transforming circuit topology into a sparse matrix system:

\[
\mathbf{G}(\mathbf{x}) \mathbf{x} = \mathbf{b}
\]

\textbf{2. Newton-Raphson iteration}\textbackslash{}\textbackslash{} Used to solve nonlinear systems:

\[
\mathbf{J}(\mathbf{x}^{(k)}) \Delta \mathbf{x}^{(k)} = -\mathbf{F}(\mathbf{x}^{(k)})
\]

\textbf{3. Time-discretization methods}\textbackslash{}\textbackslash{} Processing transient responses through implicit integration methods to improve numerical stability.

Together, these methods constituted the "numerical core engine" of SPICE, enabling it to handle complex nonlinear circuit systems.

\textbf{(5) Historical significance and the foundation of the EDA paradigm}

The success of SPICE in the 1970s was not only a technical breakthrough; more importantly, it established the basic paradigm of modern EDA (Electronic Design Automation): circuit design must be built on a unified mathematical model and numerical simulation foundation rather than on empirical trial and error.

This paradigm directly promoted the formation of later standard cell libraries, Process Design Kits (PDKs), and ASIC design flows, and ultimately gave rise to the fabless industrial structure (Nagel \textbackslash{}\& Pederson, 1973; Weste \textbackslash{}\& Harris, 2010).

\textbf{(6) Historical implications for SPICE-Q}

The development of SPICE in the 1970s shows that its key success factor was not a single algorithmic breakthrough, but the combination of "transistor-level modeling + a unified numerical solution framework". This structure provides a direct analogy for SPICE-Q: quantum chips likewise require a unified modeling system beginning from the "quantum device level (JJ junction)" and system-level simulation through multiphysics solvers, thereby realizing a foundational role analogous to that of SPICE in classical electronics.

\subsection{The Logical Layer of PCells and Their Corresponding SPICE Models}

\begin{quote}
Literature index: Classical analogies for PCells, standard cells, and PDK-style parameterized design refer to SPICE/CMOS VLSI design materials [1][3][24]; the propagation of quantum-device parameters to higher-level models refers to Krantz and Koch [7][12].
\end{quote}

Parameterized Cells (PCells) are a key abstraction mechanism in modern EDA design flows. Their core idea is to establish a computable mapping between "process constraints" and "design degrees of freedom". PCells not only define geometric layout, but also map physical device structures into simulable circuit-behavior models through coupling with SPICE models (Nagel \textbackslash{}\& Pederson, 1973; Sheu et al., 1987).

Within the SPICE framework, the essence of a PCell can be represented as a functional mapping from geometric parameters to electrical-behavior parameters:

\[
\mathcal{F}_{PCell}: \mathcal{G} \rightarrow \mathcal{P}_{SPICE}
\]

where $\mathcal{G}$ denotes the geometric design space, such as dimensions, spacing, and layer stack, and $\mathcal{P}_{SPICE}$ denotes the SPICE model-parameter space, such as capacitance, resistance, coupling coefficients, and noise parameters. This mapping is usually obtained by inversion from process-calibration data and test structures.

\textbf{(1) A SPICE-based multilayer abstraction structure}

In complex integrated systems, a PCell is not only a physical device-description unit but also a foundational module for cross-layer modeling. Combining SPICE simulation with Process Design Kits (PDKs), the following hierarchical structure can be established:

\begin{itemize}
\item \textbf{Part Level (device-layout layer)}\textbackslash{}\textbackslash{}
This corresponds to the physical layout definition of an early JJ (Josephson Junction) or transistor, emphasizing geometry and material stack. SPICE is mainly used at this layer to extract parasitic parameters and local electrical responses.

\item \textbf{Component Level (device-component layer)}\textbackslash{}\textbackslash{}
This corresponds to a single simulable device, such as the Josephson junction in a transmon or a CMOS transistor. This layer describes nonlinear electrical behavior through SPICE models such as BSIM or an effective JJ model.

\item \textbf{Logical Space Level (logic-gate layer)}\textbackslash{}\textbackslash{}
At this layer, multiple devices combine to form logical function units such as AND, OR, and NOT gates. SPICE is used here to analyze delay, power consumption, and noise margins, and the behavior can be abstracted as a Boolean function:

\end{itemize}

\[
y = f(x_1, x_2, \ldots, x_n)
\]

\begin{itemize}
\item \textbf{Arithmetic Level (arithmetic-operation layer)}\textbackslash{}\textbackslash{}
The abstraction is further raised to basic arithmetic operation units such as addition and subtraction, for example:

\end{itemize}

\[
S = A \oplus B, \quad C = A \cdot B
\]

SPICE is used to evaluate multi-device coupling errors and propagation delays.

\begin{itemize}
\item \textbf{Functional IP Level (functional-IP layer)}\textbackslash{}\textbackslash{}
This layer consists of matrix operations, FFTs, or quantum-control modules. System behavior can be represented as a linear or nonlinear operator:

\end{itemize}

\[
\mathbf{y} = \mathbf{W}\mathbf{x} + \phi(\mathbf{x})
\]

\begin{itemize}
\item \textbf{Algorithm/System/SOC Level (system level)}\textbackslash{}\textbackslash{}
At the highest layer, multiple functional IP blocks are combined to form a complete SoC or quantum processing unit (QPU). Here SPICE degenerates into a statistical performance-evaluation tool used to analyze system-level error propagation and resource constraints.

\end{itemize}

This hierarchical structure reflects the progressive abstraction from "physical device" to "system function", while SPICE serves as the unified simulation foundation across all layers.

\textbf{(2) SPICE-driven cross-layer parameter propagation}

The core value of a PCell lies in enabling cross-layer parameter propagation. In this mechanism, lower-level SPICE simulation results are used to update higher-level model parameters. For example, key parameters extracted from JJ device-level SPICE simulation include:

\[
I_c, \quad C_j, \quad L_k, \quad Q_{loss}
\]

These parameters further affect delay models and error-probability models at the logic-gate layer:

\[
\tau_{gate} = f(I_c, C_j, L_k)
\]

thereby realizing a traceable modeling chain from the physical layer to the logical layer (Krantz et al., 2019).

\textbf{(3) Process-constraint-driven PCell generation}

A PCell is not a static structure but a dynamic model generated under Process Design Kit (PDK) constraints. Its generation process can be expressed as an optimization problem:

\[
\min_{\mathcal{G}} \; \mathcal{L}_{SPICE}(\mathcal{G}, \theta_{process})
\]

where $\mathcal{L}_{SPICE}$ denotes the performance loss function based on SPICE simulation, and $\theta_{process}$ denotes process statistical parameters. This optimization process ensures that the PCell remains functionally stable under manufacturing fluctuations.

\textbf{(4) Implications for quantum-chip SPICE-Q}

In the SPICE-Q framework, the PCell concept can be extended to "Quantum PCells", such as unified parameterized representations of JJ junctions, resonators, and couplers. Through a similar hierarchical structure, unified modeling from quantum devices to quantum algorithms can be realized, thereby supporting manufacturable quantum-system design. This idea directly corresponds to the cross-layer co-optimization framework required for constructing scalable FTQC systems.

\subsection{SPICE as the Foundation of Modern EDA}

\begin{quote}
Literature index: The discussion in this subsection on SPICE as EDA infrastructure, virtualization of front-end design, and fabless/foundry division of labor refers to SPICE2, the SPICE Book, and CMOS VLSI design textbooks [1][3][24].
\end{quote}

As a foundational technology of modern electronic design automation (EDA, Electronic Design Automation), SPICE made its core contribution by transforming circuit design from a "physical-experiment-driven trial-and-error process" into a "computable optimization problem based on a unified numerical model". This transition not only greatly improved chip-design efficiency, but also fundamentally changed the organizational structure of the semiconductor industry, making the fabless design model possible (Nagel \textbackslash{}\& Pederson, 1973; Nagel, 1975; Weste \textbackslash{}\& Harris, 2010).

\textbf{(1) Mechanism by which SPICE accelerated circuit simulation}

The key by which SPICE accelerates chip design is not simply an increase in computational speed, but a reduction in the number of design iterations through "model unification + numerical decoupling". Before SPICE, circuit design relied on successive prototype fabrication and experimental measurement, and its time complexity can be approximated as:

\[
T_{design} \propto N_{prototype} \times T_{fabrication}
\]

where $N_{prototype}$ denotes the number of trial-and-error iterations, and $T_{fabrication}$ is usually measured in weeks or months.

After SPICE was introduced, the design process was transformed into a numerical-solution problem:

\[
\mathbf{F}(\mathbf{x}) = 0
\]

Through Newton iteration and sparse-matrix solution, circuit behavior can be predicted directly on a computer, thereby transforming design complexity into:

\[
T_{design} \propto T_{simulation} \ll T_{fabrication}
\]

The essence of this transition is the mapping from "physical trial-and-error space" to "numerical optimization space".

\textbf{(2) Formation of the EDA automation flow}

SPICE is not only a simulator, but also the core mathematical engine of the modern EDA toolchain. It directly gave rise to the following key technical modules:

\begin{itemize}
\item \textbf{Circuit Simulation:} used to verify transistor-level behavior;
\item \textbf{Layout vs Schematic (LVS):} ensuring consistency between physical implementation and schematic;
\item \textbf{Design Rule Check (DRC):} ensuring manufacturability;
\item \textbf{Process Design Kit (PDK):} providing unified device models and parameters.
\end{itemize}

Together, these modules form the modern EDA system and shift the design flow from "experience-driven" to "model-driven".

\textbf{(3) From transistor-level simulation to system-level design acceleration}

The accurate transistor-level modeling capability of SPICE allows complex circuits to be verified at the system level before fabrication. For example, for a circuit containing $N$ transistors, the complexity of its state space is:

\[
\mathcal{O}(N^k), \quad k > 1
\]

Traditional methods require layer-by-layer physical verification, whereas SPICE reduces the computational complexity to an engineering-manageable range through sparse-matrix structure and local nonlinear approximation (Vladimirescu, 1994).

In addition, SPICE supports multiple analysis modes:

\[
\text{DC analysis}, \quad \text{AC analysis}, \quad \text{Transient analysis}
\]

These capabilities allow engineers to complete a full analysis from static operating points to dynamic behavior within a single framework, thereby greatly shortening the design cycle.

\textbf{(4) SPICE and the formation of the fabless industrial model}

The emergence of SPICE directly promoted the separation of semiconductor industrial structure, namely design-manufacturing decoupling. Before SPICE, chip design had to rely on repeated trial production in fabs. After the SPICE and PDK systems were established, design companies could complete all front-end design using simulation tools alone.

This structural change can be formalized as:

\[
\text{Chip Development} = \text{Design (EDA + SPICE)} + \text{Fabrication (Foundry)}
\]

Here, the design process is fully virtualized, allowing companies to complete high-complexity chip design without owning a fab, and thereby giving rise to the Fabless + Foundry industrial model (Weste \textbackslash{}\& Harris, 2010).

\textbf{(5) Dual impact on economic and engineering paradigms}

SPICE changed not only engineering methods but also the economic structure of the semiconductor industry. Design risk shifted from "physical trial-and-error risk" to "model-error risk", fundamentally changing the cost structure of chip development:

\[
C_{total} = C_{design} + C_{fabrication}
\]

where $C_{design}$ is significantly reduced in the SPICE-driven EDA system, while $C_{fabrication}$ is concentrated in foundries.

This separation allows design companies to focus on architecture and system innovation, while manufacturing companies focus on process optimization, thereby substantially improving the innovation efficiency of the entire industry.

\textbf{(6) Structural implications for SPICE-Q}

In the field of quantum chips, a unified EDA infrastructure analogous to SPICE is still lacking, leading to a high degree of fragmentation among design, simulation, and manufacturing. The historical experience of SPICE shows that a scalable industrial system can emerge only when a "unified simulation framework + standardized device models + closed-loop process feedback" are simultaneously present.

Therefore, the goal of SPICE-Q is not merely to upgrade simulation tools, but to reproduce the structural reconfiguration that SPICE brought to the classical semiconductor industry, enabling quantum computing to move from the "experimental science stage" into the "engineering industrialization stage".

\subsection{Design Requirements for Large-Scale Quantum Chips}

\begin{quote}
Literature index: The transition from NISQ to FTQC refers to Preskill [5], the physical-qubit overhead of the surface code refers to Fowler et al. [6], and engineering and standardization requirements for superconducting quantum systems refer to Krantz, Kjaergaard, and Versluis [7][8][14].
\end{quote}

The evolution from Noisy Intermediate-Scale Quantum (NISQ) devices to Fault-Tolerant Quantum Computing (FTQC) marks the transition of quantum computing systems from an "experimental verification stage" to an "engineering-scalable stage". The core feature of this transition is the rapid increase in the number of physical qubits and the resulting problem of system complexity [5].

In FTQC architectures, stable encoding of logical qubits requires a large number of physical qubits for error-correction redundancy, for example in the surface code structure:

\[
N_{physical} \sim \mathcal{O}(d^2)
\]

where $d$ is the code distance. For practically useful fault-tolerant computation, system scale often needs to expand from hundreds to millions of physical qubits, posing fundamental challenges to existing design and simulation systems.

\textbf{(1) Limitations of existing quantum simulation methods}

Current quantum-chip design relies on multiple separated simulation tools, such as electromagnetic simulation (HFSS-like methods), quantum Hamiltonian and open-system simulation (QuTiP-like methods), and noise-model analysis tools [34]. However, these methods lack unified interfaces, leading to the following problems:

\begin{itemize}
\item \textbf{Lack of scalability in size:} full quantum-state simulation has complexity $\mathcal{O}(2^n)$ and cannot scale to large systems [5][31];
\item \textbf{Separation between physical and information models:} no consistent mapping relation has been established between electromagnetic field distributions and quantum-state evolution;
\item \textbf{Lack of manufacturing-constraint modeling:} process deviations, such as JJ critical-current fluctuations, have not been systematically incorporated into simulation.
\end{itemize}

Therefore, existing methods cannot support FTQC-level system design.

\textbf{(2) Need for Standardization}

The second core challenge in large-scale quantum-chip design is the lack of unified standards. Unlike the classical semiconductor industry, quantum chips have not yet formed a standardized model system analogous to a Process Design Kit (PDK).

In the classical SPICE system, device behavior is described through a unified model:

\[
I = f(V, \theta_{process})
\]

By contrast, quantum systems still lack a unified formal representation, making it difficult to share design and simulation results across different experimental platforms.

The need for standardization is mainly reflected in three aspects:

\begin{itemize}
\item \textbf{Standardization of device models:} unified parameter representation for JJs, resonators, and couplers;
\item \textbf{Standardization of interfaces:} coupling interfaces between the electromagnetic domain and the quantum domain;
\item \textbf{Standardization of error models:} unified statistical descriptions of decoherence, noise, and process deviations.
\end{itemize}

This absence directly hinders the scaling of quantum systems.

\textbf{(3) Need for Manufacturing Integration}

The third key requirement for large-scale quantum chips is to explicitly incorporate the manufacturing process into the design and simulation loop. In the current system, manufacturing deviations are usually treated as "a posteriori errors" rather than design variables, resulting in significant discrepancies between design and actual performance.

At the FTQC scale, however, manufacturing fluctuations must be modeled as random variables:

\[
\theta_{device} = \theta_0 + \delta_{fabrication}
\]

where $\delta_{fabrication}$ denotes statistical perturbations introduced by the process, such as JJ critical-current fluctuations and dielectric-loss variations.

The lack of manufacturing integration can lead to:

\begin{itemize}
\item unpredictable frequency collisions;
\item uncontrollable distributions of qubit decoherence times;
\item inability to optimize system-level errors in advance.
\end{itemize}

Therefore, manufacturing data must be fed back into simulation models to form a closed-loop DTCO structure.

\textbf{(4) Necessity of SPICE-like unified simulation}

In classical electronics, SPICE solved cross-scale problems in complex circuit design by providing a unified circuit-analysis framework. Its core roles are:

\begin{itemize}
\item unifying device models and system-level simulation;
\item incorporating process parameters into the design space;
\item providing repeatable numerical prediction capability.
\end{itemize}

Thus, its historical experience can be formalized as:

\[
\text{System Complexity} \rightarrow \text{Unified Simulation Framework}
\]

The field of quantum computing faces similar problems but lacks the corresponding unified framework.

\textbf{(5) Necessity of a "SPICE moment"}

The current development stage of quantum computing is similar to the early integrated-circuit design stage in the 1970s, namely a lack of unified simulation and design toolchains. Therefore, a consensus is gradually forming in the field that quantum computing needs a "unified simulation moment (SPICE moment)" analogous to SPICE, so as to realize engineering-oriented modeling from the device level to the system level (Kjaergaard et al., 2020).

The core requirements of this "SPICE moment" include:

\begin{itemize}
\item unified multiphysics modeling (electromagnetics + quantum dynamics);
\item a closed loop for manufacturing feedback;
\item standardized device-model libraries;
\item system-level error-propagation analysis.
\end{itemize}

SPICE-Q is proposed precisely in this context, to fill the critical gap in the transition of quantum computing from experimental science to an engineering industrial system.

\subsection{The Transition of SPICE-Q from Physical Design to Engineering Design}

\begin{quote}
Literature index: Unified multiphysics modeling is based on cQED reviews [9][10][11]; superconducting-qubit engineering, transmon parameter sensitivity, and manufacturing scalability refer to Krantz, Kjaergaard, Koch, and Verjauw [7][8][12][25].
\end{quote}

Scaling superconducting quantum processors to the million-qubit level requires a fundamental transition in the quantum-chip design paradigm: from a physical-design method centered on "isolated electromagnetic simulation and device-level optimization" to an engineering-design system centered on "unified multiphysics modeling and Design-Technology Co-Optimization (DTCO)". The essence of this transition is to reconstruct quantum hardware from an experimental physics system into a predictable, iterable, and manufacturable engineering system (Kjaergaard et al., 2020; Krantz et al., 2019).

Against this background, SPICE-Q is proposed as a foundational simulation and modeling framework. Its goal is to bridge the systemic gap between quantum-circuit design and large-scale manufacturing, thereby providing a unified engineering foundation for quantum chips at FTQC scale.

\textbf{(1) From isolated electromagnetic simulation to unified multiphysics modeling}

Traditional quantum-chip design relies mainly on electromagnetic simulation tools, such as HFSS-like methods, to optimize microwave structures. Its core concerns are resonant frequency, coupling strength, and parasitic loss. However, this approach has a structural limitation: it lacks a unified modeling interface between the electromagnetic domain and the quantum-information domain.

In the SPICE-Q framework, the system is reconstructed as a cross-domain coupled multiphysics system, whose basic structure can be expressed as:

\[
\mathcal{H}_{total} = \mathcal{H}_{quantum} + \mathcal{H}_{EM} + \mathcal{H}_{coupling}
\]

where $\mathcal{H}_{quantum}$ describes the Hamiltonian dynamics of Josephson junctions and qubits, $\mathcal{H}_{EM}$ describes microwave and transmission-line electromagnetic fields, and $\mathcal{H}_{coupling}$ describes the energy-exchange mechanism between the two.

This unified modeling approach allows physical space and information space to be co-optimized within the same simulation framework, thereby avoiding the error-accumulation problem caused by traditional "separated simulation".

\textbf{(2) Necessity of the DTCO paradigm in quantum chips}

Design-Technology Co-Optimization (DTCO) has become a core method for process development at advanced nodes (7 nm and below) in the classical semiconductor industry. Its essence is to drive process optimization through design feedback while allowing process constraints to shape the design space in return.

In quantum chips, the necessity of DTCO is even more pronounced because device performance is highly sensitive to process fluctuations. For example, the critical current of a Josephson junction can be expressed as:

\[
I_c = I_{c0} + \delta I_c(\theta_{process})
\]

where $\delta I_c$ denotes random fluctuations introduced by oxide-layer thickness, interface roughness, and material nonuniformity.

These fluctuations directly affect qubit frequencies:

\[
\omega_q \approx \sqrt{8 E_J E_C} - E_C
\]

where $E_J \propto I_c$. Therefore, process perturbations map directly onto uncertainty in the qubit energy-level structure and further affect system-level error-correction performance.

Accordingly, SPICE-Q takes DTCO as its core design paradigm, allowing the process statistical distribution $\theta_{process}$ to explicitly enter the design-optimization process:

\[
\min_{\mathcal{D}} \; \mathbb{E}_{\theta_{process}} \left[ \mathcal{L}_{system}(\mathcal{D}, \theta_{process}) \right]
\]

\textbf{(3) Bottom-up modeling method: from JJ to system-level architecture}

SPICE-Q emphasizes a "bottom-up" modeling method, namely beginning with the physical modeling of the most basic Josephson Junction (JJ) and progressively constructing system-level behavioral models.

At the device layer, a JJ can be described by the RCSJ model:

\[
C \frac{d^2 \phi}{dt^2} + \frac{1}{R} \frac{d \phi}{dt} + I_c \sin(\phi) = I_{bias}
\]

This equation constitutes the foundational nonlinear dynamical unit of the SPICE-Q device layer.

On this basis, the parameterized-cell (PCell) mechanism extends JJs into intermediate-layer structures such as resonators, couplers, and qubits, and further maps them to the logic layer and system layer.

This hierarchical structure ensures that the behavior of each layer remains traceable to physical device parameters, thereby avoiding the failure of "black-box models" in large-scale systems.

\textbf{(4) Core features of the engineering transition}

The engineering-design paradigm represented by SPICE-Q has three key features:

\begin{itemize}
\item \textbf{Predictability:} system performance can be predicted in advance by statistical models rather than depending on experimental debugging;
\item \textbf{Scalability:} cross-scale modeling from a single JJ or single qubit to multi-qubit arrays, while retaining composable interfaces for larger systems;
\item \textbf{Manufacturability:} process constraints explicitly enter the design-optimization process.
\end{itemize}

This transition marks the shift of quantum-chip design from "physical-experiment-driven" to "engineering-system-driven".

\textbf{(5) SPICE-Q as the unified vehicle for the DTCO framework}

SPICE-Q is not only a simulation tool, but also the implementation vehicle for the DTCO methodology. In this framework, design, simulation, and manufacturing form a closed loop:

\[
\text{Design} \rightarrow \text{SPICE-Q Simulation} \rightarrow \text{Fabrication} \rightarrow \text{Measurement} \rightarrow \text{Model Update}
\]

This closed-loop mechanism allows manufacturing data to be continuously fed back to the model layer, thereby continually correcting device-parameter distributions and system-level error models.

Ultimately, the goal of SPICE-Q is to establish a unified infrastructure analogous to the role of classical SPICE in the semiconductor industry, enabling quantum-chip design to transition from experimental science to a discipline of systems engineering.

\subsection{Impact of SPICE-Q on the Quantum-Chip Manufacturing Cycle}

\begin{quote}
Literature index: The discussion of predictive manufacturing, JJ/frequency distributions, yield, and manufacturing feedback refers to the literature by Krantz, Koch, Morvan, Verjauw, and others [7][12][15][25].
\end{quote}

By incorporating design, simulation, and manufacturing processes into a unified multiphysics DTCO closed-loop framework, SPICE-Q fundamentally changes the traditional workflow of quantum-chip manufacturing, transforming it from "experience-driven process iteration" into "model-driven predictive manufacturing". The core value of this transition lies in significantly shortening design convergence time and improving the yield and consistency of large-scale quantum processors (Krantz et al., 2019; Kjaergaard et al., 2020).

\textbf{(1) From trial-and-error manufacturing to predictive manufacturing}

In traditional quantum-chip manufacturing workflows, device performance depends strongly on process-window control, such as oxide-layer thickness, junction-area errors, and thin-film uniformity. However, these variables are usually empirically tuned through multi-cycle fabrication, and the convergence process can be expressed as:

\[
\theta^{(k+1)} = \theta^{(k)} + \Delta \theta_{exp}
\]

where $\theta$ denotes the set of process parameters, and $\Delta \theta_{exp}$ comes from experimental feedback.

After SPICE-Q is introduced, this process is reconstructed as a model-based optimization problem:

\[
\theta^{*} = \arg \min_{\theta} \; \mathbb{E}_{\theta_{process}} \left[ \mathcal{L}_{device}(\theta) \right]
\]

where the loss function $\mathcal{L}_{device}$ comes from the joint simulation results of SPICE-Q for qubit frequency, coupling strength, and decoherence time.

This transition upgrades the manufacturing process from "empirical iteration" to "simulation-driven optimization".

\textbf{(2) Mechanism for improving manufacturing yield}

In large-scale quantum chips, yield is mainly affected by frequency collisions, parameter drift, and deviations in decoherence distributions. SPICE-Q explicitly incorporates these problems into the design space through statistical modeling.

Yield can be expressed as:

\[
Y = P(\forall i, \; \omega_i \notin \mathcal{B}_{collision})
\]

where $\omega_i$ denotes the frequency of the $i$-th qubit, and $\mathcal{B}_{collision}$ denotes the permitted frequency-avoidance window.

Through Monte Carlo simulation in SPICE-Q, the overlap probability of frequency distributions can be evaluated in advance at the design stage, thereby optimizing JJ parameter distributions and routing topology.

\textbf{(3) Acceleration of the manufacturing cycle by the design-manufacturing closed loop}

The core acceleration mechanism of SPICE-Q comes from the fact that the DTCO closed loop reduces the number of manufacturing experiments. In the traditional flow, design convergence time is proportional to the number of experimental rounds:

\[
T_{total} \propto N_{fab} \cdot T_{cycle}
\]

where $N_{fab}$ is the number of manufacturing iterations.

Under the SPICE-Q framework, because simulation can predict the impact of manufacturing deviations in advance, $N_{fab}$ is significantly reduced:

\[
N_{fab}^{SPICE-Q} \ll N_{fab}^{traditional}
\]

thereby reducing part of the physical trial-fabrication cycles and shortening the design convergence cycle.

\textbf{(4) Cross-layer error prediction and process-robust design}

SPICE-Q optimizes not only average performance, but also system robustness to process fluctuations. For example, the sensitivity of qubit frequency to JJ critical-current fluctuations can be expressed as:

\[
\frac{\partial \omega_q}{\partial I_c} \propto \frac{1}{\sqrt{I_c}}
\]

Therefore, by adjusting device-parameter distributions at the design stage, the sensitivity of the system to manufacturing noise can be reduced, thereby improving overall yield.

This "robust design optimization" is a key capability not covered by traditional SPICE and is also one of the central contributions of SPICE-Q to manufacturing acceleration.

\textbf{(5) Significance of SPICE-Q as manufacturing infrastructure}

SPICE-Q is not merely a simulation tool, but also an infrastructure layer connecting design and manufacturing. In this framework, manufacturing data are no longer passive outcomes but input variables that actively participate in design optimization:

\[
\text{Manufacturing Data} \rightarrow \text{SPICE-Q Model Update} \rightarrow \text{Design Refinement}
\]

This mechanism transforms quantum-chip manufacturing from an "experimental science workflow" into a "data-driven engineering system".

Ultimately, SPICE-Q provides a methodological foundation for manufacturable design of larger-scale quantum processors, enabling quantum-computing hardware to move gradually from laboratory device development toward engineering, statistical, and industrial production.

\section{SPICE-Q Model Composition}

\begin{quote}
Literature index: The basic framework of microwave photons, resonators, qubits, and cQED hybrid systems in this chapter refers to Blais 2004/2021, Wallraff 2004, and Krantz 2019 [7][9][10][11]; electromagnetic-to-Hamiltonian parameter extraction refers to EPR/pyEPR, SQuADDS, SQcircuit, and scqubits workflows [26][27][28][33]; open-system dynamics solving refers to QuTiP [34]; and the background on manufacturability and manufacturing feedback refers to the work of Verjauw, Van Damme, and others [25][29].
\end{quote}

SPICE-Q differs from traditional SPICE (Simulation Program with Integrated Circuit Emphasis). Traditional SPICE uses voltage, current, nodes, and device models as its core variables, whereas a design workflow for superconducting quantum chips must simultaneously handle geometric layout, three-dimensional electromagnetic modes, equivalent circuit parameters, Josephson nonlinearity, quantum Hamiltonian parameters, cryogenic noise models, and manufacturing-measurement feedback. In superconducting quantum circuits, the interactions among microwave resonators, transmission lines, and qubits are usually described by cQED models, while quantum-state evolution is jointly determined by effective Hamiltonians, master equations, and open-system dynamics [7][9][10][11][34].

Therefore, a more rigorous SPICE-Q abstraction is not simply a "physical-space-information-space duality", but a verifiable and traceable parameter-mapping chain:

\[
\text{Process/PDK constraints} \rightarrow \text{Layout} \rightarrow \text{EM modes} \rightarrow \{C,L,Z(\omega),S_{ij},p_{mj}\} \rightarrow \mathcal{H}_{\mathrm{eff}} \rightarrow \text{device metrics} \rightarrow \text{model update}
\]

Here, $p_{mj}$ denotes the energy participation ratio (EPR), which can be used to connect electromagnetic eigenmodes with Josephson nonlinearity, frequency, anharmonicity, and dispersive coupling [27]. This chain has already been partially implemented in existing tools: Qiskit Metal can undertake parameterized layout generation; SQuADDS connects target Hamiltonian parameters with a database of simulated and partially experimentally validated device designs [26]; pyEPR/EPR methods transform mode participation ratios from three-dimensional electromagnetic simulations such as HFSS into quantum Hamiltonian parameters [27]; tools such as SQcircuit and scqubits compute spectra, coherence, matrix elements, and coupling operators from circuit descriptions [28][33]; and QuTiP can serve as an open-quantum-system dynamics solver for time evolution and Lindblad channels [34].

Therefore, SPICE-Q in this paper should be understood as a system-level integration of these distributed workflows. It is not a single software package intended to replace HFSS, Qiskit Metal, pyEPR, SQcircuit, scqubits, QuTiP, or SQuADDS. Rather, it defines a unified interface and data closed loop from process constraints, layout, EM simulation, quantum-parameter extraction, and open-system dynamics evaluation to manufacturing-feedback updates. Its model composition includes at least four mutually coupled layers: physical-space simulation, information-space simulation, a SPICE-Q netlist capable of expressing cross-domain relationships, and a solver/feedback layer responsible for parameter extraction, dynamics solving, and model calibration.

\subsection{Physical-Space Simulation}

\begin{quote}
Literature index: Microwave transmission, S-parameters, CPW/resonators, and full-wave electromagnetic simulation refer to Pozar, Jin, Blais, and Wallraff [9][11][21][22]; the transmission of EM results to EPR/Hamiltonian parameters refers to SQuADDS and EPR quantization work [26][27].
\end{quote}

In the SPICE-Q framework, physical-space simulation mainly describes the propagation of microwave signals in superconducting quantum chips, including transmission-line networks, readout resonators, coupling structures, package interconnects, and the energy distribution and loss mechanisms in the three-dimensional electromagnetic environment surrounding the chip. As information carriers for control and readout, microwave signals propagate through on-chip transmission lines to resonators and further undergo dispersive or drive coupling with qubits. These processes form the physical basis for readout, control, and crosstalk analysis in cQED architectures [9][10][11].

Such processes remain, in essence, within the domain of classical electromagnetism. The core simulation objects are therefore electromagnetic field distributions $\mathbf{E}(\mathbf{r},t)$ and $\mathbf{H}(\mathbf{r},t)$, as well as the eigenfrequencies, impedances, coupling rates, loss participation ratios, scattering parameters, and parasitic modes derived from them. In engineering implementation, such problems are usually solved using finite-element methods, method of moments, or related full-wave electromagnetic simulation tools. In workflows closer to quantum-chip design automation, field solutions must further enter EPR/pyEPR or SQuADDS-style parameter-extraction procedures to form low-dimensional parameters usable for Hamiltonian solving [21][22][26][27]. Their mathematical foundation is Maxwell's equations:

\[
\nabla \times \mathbf{E} = -\frac{\partial \mathbf{B}}{\partial t}, \quad
\nabla \times \mathbf{H} = \mathbf{J} + \frac{\partial \mathbf{D}}{\partial t}
\]

Under constraints from material parameters, metal boundaries, port excitations, and package boundary conditions, one can obtain propagation modes, standing-wave distributions, and energy loss of microwaves in complex three-dimensional interconnect structures. For SPICE-Q, the key is not to stop at field maps or a single S-parameter, but to transform these results into parameters usable by subsequent information-space modeling, such as $\omega_r$, $Z(\omega)$, $Q$, $p_{mj}$, $g_{ij}$, and crosstalk matrices.

\textbf{(1) Microwave-signal conduction and electromagnetic leakage}

In actual quantum chips, when microwave signals propagate through transmission lines, airbridges, TSVs, flip-chip interconnects, and package cavities, electromagnetic leakage, coupling to non-target modes, and crosstalk can occur. These effects can be described by scattering parameters (S-parameters):

\[
\mathbf{S} =
\begin{bmatrix}
S_{11} & S_{12} \\
S_{21} & S_{22}
\end{bmatrix}
\]

Here, $S_{21}$ denotes the transmission efficiency from input port to output port, and $S_{11}$ denotes reflection loss. Electromagnetic leakage usually appears as coupling to non-target modes, whose impact can be further projected into undesired interaction terms between qubits or resonators:

\[
\mathcal{H}_{leak} \propto g_{ij}^{(leak)} (a_i^\dagger a_j + a_j^\dagger a_i)
\]

Therefore, physical-space simulation is used not only to optimize signal integrity, but also to directly affect the structure of the quantum-layer Hamiltonian and the system-level crosstalk budget.

\textbf{(2) Spectrum and distortion analysis of microwave measurement and control links}

In addition to propagation paths, microwave signals are affected during input and output by frequency-response limitations, nonlinear distortion, phase noise, and link mismatch. In qubit readout and control, signals are usually modulated within the GHz band, and their spectral structure can be expressed as:

\[
X(\omega) = \int_{-\infty}^{\infty} x(t)e^{-i\omega t}dt, \quad Y(\omega)=H(\omega)X(\omega)
\]

where $H(\omega)$ is the frequency-response function of the transmission link. Nonideal responses introduce amplitude distortion and phase drift, thereby affecting readout fidelity, gate-pulse calibration, and residual coherent error. Nonlinear characteristics of measurement-and-control components such as amplifiers and mixers can also introduce higher-order harmonics and intermodulation distortion, which can be approximately represented by a polynomial:

\[
y(t)=\alpha x(t)+\beta x^2(t)+\gamma x^3(t)
\]

These distortion terms reduce gate-operation accuracy in high-precision quantum control and increase the risk of error propagation. SPICE-Q needs to connect such classical microwave-link parameters with drive Hamiltonians, readout errors, and decoherence channels in information space.

\textbf{(3) Engineering significance of physical-space simulation}

The core role of physical-space simulation is to establish a mapping from electromagnetic structures to quantum-control performance, allowing microwave-engineering parameters such as impedance matching, loss, crosstalk, port isolation, and package modes to directly affect quantum system-level metrics such as gate fidelity, readout fidelity, Purcell limits, and decoherence time. Therefore, this simulation layer is not merely a classical electromagnetic problem; it is also an important foundational layer of SPICE-Q's unified multiphysics modeling.

\subsection{Information-Space Simulation}

\begin{quote}
Literature index: Two-level/multilevel approximations, Hamiltonians, Lindblad master equations, and cQED information-space modeling refer to Krantz, Kjaergaard, Blais, and Koch [7][8][9][12]; energy levels, anharmonicity, and circuit-quantization solution workflows refer to the EPR, SQcircuit, and scqubits literature [27][28][33]; open-system dynamics solving refers to QuTiP [34].
\end{quote}

In the SPICE-Q framework, information-space simulation is used to describe the dynamical evolution of qubits in an abstract quantum-state space. Unlike electromagnetic field simulation in physical space, this layer no longer directly handles continuous electromagnetic field distributions. Instead, it abstracts the system into quantum states, density matrices, and effective Hamiltonians in a finite-dimensional Hilbert space. In ordinary cases, superconducting qubits can be approximated as two-level systems in the computational subspace. In higher-accuracy modeling, they must be extended to three-level or multilevel systems to account for transmon anharmonicity, leakage states, matrix-element selection rules, and drive-induced errors [7][8][12][33].

Under the two-level approximation, a quantum state can be expressed as:

\[
|\psi(t)\rangle = \alpha(t)|0\rangle + \beta(t)|1\rangle
\]

Its dynamics are governed by the Schrodinger equation:

\[
i\hbar \frac{d}{dt}|\psi(t)\rangle = \mathcal{H}(t)|\psi(t)\rangle
\]

where $\mathcal{H}(t)$ is the system's effective Hamiltonian, including intrinsic-frequency terms, drive terms, and coupling terms. In SPICE-Q, this Hamiltonian is usually jointly mapped from device-level parameters and EM parameters, such as JJ critical current, total capacitance, resonant frequency, impedance, energy participation ratio, and coupling strength. Thus, information space is not an abstraction layer independent of physical structure; rather, it is an engineering model obtained after quantization and dimensional reduction of physical space.

\textbf{(1) Quantum fluctuations, noise spectra, and open-system dynamics}

At the information-space layer, quantum fluctuations and the uncertainty principle form the basis for noise modeling and decoherence analysis. For any two operators $\hat{A}$ and $\hat{B}$, their uncertainty relation satisfies:

\[
\Delta A \cdot \Delta B \geq \frac{1}{2} |\langle [\hat{A}, \hat{B}] \rangle|
\]

For typical superconducting-qubit systems, common conjugate variables include the phase operator $\hat{\phi}$ and the charge operator $\hat{n}$, satisfying:

\[
[\hat{\phi}, \hat{n}] = i
\]

This relation affects energy-level stability, noise sensitivity, and coherence time. In SPICE-Q information-space simulation, quantum fluctuations are usually not treated as an isolated module, but enter the model through noise spectra, collapse operators, and open-system dynamics. For example, decoherence processes can be described by the Lindblad master equation:

\[
\frac{d\rho}{dt} = -\frac{i}{\hbar}[\mathcal{H}, \rho] + \sum_k \left( L_k \rho L_k^\dagger - \frac{1}{2}\{L_k^\dagger L_k, \rho\} \right)
\]

where $\rho$ denotes the density matrix, and $L_k$ denotes different types of noise operators, such as energy relaxation, phase decoherence, thermal excitation, and leakage-related channels. In engineering implementation, such master equations are usually handled by open-system solvers such as QuTiP. What SPICE-Q needs to do is organize parameters from EM/EPR/circuit quantization, noise spectra from test data, and drive terms from control systems into consistent inputs, rather than reinventing all low-level time-evolution algorithms [34].

\textbf{(2) Multilevel extension and leakage-state modeling}

In more accurate SPICE-Q modeling, a qubit is not a strictly two-level system, but contains high-energy leakage states. The system can then be extended to a three-level or multilevel model:

\[
|\psi\rangle = \alpha|0\rangle + \beta|1\rangle + \gamma|2\rangle + \cdots
\]

The existence of leakage states leads to accumulated logic-gate errors. Their dynamics are still described by an extended Hamiltonian and simulated through numerical solvers for time evolution. For weakly anharmonic systems such as transmons, the choice of multilevel truncation affects the prediction accuracy of anharmonicity, cross-Kerr coupling, drive crosstalk, and readout error [12][28].

\textbf{(3) Computational-complexity reduction significance of information-space simulation}

The core advantage of information-space modeling lies in significantly reducing computational complexity through physical abstraction. For a system of $n$ qubits, the dimension of the full state space is $2^n$; if the multilevel structure of each device is considered, the dimension grows even faster. SPICE-Q cannot rely on direct simulation of the complete many-body state. Instead, it must approximately represent the system through local Hamiltonian decomposition, effective coupling models, and sparse interaction graphs:

\[
\mathcal{H}_{total} \approx \sum_i \mathcal{H}_i + \sum_{\langle i,j \rangle} \mathcal{H}_{ij}
\]

This structure allows large-scale systems to be approximately simulated in a modular manner, thereby alleviating the exponential computational blowup of full quantum-state simulation [31].

\textbf{(4) Mapping relationship between information space and physical space}

Information space does not exist independently; it is jointly mapped from physical-space parameters, device parameters, and measurement feedback:

\[
\mathcal{H}_{information} = \mathcal{F}(E_J,E_C,\omega_r,Z(\omega),p_{mj},g_{ij},Q_{loss})
\]

Here, $E_J$ and $E_C$ come respectively from Josephson energy and charging energy, $\omega_r$ denotes the resonant frequency, $p_{mj}$ denotes the energy participation ratio, $g_{ij}$ denotes coupling strength, and $Q_{loss}$ denotes the quality factor or loss-related parameters. This mapping relation is the core basis for SPICE-Q to realize cross-layer unified modeling, allowing quantum systems to undergo predictable design, error budgeting, and manufacturing-feedback calibration at the engineering level.

\textbf{(5) Engineering significance of information space}

Information-space simulation is used not only to describe quantum-state evolution, but also to quantify error propagation, noise structure, readout response, and logic-gate fidelity. It is the core bridge layer in SPICE-Q connecting physical devices, microwave measurement and control, and quantum-algorithm requirements.

\subsection{Model Boundaries of Traditional SPICE}

\begin{quote}
Literature index: Voltage/current variables, lumped-parameter assumptions, MNA equations, and high-frequency extension boundaries of traditional SPICE refer to Nagel, Ho-Ruehli-Brennan, Vladimirescu, and Pozar [1][2][3][21].
\end{quote}

Traditional SPICE (Simulation Program with Integrated Circuit Emphasis) is mainly used to simulate voltage and current behavior in circuits. Its basic modeling objects are lumped elements, including resistors, capacitors, inductors, and nonlinear transistor models. Under this framework, circuit behavior is usually represented through Modified Nodal Analysis (MNA) as:

\[
\mathbf{G}(\mathbf{x})\mathbf{x} + \mathbf{C}\frac{d\mathbf{x}}{dt} = \mathbf{b}(t)
\]

where $\mathbf{G}$ denotes the conductance matrix, $\mathbf{C}$ denotes the capacitance matrix, and $\mathbf{x}$ denotes node state variables. This method has good numerical stability and engineering applicability in low- and medium-frequency circuits [1][2][3].

However, the core assumption of traditional SPICE is the "lumped-parameter approximation", namely that component dimensions are much smaller than the signal wavelength. Under high-frequency microwave conditions at the GHz scale, this assumption gradually fails, causing distributed effects and parasitic coupling to become significant and requiring extension of the model to electromagnetic-consistency analysis.

\textbf{(1) Extraction problem for capacitance and inductance matrices}

In high-density three-dimensional interconnect structures, capacitance and inductance are no longer scalar parameters, but must be described in matrix form to represent spatial coupling relations:

\[
\mathbf{C}_{ij} = C(\mathbf{r}_i, \mathbf{r}_j), \quad
\mathbf{L}_{ij} = L(\mathbf{r}_i, \mathbf{r}_j)
\]

This matrix structure reflects nonlocal electromagnetic coupling effects. In TSV (Through-Silicon Via) and flip-chip structures in particular, parasitic capacitance and mutual inductance can significantly affect signal integrity.

Such matrix-based modeling is an important foundation for extending traditional SPICE toward electromagnetically consistent SPICE (Vladimirescu, 1994).

\textbf{(2) High-frequency extensions and microwave SPICE models}

To support high-frequency circuit analysis, SPICE has developed several extended forms, such as RF-SPICE and harmonic balance methods, for handling distributed-parameter networks. In these extensions, circuits are no longer described only by ODEs, but require the introduction of transmission-line equations:

\[
\frac{\partial V(x,t)}{\partial x} = -L \frac{\partial I(x,t)}{\partial t}, \quad
\frac{\partial I(x,t)}{\partial x} = -C \frac{\partial V(x,t)}{\partial t}
\]

This model can describe microwave propagation in chip interconnect structures, but the computational complexity increases significantly.

\textbf{(3) Preliminary introduction of quantum effects}

In some extended SPICE studies, researchers have attempted to incorporate quantum effects into classical circuit-simulation frameworks, for example by introducing Josephson-junction models in superconducting circuits:

\[
I = I_c \sin(\phi)
\]

as well as the quantized flux relation:

\[
\frac{\hbar}{2e}\frac{d\phi}{dt} = V
\]

These extensions allow SPICE to partially describe superconducting-device behavior, but they remain limited to semiclassical approximation frameworks and cannot directly cover multilevel Hamiltonians, open-system decoherence, and quantum-measurement feedback [7][8][12].

\textbf{(4) Asymmetric noise modeling in ultralow-temperature superconducting circuits}

In low-temperature quantum circuits at the mK scale, noise behavior exhibits significant asymmetry and nonequilibrium features. The traditional Johnson-Nyquist noise model,

\[
S_V(f) = 4k_BTR
\]

is no longer fully applicable under ultralow-temperature conditions, and quantum correction terms must be introduced, such as zero-point fluctuation noise:

\[
S(\omega) \propto \hbar \omega \coth\left(\frac{\hbar \omega}{2k_B T}\right)
\]

These noise mechanisms directly affect qubit decoherence time and are one of the important motivations for extending SPICE-Q.

\textbf{(5) Summary of the limitations of classical SPICE}

The core limitations of traditional SPICE can be summarized in three points:

\begin{itemize}
\item it cannot fully handle high-frequency distributed electromagnetic effects;
\item it lacks a unified descriptive capability for quantum dynamics;
\item it provides insufficient modeling of superconducting and cryogenic noise mechanisms.
\end{itemize}

Therefore, in quantum-chip design, a multiphysics unified framework such as SPICE-Q must be introduced to realize a continuous modeling transition from classical circuits to quantum systems.

\subsection{Extension Boundaries of Traditional SPICE Netlists}

\begin{quote}
Literature index: Traditional SPICE netlists, node/element/parameter representations, noise modules, and microwave-port extensions refer to SPICE2, the SPICE Book, MNA, and microwave-engineering literature [1][2][3][21].
\end{quote}

Traditional SPICE circuit descriptions rely on a Netlist structure, whose essence is to map circuit topology into a set of node-connection relations and element parameters. In this representation, a circuit system is abstracted as a graph $\mathcal{G}=(\mathcal{N},\mathcal{E})$, where $\mathcal{N}$ denotes the node set and $\mathcal{E}$ denotes element-connection relations. A standard SPICE Netlist usually contains resistors, capacitors, inductors, source terms, and transistor models, and its basic expression can be summarized as:

\[
I_k=f(V_i-V_j,\theta_k)
\]

where $I_k$ denotes the current response of an element, and $\theta_k$ denotes the device-parameter set. In matrix form, the system is ultimately transformed into a nonlinear differential-algebraic equation system (DAE), solved by Newton iteration, sparse-matrix solution, and time-integration methods [1][2][3].

\textbf{(1) Classical nodes and microwave ports}

Nodes in traditional netlists mainly represent voltage variables, while branches represent elements or controlled-source relations. After entering RF and microwave settings, the netlist must explicitly define input/output ports to describe signal injection, reflection, and transmission paths. Port behavior is usually represented by scattering parameters (S-parameters):

\[
b=\mathbf{S}a
\]

where $a$ and $b$ denote incident-wave and reflected-wave vectors, respectively. This form is widely used in RF-SPICE and electromagnetic co-simulation, but its semantics remain those of classical network parameters rather than quantum states themselves [21].

\textbf{(2) Noise blocks and requirements for cryogenic extension}

In traditional SPICE, noise is usually embedded in the netlist as an additional random process or small-signal noise source, mainly describing thermal noise, shot noise, and device noise. Its mathematical description can be given by a spectral-density function:

\[
S(\omega)=\int_{-\infty}^{\infty}\langle \xi(t)\xi(0)\rangle e^{-i\omega t}dt
\]

This representation can enter a circuit-simulation system as a composable module. However, in superconducting quantum chips, cryogenic nonequilibrium noise, quantum fluctuations, decoherence, and leakage states must be further mapped into open-quantum-system models. In other words, traditional netlists can express microwave ports and classical-noise boundaries, but cannot directly express local Hilbert spaces, Hamiltonian blocks, and Lindblad channels.

\textbf{(3) Transition from traditional netlists to SPICE-Q netlists}

The extension of traditional netlists by SPICE-Q should not be understood as simply attaching several quantum variables to nodes. Rather, the netlist layer must simultaneously retain classical microwave networks and quantum-dynamical networks. Classical nodes, ports, and device parameters record layout and electromagnetic structures, while quantum nodes, Hamiltonian blocks, and noise/decoherence channels record information-space models. Traceable mappings between the two are established through parameters such as $Z(\omega)$, $S_{ij}$, $p_{mj}$, $E_J$, $E_C$, and $g_{ij}$.

\subsection{Cross-Domain Representation of SPICE-Q Netlists}

\begin{quote}
Literature index: Dual-domain netlists, Jaynes-Cummings coupling terms, quantum nodes, and Hamiltonian modules refer to Blais, Wallraff, and Krantz [7][9][10][11], and are connected with EPR parameter extraction and SQcircuit-style circuit-quantization workflows [27][28].
\end{quote}

The SPICE-Q Netlist extends the traditional SPICE structure so that it can simultaneously express classical electromagnetic networks and quantum-dynamical systems. The core distinction is that a traditional Netlist mainly describes voltage/current relations and local element constitutive equations, whereas a SPICE-Q Netlist must simultaneously include the electromagnetic domain, the Hamiltonian domain, and the parameter mappings between them. A compact abstract form is:

\[
\mathcal{N}_{Q}=\{\mathcal{N}_{EM},\mathcal{N}_{H},\mathcal{P}_{map}\}
\]

where $\mathcal{N}_{EM}$ denotes the microwave-propagation, port, S-parameter, and geometry-related network; $\mathcal{N}_{H}$ denotes the Hamiltonian network composed of qubits, resonators, couplers, and drive terms; and $\mathcal{P}_{map}$ records mapping relations from EM results to Hamiltonian parameters, such as impedance, eigenfrequency, energy participation ratio, Josephson energy, and coupling strength.

\textbf{(1) Cross-domain coupling terms}

The key feature of a SPICE-Q Netlist is the introduction of cross-domain coupling operators that connect microwave signals with quantum-state evolution. Under the cQED approximation, a typical coupling term can be written as:

\[
\mathcal{H}_{coupling}=g(\omega)(a^\dagger\sigma^-+a\sigma^+)
\]

where $a^\dagger,a$ denote microwave-field or resonator-mode operators, $\sigma^\pm$ denote qubit transition operators, and $g(\omega)$ is the frequency-dependent coupling strength [9][10]. In an engineering model, $g(\omega)$ should not be treated as an isolated fitting constant, but should be traced as much as possible to geometry, electromagnetic modes, and EPR/circuit-quantization parameters [27][28].

\textbf{(2) Mapping from Netlist to simulation solving}

A SPICE-Q Netlist no longer maps only to a traditional DAE system, but to a staged hybrid solution problem:

\[
\mathcal{N}_{Q}\rightarrow\{\text{EM extraction},\text{circuit quantization},\text{Hamiltonian/open-system solving}\}
\]

In this workflow, the EM Solver handles electromagnetic-field propagation and port responses; EPR or circuit-quantization steps transform field modes and Josephson nonlinearity into an effective Hamiltonian; and the Hamiltonian/Open-system Solver further handles quantum-state evolution, decoherence, and gate errors. The coupling layer is responsible for parameter transfer, consistency checks, and manufacturing-feedback updates, rather than requiring all physical quantities to be solved simultaneously in a single real-time solver.

\textbf{(3) Engineering significance}

The introduction of SPICE-Q Netlists gives quantum-chip design a hierarchical capability similar to classical IC design. The same structure can express composable quantum modules, record cross-physical-domain parameter sources, and provide a unified data interface for DTCO closed-loop optimization. This structure marks a key transition of quantum-chip design from "physical device-level modeling" to "system-level engineering abstraction".

\subsection{Supplement on Microwave Ports and Noise Blocks in Traditional Netlists}

\begin{quote}
Literature index: The traditional netlists, S-parameter ports, and noise-block content supplemented in this subsection refer to the SPICE Book, MNA, and Pozar's microwave-network theory [2][3][21].
\end{quote}

A traditional SPICE netlist is the core representation of circuit simulation, and its essence is a structured description of circuit topology and component parameters. In classical electronic design, a netlist defines circuit behavior through node-connection relationships and is analyzed by a solver in the time domain or frequency domain. Its basic form can be written as:

\[
\text{Netlist}=\{\text{Nodes},\text{Elements},\text{Parameters}\}
\]

Within this framework, each device, such as a resistor, capacitor, or transistor, is defined through a local constitutive relation, for example:

\[
I=C\frac{dV}{dt}, \quad V=L\frac{dI}{dt}
\]

This representation is highly efficient in low- and intermediate-frequency electronic systems. When the system enters the GHz microwave band and includes packaging, transmission-line, and port effects, the netlist requires additional extensions to handle distributed effects and parasitic coupling.

\textbf{(1) Microwave-port modeling}

In RF and microwave extensions of SPICE, circuits are no longer described only by voltage and current nodes; instead, the concept of ports is introduced to describe the relationship between incident and reflected waves. A typical representation is given by scattering parameters (S-parameters):

\[
b_i=\sum_j S_{ij}a_j
\]

where $a_j$ denotes an incident wave, $b_i$ denotes a reflected wave, and $S_{ij}$ denotes transmission characteristics between ports. Microwave-port models allow SPICE/RF network analysis to partially handle GHz-scale signal-propagation problems, but they usually still rely on linear or weakly nonlinear assumptions. For quantum chips, these port parameters must be further mapped into readout links, Purcell limits, crosstalk, and drive Hamiltonians [21].

\textbf{(2) Noise blocks}

In traditional SPICE, noise is usually superposed on circuit variables as an additional random process, mainly including thermal noise and shot noise:

\[
S_I=2qI, \quad S_V=4k_BTR
\]

where $S_I$ and $S_V$ denote current and voltage noise spectral densities, respectively. At the netlist level, a noise module can be embedded as an independent block in the system for analyzing signal integrity and bit-error rate. However, in superconducting quantum chips, cryogenic noise, quantum fluctuations, nonequilibrium excitations, and decoherence channels must further enter Lindblad or equivalent open-system models.

\textbf{(3) Limitations of traditional Netlists}

The structure of a traditional SPICE netlist is essentially a combination of "circuit topology + local physical models". Its advantages lie in modularity and extensibility, but it is difficult for it to directly express distributed electromagnetic-field coupling, multiphysics joint constraints, and state-dependent dynamic Hamiltonian systems. Therefore, in quantum-chip design, a higher-level SPICE-Q netlist abstraction must be introduced to connect classical microwave networks and quantum information space.

\subsection{SPICE-Q Netlists}

\begin{quote}
Literature index: Quantum nodes, Hamiltonian blocks, and microwave-to-Hamiltonian mappings refer to cQED reviews, transmon papers, and superconducting-qubit engineering reviews [7][8][9][12]; energy participation ratios and circuit-quantization parameter transfer refer to the EPR, SQcircuit, and scqubits literature [27][28][33]; open-system solver interfaces refer to QuTiP [34].
\end{quote}

A SPICE-Q netlist is an extended form based on the traditional SPICE netlist. Its core change is that the system is no longer composed only of voltage and current nodes, but introduces quantum nodes, Hamiltonian blocks, a microwave coupling graph, and noise/measurement blocks, thereby extending circuit simulation into the quantum-information domain. Its abstract representation is:

\[
\text{SPICE-Q Netlist}=\{\text{Quantum Nodes},\text{Hamiltonian Blocks},\text{Microwave Coupling Graph},\text{Noise/Measurement Blocks}\}
\]

\textbf{(1) Quantum Nodes}

Quantum nodes no longer represent simple classical circuit node voltages, but represent local quantum states, Josephson phase/charge degrees of freedom, or truncated local Hilbert spaces. For example:

\[
|\psi_i\rangle\in\mathcal{H}_i
\]

Each node can be represented by a state vector or density matrix and carries traceable physical-parameter sources such as $E_J$, $E_C$, $\omega_i$, anharmonicity, and noise sensitivity. The significance of this treatment is that it preserves the modularity advantages of traditional netlists while avoiding the misrepresentation of quantum states as subsidiary fields of ordinary voltage nodes.

\textbf{(2) Hamiltonian Blocks}

Hamiltonian blocks in SPICE-Q are used to describe local physical subsystems, such as qubits, resonators, couplers, drive lines, and readout links. The total Hamiltonian can be written as:

\[
\mathcal{H}_{total}=\sum_i\mathcal{H}_{Q_i}+\sum_j\mathcal{H}_{R_j}+\sum_{i,j}\mathcal{H}_{coupling}+\mathcal{H}_{drive}
\]

where $\mathcal{H}_{Q_i}$ describes qubits, $\mathcal{H}_{R_j}$ describes resonator modes, $\mathcal{H}_{coupling}$ describes capacitive/inductive/dispersive coupling, and $\mathcal{H}_{drive}$ describes control pulses and readout drives. This structure allows SPICE-Q to embed quantum-dynamics evolution rules at the netlist level and to interface with SQcircuit, scqubits, or customized Hamiltonian solvers [28][33]. If the netlist also contains Lindblad channels, measurement operators, and time-dependent drive terms, it should also be able to export structured inputs readable by open-system solvers such as QuTiP [34].

**(3) Microwave $\rightarrow$ Quantum Mapping**

The key characteristic of a SPICE-Q netlist is the establishment of a mapping relation from "microwave network $\rightarrow$ quantum Hamiltonian":

\[
\mathcal{F}:\{S_{ij},Z(\omega),p_{mj},\omega_m\}\rightarrow\mathcal{H}_{eff}
\]

Here, microwave-port parameters, eigenfrequencies, impedance, and energy participation ratios jointly determine the coupling strength, frequency shifts, drive terms, and loss channels of the quantum system. Through this mapping, classical electromagnetic structures are no longer merely layout-verification results, but become parameter sources for information-space quantum models [26][27].

\textbf{(4) Engineering significance of the SPICE-Q Netlist}

This netlist structure gives quantum-chip design a scalability similar to classical IC design, enabling designers to directly perform quantum-circuit planning, error analysis, cross-domain consistency checking, and process co-optimization at the system level. It also provides a natural entry point for manufacturing feedback: cryogenically measured frequencies, couplings, $T_1/T_2$, and readout metrics can be written back into netlist parameters for the next round of model calibration.

\subsection{SPICE-Q Solvers and Cross-Domain Parameter Extraction}

\begin{quote}
Literature index: Classical EM solving, quantum Hamiltonian solving, and hybrid cQED solver frameworks refer respectively to finite-element electromagnetics, Krantz/Kjaergaard, and Blais reviews [7][8][9][22]; actual EM-to-Hamiltonian parameter extraction refers to EPR/pyEPR, SQuADDS, SQcircuit, and scqubits workflows [26][27][28][33]; open-system time evolution and master-equation solving refer to QuTiP [34].
\end{quote}

The SPICE-Q solvers layer is the computational core of the entire simulation framework. Its goal is to coordinate parameter transfer, model calibration, and dynamics solving between physical space (electromagnetic domain) and information space (quantum information domain) under a unified data interface. Unlike traditional SPICE, which mainly relies on numerical integration to solve nonlinear circuit equations, SPICE-Q is closer to a multi-solver collaborative pipeline: electromagnetic solvers are responsible for fields and port parameters; EPR/circuit quantization is responsible for low-dimensional Hamiltonian construction; spectrum/matrix-element solvers are responsible for intrinsic device properties; and open-system solvers are responsible for decoherence, drive response, and gate-error evaluation [26][27][28][33][34].

This system can be abstracted as a hierarchical solver structure:

\[
\mathcal{S}_{SPICE-Q} = \{\mathcal{S}_{EM}, \mathcal{S}_{Q}, \mathcal{S}_{HYB}\}
\]

which correspond respectively to classical electromagnetic solvers, quantum Hamiltonian solvers, and hybrid coupled solvers.

\textbf{(1) Classical Solvers}

Classical solvers are mainly responsible for electromagnetic-field propagation problems in physical space. Their core includes numerical solution of Maxwell's equations and frequency-domain/time-domain finite-element analysis. In engineering implementation, such solvers are usually implemented using HFSS, FDTD, or MoM methods (Jin, 2014).

Their basic governing equations are:

\[
\nabla \times \mathbf{E} = -\frac{\partial \mathbf{B}}{\partial t}, \quad
\nabla \times \mathbf{H} = \mathbf{J} + \frac{\partial \mathbf{D}}{\partial t}
\]

In SPICE-Q, the output of this layer includes not only field distributions but also frequency-dependent coupling parameters:

\[
\{Z(\omega), S_{ij}(\omega)\}
\]

These parameters are then passed as inputs to the quantum-solver layer.

\textbf{(2) Quantum Solvers}

Quantum solvers are responsible for state-evolution problems in information space. Their core is solving the time-dependent Schrodinger equation or density-matrix dynamics:

\[
i\hbar \frac{d}{dt}|\psi(t)\rangle = \mathcal{H}(t)|\psi(t)\rangle
\]

or using the Lindblad master equation in open quantum systems:

\[
\frac{d\rho}{dt} = -\frac{i}{\hbar}[\mathcal{H}, \rho] + \sum_k \mathcal{D}[L_k]\rho
\]

where $\mathcal{D}[L_k]$ denotes a dissipative operator.

The key task of this layer is to map device-level parameters such as coupling strength and drive frequency into quantum-evolution dynamics, thereby evaluating gate fidelity, coherence time, and error-propagation behavior [7][8].

\textbf{(3) Hybrid Solvers}

The hybrid-solver layer of SPICE-Q is used to describe the coupling relationship between physical space and information space. More precisely, this layer can perform joint dynamical solution in small-scale problems, and can also undertake parameter transfer, consistency constraints, and model-calibration functions in engineering workflows. Its basic coupling structure can be written as:

\[
\mathcal{H} = \mathcal{H}_{Q} + \mathcal{H}_{EM} + \mathcal{H}_{int}
\]

where the interaction term $\mathcal{H}_{int}$ dynamically connects the microwave field and the quantum state, for example:

\[
\mathcal{H}_{int} = g(\mathbf{r}) (a^\dagger \sigma^- + a \sigma^+)
\]

This structure originates from the cQED (circuit quantum electrodynamics) model and is the theoretical foundation of superconducting quantum-chip design [9][10][11].

\textbf{(4) Pipeline of EM extraction, EPR quantization, and system-level simulation}

In engineering implementation, the workflow that is more consistent with the existing literature is not to let HFSS and the Hamiltonian solver perform fully bidirectional real-time coupling, but to use staged parameter transfer. First, a geometric model is generated from a parameterized layout; then a three-dimensional EM solver extracts eigenfrequencies, impedances, field distributions, S-parameters, and energy participation ratios; next, EPR/pyEPR or equivalent circuit-quantization methods construct the effective Hamiltonian; subsequently, SQcircuit, scqubits, or custom solvers evaluate spectra, anharmonicity, matrix elements, and coupling strengths; finally, QuTiP or similar open-system solvers evaluate decoherence channels, time-dependent drives, and gate errors [26][27][28][33][34].

This process can be formalized as:

\[
\mathcal{G}_{layout}
\xrightarrow{\mathrm{EM}}
\Theta_{EM}=\{\omega_m,Z_m,p_{mj},S_{ij}\}
\xrightarrow{\mathrm{quantization}}
\mathcal{H}_{eff}(\Theta_{EM},E_J,E_C)
\xrightarrow{\mathrm{dynamics}}
\mathcal{M}_{device}
\]

where $\mathcal{M}_{device}$ includes metrics such as frequency, anharmonicity, coupling matrices, Purcell limits, readout parameters, and noise sensitivity. The role of SPICE-Q is to add standardized interfaces and manufacturing feedback to this pipeline, rather than requiring all physical layers to be solved simultaneously in the same numerical solver.

\textbf{(5) Engineering significance of the solver layer}

The essence of the SPICE-Q solver system is to extend traditional "single-domain numerical simulation" into a "cross-physical-domain co-optimization system". By unifying HFSS-like electromagnetic solvers and Hamiltonian quantum solvers within a single framework, SPICE-Q can directly predict the impact of manufacturing errors on quantum performance at the design stage, thereby significantly reducing experimental iteration cost and improving the manufacturability and scalability of large-scale quantum chips.

\section{SPICE-Q Device-Level Models}

\begin{quote}
Literature index: The physical foundations of device-level modeling in this chapter come from reviews of Josephson junctions/transmons, cQED, and superconducting-qubit engineering[7][8][9][12]; the mapping from device parameters to Hamiltonian quantities through energy participation ratios and circuit quantization follows the EPR and SQcircuit literature[27][28]; the background on manufacturability, material loss, and cross-wafer statistics follows the work of Verjauw, Place, Van Damme, and others[19][25][29].
\end{quote}

Device-level modeling in SPICE-Q is the key bridge connecting physical-space simulation with information-space simulation. Its core objective is to map manufacturable micro- and nanoscale structural parameters, such as geometric dimensions, material parameters, and process variations, into computable quantum-Hamiltonian parameters, thereby realizing a unified chain of "device $\rightarrow$ electromagnetic response $\rightarrow$ circuit quantization $\rightarrow$ quantum dynamics." At this level, each device is not only a circuit element, but also a coupled multiphysics system: Josephson junctions determine nonlinearity and frequency sensitivity; resonators and couplers determine readout, coupling, and crosstalk paths; routing and 3D interconnects determine package modes, parasitic coupling, and system scalability. The value of device-level models does not lie in replacing HFSS, EPR, or Hamiltonian solvers, but in organizing the geometric, electromagnetic, quantum, and manufacturing-statistical parameters produced by these tools into traceable and calibratable design objects.

\subsection{Josephson Junction}

\begin{quote}
Literature index: The JJ current-phase relation, RCSJ/effective circuits, the $E_J$-$I_c$ mapping, and process sensitivity mainly follow Koch, Krantz, Kjaergaard, and Verjauw[7][8][12][25]; quantization from device parameters to effective Hamiltonians and energy-participation-ratio mappings follow the EPR and SQcircuit literature[27][28].
\end{quote}

The Josephson Junction (JJ) is the most fundamental and most critical nonlinear element in the SPICE-Q device layer. Its physical origin is the Cooper-pair tunneling effect in a superconductor-insulator-superconductor structure. In the SPICE-Q framework, a JJ is not merely a circuit element, but a core mapping node that connects electromagnetic space with quantum-information space; its critical current, capacitance, parasitic environment, and process discreteness jointly determine qubit frequency, anharmonicity, coupling strength, and the probability of frequency collisions[7][8][12][25].

\textbf{(1) Electromagnetic Environment Modeling and Geometric Dependence}

At the device-physics level, the nonlinearity of a JJ is mainly determined by the Josephson relation, but its effective capacitance, surrounding metal structures, shunt capacitor, package boundaries, and readout/control lines jointly affect the final qubit frequency and coupling parameters. Therefore, in SPICE-Q, the JJ should not be modeled in isolation, but should enter the parameter-extraction workflow together with its local electromagnetic environment. Three-dimensional electromagnetic simulation is mainly used to extract total capacitance, parasitic inductance, energy participation ratios, and fringe-field distributions, rather than to replace the quantum nonlinear model of the JJ itself.

The local electromagnetic problem can be constrained by Maxwell's equations and material boundary conditions:

\[
\nabla \times \mathbf{E} = -\frac{\partial \mathbf{B}}{\partial t}, \quad
\nabla \times \mathbf{H} = \mathbf{J} + \frac{\partial \mathbf{D}}{\partial t}
\]

By solving this set of equations, one can obtain the total capacitance $C_\Sigma$ and parasitic coupling parameters contributed by the JJ and its neighboring structures, thereby affecting the transition frequency under the transmon approximation:

\[
\omega_{01} \approx \frac{\sqrt{8E_JE_C}-E_C}{\hbar}, \quad E_C = \frac{e^2}{2C_\Sigma}, \quad E_J=\frac{\Phi_0 I_c}{2\pi}
\]

Thus, a JJ is not an isolated component, but a coupled system that strongly depends on surrounding wiring, capacitor pads, substrates, and three-dimensional package structures.

\textbf{(2) JJ as a Basic PCell Unit}

In the hierarchical design system of SPICE-Q, the JJ is a basic primitive for constructing parameterized cells (PCells). By parameterizing the junction area, oxide-layer thickness, critical-current density, material stack, and local parasitic environment, reusable design cells can be generated for higher-level circuit construction. This process can be abstracted as a mapping function:

\[
\text{PCell}_{JJ}=\mathcal{F}(A,d,\epsilon_r,J_c,C_\Sigma,\sigma_{I_c})
\]

where $A$ denotes the junction area, $d$ denotes the oxide-layer thickness, $J_c$ denotes the critical-current density per unit area, $C_\Sigma$ denotes the total capacitance including the junction capacitance and surrounding parasitic capacitance, and $\sigma_{I_c}$ denotes manufacturing-statistical discreteness. This parameterized structure allows the JJ to serve as a programmable basic module in an EDA system while preserving traceability to fabrication batches and cryogenic measurement data.

\textbf{(3) Fixed JJs, SQUID Structures, and Frequency Tunability}

In engineering implementations, rather than simply distinguishing between "fixed JJs" and "tunable JJs," a more accurate statement is that a single JJ usually provides a fixed Josephson energy, whereas a SQUID loop composed of two JJs can tune the effective Josephson energy through an externally applied magnetic flux. For an approximately symmetric SQUID, the effective critical current can be written as:

\[
I_c(\Phi) \approx I_{c,\max}\left|\cos\left(\frac{\pi\Phi}{\Phi_0}\right)\right|
\]

This structure allows the qubit frequency or equivalent coupler parameters to be tuned, thereby supporting frequency avoidance, tunable coupling, and specific two-qubit gate operations. However, flux tunability also introduces sensitivity to magnetic noise. Therefore, a SPICE-Q model must simultaneously record the tunable range, sweet-spot location, flux-line crosstalk, and decoherence cost[7][8][14].

\textbf{(4) Area Dependence and Process-Control Issues}

The key process variables of a JJ are the junction area $A$ and the critical-current density $J_c$, which jointly determine the critical current:

\[
I_c = J_c A
\]

Under the approximation of fixed $E_C$, the first-order sensitivity of $\omega_{01}$ to $I_c$ mainly comes from $E_J\propto I_c$. Neglecting anharmonicity corrections, it can be approximately written as:

\[
\frac{\delta\omega_{01}}{\omega_{01}} \approx \frac{1}{4}\frac{\delta I_c}{I_c}
\]

This relation shows that statistical fluctuations in JJ area, oxide-layer thickness, and critical-current density are directly mapped into the qubit frequency distribution. In large-scale fixed-frequency processors, this frequency discreteness increases the probability of frequency collisions and is a key bottleneck for yield optimization and frequency allocation[12][15][25].

\textbf{(5) Weak Capacitance and High-Precision Modeling of Parasitic Effects}

In SPICE-Q, fF-level capacitance variations in a JJ and its neighboring structures must be modeled with high precision, because they change $E_C$, anharmonicity, and coupling strength. The equivalent capacitance is usually expressed as:

\[
C_\Sigma = C_J + C_{shunt} + C_{parasitic}
\]

where $C_J$ comes from the junction itself, $C_{shunt}$ comes from the designed capacitor pad, and $C_{parasitic}$ comes from wiring, substrates, package boundaries, and neighboring devices. Errors in this term further affect the quantum Hamiltonian:

\[
\mathcal{H}=4E_C(\hat n-n_g)^2-E_J\cos\hat\phi
\]

Therefore, accurate extraction of JJ capacitance is not a single geometric calculation problem, but a problem jointly calibrated by EM simulation, EPR/circuit quantization, and cryogenic frequency measurement[27][28].

\textbf{(6) Engineering Significance of JJ Simulation}

The core value of JJ simulation is to establish a full-chain mapping from micro- and nanoscale structures to quantum behavior, allowing process parameters to directly enter quantum system-level simulation frameworks. This capability upgrades the JJ from a single device into a "cross-physical-domain interface unit" and makes it a foundational node in the multiscale modeling system of SPICE-Q.

\subsection{Resonator}

\begin{quote}
Literature index: CPW/lumped resonators, readout resonators, cQED coupling, the Purcell effect, and resonator parameter extraction follow Blais 2004/2021, Wallraff 2004, Pozar, and Krantz[7][9][10][11][21]; the mapping from EM modes to Hamiltonian parameters follows the EPR and SQuADDS workflows[26][27].
\end{quote}

The resonator is a key intermediate structure in the SPICE-Q device layer that connects microwave physical space with quantum-information space. Its core functions include qubit readout, quantum-state storage, frequency multiplexing, and energy exchange with control lines. In superconducting quantum circuits, resonators typically operate in the GHz band. Their behavior is jointly determined by electromagnetic boundary conditions, dielectric loss, port coupling, and geometry, and therefore requires a combined description using full-wave/quasistatic EM simulation, equivalent-circuit extraction, and cQED Hamiltonians[9][10][11][21].

\textbf{(1) EM Modeling and Electromagnetic Eigenmode Analysis}

In the physical-space simulation of SPICE-Q, a resonator is first modeled using HFSS, Sonnet, or a similar EM solver to extract its eigenfrequencies, field distributions, port responses, and loss participation ratios. The basic governing equations remain Maxwell's equations:

\[
\nabla \times \mathbf{E} = -\frac{\partial \mathbf{B}}{\partial t}, \quad
\nabla \times \mathbf{H} = \mathbf{J} + \frac{\partial \mathbf{D}}{\partial t}
\]

By imposing material, metal, port, and package boundary conditions, one can obtain the eigenfrequency $\omega_r$ of the resonator and the field-distribution mode $\mathbf{E}_n(\mathbf{r})$. For a lumped resonator, the ideal frequency can be written as:

\[
\omega_r = \frac{1}{\sqrt{LC}}
\]

For distributed structures, such as CPW resonators, the frequency further depends on geometric length, boundary conditions, and the effective dielectric constant.

\textbf{(2) CPW and Lumped-Element Resonator Models}

In practical devices, resonators usually include two classes: coplanar waveguide resonators (CPW resonators) and lumped resonators. The resonant frequency of a CPW half-wave resonator can be approximately expressed as:

\[
f_n = \frac{n v_p}{2L}, \quad v_p = \frac{c}{\sqrt{\epsilon_{eff}}}
\]

where $\epsilon_{eff}$ is the effective dielectric constant. If a quarter-wavelength resonator is used, the frequency condition must be correspondingly rewritten in terms of odd modes. A lumped resonator, by contrast, can be modeled as an LC oscillator system and is more suitable for compact designs, high-density readout, and local coupling structures. SPICE-Q must record geometric parameters, boundary conditions, and applicable approximations simultaneously, avoiding the mistaken representation of all resonators as the same LC or CPW model.

\textbf{(3) Mapping from Frequency to Geometric Parameters}

One of the key capabilities of SPICE-Q is to establish a computable mapping function of "geometric parameters $\rightarrow$ resonant frequency/impedance/coupling rate," thereby supporting large-scale frequency planning, readout multiplexing, and frequency collision avoidance:

\[
\omega_r = \mathcal{F}(L,w,g,\epsilon_r,t,C_{parasitic},Z_0,\text{boundary})
\]

where $L$ is the resonator length, $w$ is the line width, $g$ is the gap, $\epsilon_r$ is the dielectric constant, $t$ is the metal/dielectric thickness, $C_{parasitic}$ is the parasitic capacitance, and $Z_0$ is the target impedance. This mapping transforms the frequency-allocation problem from empirical design into a constrained optimization problem.

\textbf{(4) Parameter Extraction and Equivalent-Model Construction}

Through EM simulation or experimental measurement, key resonator parameters can be extracted, including the internal quality factor $Q_i$, external quality factor $Q_c$, total linewidth $\kappa$, effective impedance $Z_{eff}$, coupling strength $g$, and loss participation ratios. Common relations are:

\[
Q = \frac{\omega_r}{\kappa}, \quad \kappa=\kappa_i+\kappa_c
\]

where $\kappa_i$ describes internal loss, and $\kappa_c$ describes coupling to the readout line or external port. These parameters are subsequently mapped into equivalent elements in the SPICE-Q netlist and into cQED Hamiltonian parameters, thereby realizing conversion from an electromagnetic model to a quantum model[26][27].

\textbf{(5) Role of Resonators in Quantum Systems}

At the quantum-system level, resonators are not only used for readout, but can also serve as media for information exchange between qubits. Their coupling to qubits can be described by the cQED model:

\[
\mathcal{H}_{int}=g(a^\dagger\sigma^-+a\sigma^+)
\]

Under dispersive readout conditions, the resonator frequency shifts depending on the qubit state and can be approximately written as $\omega_r\rightarrow\omega_r\pm\chi$. At the same time, coupling to the readout line also introduces a Purcell limit, causing the qubit to radiate into the external environment through the resonator. Therefore, the resonator model in SPICE-Q must simultaneously serve readout speed, readout fidelity, Purcell protection, crosstalk control, and frequency-multiplexing design[7][9][10].

\textbf{(6) Engineering Significance: Frequency Engineering in Large-Scale Systems}

In large-scale quantum chips, hundreds to thousands of resonators must avoid frequency overlap, coupling to package modes, and congestion in the readout chain. By establishing a unified mapping of "geometry $\rightarrow$ frequency $\rightarrow$ impedance/coupling $\rightarrow$ readout metrics," SPICE-Q makes frequency planning a computable problem:

\[
\min \sum_{i \neq j} \mathbb{I}(|\omega_i-\omega_j|<\Delta)
\]

where $\Delta$ is the frequency safety spacing. Therefore, resonator modeling is not only a device-simulation problem, but also one of the core constraints in system-level scalability design.

\subsection{Coupler}

\begin{quote}
Literature index: Capacitive/inductive/tunable couplers, SQUID modulation, crosstalk, and coupling-matrix design follow Kjaergaard, Krantz, Blais, and Versluis[7][8][9][14]; multi-qubit frequency allocation and coupling-graph optimization follow the work of Morvan and others[15].
\end{quote}

The coupler is the core device in the SPICE-Q system for realizing information exchange and entanglement generation between qubits. Its role is to establish a controllable electromagnetic coupling path in physical space and to correspond, in information space, to interaction terms in the Hamiltonian. In superconducting quantum chips, couplers typically operate in an mK cryogenic environment, and their behavior is jointly affected by quantum fluctuations, residual material loss, microwave-mode structure, package modes, and control-line crosstalk[7][8][9][14].

\textbf{(1) Physical Characteristics of Couplers at Cryogenic Temperatures}

In a cryogenic superconducting operating environment, the electromagnetic response of a coupler can be approximately described using low-loss superconducting boundary conditions. However, practical devices still contain nonideal factors such as dielectric loss, surface loss, quasiparticles, flux noise, and control-line leakage. Therefore, a coupler cannot be modeled only as an ideal mutual capacitance or mutual inductance; it must also record residual loss terms $\kappa$, parasitic coupling paths, and their impact on coherence time. Its local electromagnetic field is still constrained by Maxwell's equations and port boundary conditions:

\[
\nabla \times \mathbf{E} = -\frac{\partial \mathbf{B}}{\partial t}, \quad
\nabla \times \mathbf{H} = \mathbf{J}+\frac{\partial \mathbf{D}}{\partial t}
\]

\textbf{(2) Capacitive Coupling and Inductive Coupling}

Capacitive coupling realizes interactions between qubits through electric-field energy exchange. Its strength is jointly related to the mutual capacitance $C_{12}$, the total capacitances of the respective qubits, and the mode frequencies, and can be roughly written as:

\[
g_{cap} \propto \frac{C_{12}}{\sqrt{C_1C_2}}
\]

Inductive coupling realizes interactions through shared magnetic flux, and its strength is related to the mutual inductance $M$ and local inductances:

\[
g_{ind} \propto \frac{M}{\sqrt{L_1L_2}}
\]

These expressions represent only first-order engineering intuition. In an actual SPICE-Q model, coupling strengths should be jointly determined by three-dimensional EM simulation, equivalent-circuit extraction, and Hamiltonian quantization. Capacitive coupling is sensitive to neighboring metals, substrates, and package modes, whereas inductive coupling is usually more sensitive to external magnetic noise. Therefore, the two types of coupling have different boundaries in noise budgeting and layout constraints.

\textbf{(3) Tunable Coupler}

A tunable coupler realizes dynamic adjustment of coupling strength by introducing a SQUID, a tunable intermediate qubit, or a tunable resonator structure. For practical tunable couplers, the coupling cannot always be fully described by a single $g_0\cos(\pi\Phi/\Phi_0)$. A more common engineering abstraction is to write the effective coupling as a combination of direct coupling and coupling mediated through an intermediate mode, for example:

\[
g_{eff}(\Phi) \approx g_{12}+g_{1c}g_{2c}\,\mathcal{F}(\omega_1,\omega_2,\omega_c(\Phi))
\]

where $\omega_c(\Phi)$ denotes the frequency of the tunable intermediate mode, and $g_{1c}$ and $g_{2c}$ denote the coupling strengths from the qubits to the coupler. This structure allows quantum-gate operations to switch between effective "on/off coupling" regions, thereby reducing idle crosstalk and improving gate fidelity. However, it also introduces flux noise, control-line crosstalk, and calibration complexity. Therefore, the tunable range, residual coupling in the off state, and noise sensitivity must be explicitly recorded in the SPICE-Q model[8][14].

\textbf{(4) Coupler Modeling from the SPICE-Q Perspective}

In conventional SPICE, couplers are usually represented only by equivalent capacitances or inductances. In SPICE-Q, they should be extended into cross-domain coupling units. Physical space records S-parameters, mutual capacitance/mutual inductance, port isolation, and parasitic modes obtained from EM solution; information space records exchange coupling, ZZ coupling, drive-dependent terms, and residual crosstalk in the Hamiltonian; noise space records decoherence, flux noise, control-line leakage, and random parameter drift. A unified expression can be written as:

\[
\mathcal{H}_{int}=\sum_{i,j}g_{ij}(\mathbf{r},\Phi)(a_i^\dagger a_j+a_j^\dagger a_i)+\sum_{i,j}\zeta_{ij}n_in_j
\]

where $g_{ij}$ describes exchange-type coupling, and $\zeta_{ij}$ can represent residual ZZ or frequency-dependent coupling terms. This expression uniformly maps geometric structures, material parameters, and quantum dynamics.

\textbf{(5) Engineering Significance: Coupling-Matrix Design in Large-Scale Systems}

In large-scale quantum chips, couplers form a sparse but highly structured coupling matrix $\mathbf{G}$. System-level optimization should not simply minimize all $g_{ij}$ values, but should balance sufficiently strong target coupling, sufficiently weak non-target coupling, satisfiable frequency avoidance, and acceptable control complexity:

\[
\min \left(\sum_{(i,j)\notin E}|g_{ij}|^2+\lambda_1\text{crosstalk}+\lambda_2\text{calibration cost}\right)
\]

Through joint EM extraction and Hamiltonian simulation, SPICE-Q realizes the traceable generation of "coupling geometry $\rightarrow$ parameterized coupling matrix $\rightarrow$ quantum logic connectivity graph," thereby supporting scalable quantum-architecture design.

\subsection{Microwave Routing and 3D Interconnect}

\begin{quote}
Literature index: The effects of microwave routing, airbridges, TSVs, flip-chip, and 3D packaging on coherence/crosstalk follow Rosenberg, Vahidpour, Dunsworth, Pozar, and Wendin[16][17][18][20][21]; the requirements for scalable surface-code control and 3D integration follow the work of Versluis and others[14].
\end{quote}

In the SPICE-Q framework, Microwave Routing is not only a physical wiring problem that connects different quantum devices, but also a strongly coupled multiphysics optimization problem. Its core objective is to design high-density signal-transmission paths that are low-loss, low-crosstalk, packageable, and testable while maintaining quantum coherence. In superconducting quantum chips, routing structures change port impedance, package modes, parasitic coupling networks, and decoherence channels. Therefore, they must be modeled in a unified manner through SPICE-Q[16][17][18][20][21].

\textbf{(1) Microwave Routing}

The propagation of microwave signals on a chip is usually a distributed electromagnetic-wave problem. Its propagation loss can be expressed as:

\[
\alpha(\omega)=\alpha_c+\alpha_d+\alpha_r
\]

where $\alpha_c$ is conductor loss, $\alpha_d$ is dielectric loss, and $\alpha_r$ is radiation loss. In SPICE-Q, this path is not determined only by geometry; it is also affected by neighboring quantum devices, electromagnetic packaging, and control/readout ports. Therefore, EM simulation, S-parameter extraction, and system-level coupling-matrix analysis must be combined.

\textbf{(2) Airbridge Structure Design}

Airbridges are used to cross ground-discontinuity regions in CPW structures, suppress slotline modes, reduce parasitic-mode excitation, and improve ground continuity. Their effect can be approximately represented by changes in port response:

\[
\Delta S_{ij}=S_{ij}^{with\ bridge}-S_{ij}^{without\ bridge}
\]

In SPICE-Q, an airbridge is not only a structural compensation unit, but also an important design variable for controlling high-frequency mode leakage, local parasitic capacitance, and cross-line crosstalk. Its geometry, material, and position need to be jointly optimized with resonators, readout lines, and package modes[18].

\textbf{(3) TSV Routing (Through-Silicon Via Routing)}

TSVs are used to realize vertical interconnects in three-dimensional quantum chips and are important candidate technologies for high-density wiring and modular packaging. Their electromagnetic characteristics can be represented by a frequency-dependent equivalent network:

\[
Z_{TSV}(\omega)=R(\omega)+j\omega L(\omega)+\frac{1}{j\omega C(\omega)}
\]

where parasitic inductance and parasitic capacitance significantly affect signal integrity, port isolation, and package-mode coupling in the GHz band. SPICE-Q performs system-level modeling of crosstalk paths introduced by TSVs through coupled electromagnetic-circuit simulation[17].

\textbf{(4) Flip-Chip Routing}

Flip-chip structures realize connections between upper and lower chips through bump bonding. Their main advantage lies in high-density interconnects and modular design capability. However, this structure introduces additional parasitic capacitance, interface loss, mode conversion, and local current crowding:

\[
C_{bump}\propto\frac{\epsilon A}{d}
\]

where $A$ is the contact area and $d$ is the spacing. In SPICE-Q, flip-chip interconnects should be modeled as cross-layer coupling edges. Their influence is not merely an additional capacitance, but also includes changes in S-parameters, package-cavity modes, readout-chain loss, and possible corrections to Hamiltonian parasitic coupling[16].

\textbf{(5) System-Level Optimization Problem for Routing}

In large-scale quantum chips, routing is essentially a multi-objective optimization problem, whose objective function can be expressed as:

\[
\min \left(\sum\alpha_{loss}+\lambda_1\text{crosstalk}+\lambda_2\text{mode\ crowding}+\lambda_3\text{calibration\ complexity}\right)
\]

By combining electromagnetic propagation models in physical space with coupling matrices in information space, SPICE-Q realizes a closed-loop design of "routing $\rightarrow$ S-parameters/parasitic networks $\rightarrow$ Hamiltonian correction $\rightarrow$ system-performance prediction."

\textbf{(6) Engineering Significance}

Routing is no longer a back-end auxiliary issue in SPICE-Q, but a core constraint layer that determines the scalability of quantum chips. By uniformly modeling microwave propagation, structural parasitic effects, package modes, and quantum-coupling relations, SPICE-Q transforms routing from empirical design into a computable optimization problem, providing a foundation for interconnects, readout multiplexing, and package co-design in larger-scale superconducting quantum processors.

\subsection{Single-Qubit Models}

\begin{quote}
Literature index: Transmon, Xmon, flux/tunable qubits, two-level/multilevel Hamiltonians, and coherence metrics follow Koch, Barends, Krantz, Kjaergaard, and Place[7][8][12][13][19]; circuit quantization and workflows for solving arbitrary superconducting circuits follow the EPR and SQcircuit literature[27][28].
\end{quote}

The single qubit is the most fundamental information-processing unit in the SPICE-Q system. In essence, it is a quantum anharmonic oscillator system jointly defined by Josephson nonlinearity and electromagnetic constraints. At this level, the system is mapped from physical space, including geometric structures, electromagnetic field distributions, material loss, and port boundaries, into information space, including quantum-state evolution, effective Hamiltonians, and open-system dynamics, thereby realizing the abstraction from "device $\rightarrow$ qubit"[7][8][12].

\textbf{(1) General Qubit Hamiltonian Description}

The low-energy computational subspace of a single qubit is commonly described by an effective two-level Hamiltonian:

\[
\mathcal{H}=\frac{1}{2}\hbar\omega_q\sigma_z+\mathcal{H}_{drive}+\mathcal{H}_{noise}
\]

where $\omega_q$ is the qubit eigenfrequency, $\mathcal{H}_{drive}$ represents the external microwave drive, and $\mathcal{H}_{noise}$ represents environmental noise and decoherence terms. In SPICE-Q, these parameters are not abstractly specified; they are jointly determined by the JJ, resonator, coupling structure, routing environment, and cryogenic-measurement feedback.

\textbf{(2) Transmon Qubit}

The transmon is currently the most widely used superconducting-qubit architecture. Its Hamiltonian is:

\[
\mathcal{H}_{transmon}=4E_C(\hat n-n_g)^2-E_J\cos\hat\phi
\]

where $E_C$ is the charging energy and $E_J$ is the Josephson energy. Its energy-level structure is approximately that of a weakly anharmonic oscillator. If $E_J$ and $E_C$ are expressed in energy units, the transition angular frequency can be written as:

\[
\omega_{01}\approx\frac{\sqrt{8E_JE_C}-E_C}{\hbar}
\]

The core advantage of the transmon is that it substantially reduces sensitivity to charge noise in the $E_J/E_C\gg1$ regime, but this robustness comes at the cost of reduced anharmonicity. Therefore, the SPICE-Q model must simultaneously record frequency, anharmonicity, drive leakage, $T_1/T_2$, and material/interface loss parameters[7][12][19].

\textbf{(3) Xmon and Planarized Transmon Variants}

Xmon is a planar geometric variant of the transmon. By using a cross-shaped or multi-arm capacitor structure, it improves the designability of coupling ports and enhances wiring and coupling-control capability in two-dimensional arrays[13]. In SPICE-Q, Xmon should not merely be regarded as a differently named qubit, but should be modeled as a quantum node with a direction-dependent coupling matrix and port definitions:

\[
\mathcal{H}_{Xmon}=\sum_i g_i(\mathbf{r})(a_i^\dagger\sigma^-+a_i\sigma^+)+\mathcal{H}_{transmon}
\]

This structure emphasizes geometric controllability, clarity of coupling ports, and array scalability in scalable planar quantum chips.

\textbf{(4) Flux Qubit and Tunable Qubits}

A flux qubit is based on loop-shaped superconducting current states, and its effective Hamiltonian depends on the external magnetic flux:

\[
\mathcal{H}_{flux}=-\frac{1}{2}\Delta\sigma_x-\frac{1}{2}\epsilon(\Phi)\sigma_z, \quad \epsilon(\Phi)\propto\Phi-\Phi_0/2
\]

The advantages of a flux qubit lie in strong nonlinearity and tunability, but it is highly sensitive to magnetic noise. Therefore, environmental flux noise and control-line crosstalk models must be explicitly included in SPICE-Q. A tunable transmon usually realizes frequency tuning by replacing a single JJ with a SQUID, whose effective Josephson energy varies with magnetic flux. Under the symmetric SQUID approximation, the frequency can be roughly expressed as:

\[
\omega_q(\Phi)\approx\frac{\sqrt{8E_CE_J(\Phi)}-E_C}{\hbar}, \quad E_J(\Phi)\propto \left|\cos\left(\pi\Phi/\Phi_0\right)\right|
\]

This structure allows dynamic avoidance of frequency collisions and supports partially controllable two-qubit gate operations, but it also brings flux noise, frequency drift, and calibration complexity.

\textbf{(5) Information-Space Simulation: From Devices to Quantum Dynamics}

In the SPICE-Q framework, single-qubit simulation no longer relies on isolated empirical parameters, but is realized through the following chain:

\[
\text{Geometry} \rightarrow \text{EM fields} \rightarrow \text{Circuit parameters} \rightarrow \mathcal{H}_{eff} \rightarrow \rho(t)
\]

For an open system, the density-matrix evolution can be written as:

\[
\frac{d\rho}{dt}=-\frac{i}{\hbar}[\mathcal{H},\rho]+\mathcal{L}(\rho)
\]

where $\mathcal{L}(\rho)$ denotes the Lindblad dissipation term. In practical solution, the two-level, three-level, or multilevel truncation must also be selected according to the device anharmonicity. For strong driving, leakage errors, or readout processes, a simple two-level model is often insufficient, and higher energy levels together with measurement-chain parameters must be retained[27][28].

\textbf{(6) Engineering Significance: The Single Qubit as a SPICE-Q Primitive}

The single qubit is the smallest computable information unit in the entire SPICE-Q system. Its parameter accuracy directly affects gate fidelity, coherence time, readout fidelity, multi-qubit frequency allocation, and the stability of array scaling. Therefore, through accurate modeling of the single qubit, SPICE-Q establishes a unified foundation from device physics to system-level quantum computation.

\section{Standardized Manufacturing System}

\begin{quote}
Literature index: The discussion in this chapter on process repeatability, manufacturable workflows, material/interface loss, and 3D packaging constraints follows Verjauw, Place, Rosenberg, Vahidpour, Dunsworth, and Krantz[7][16][17][18][19][25]; 300 mm CMOS-compatible superconducting-qubit fabrication and cross-wafer statistics follow the Nature 2024 work by Van Damme et al.[29]; modeling of device parameters, frequency, coherence, and manufacturing feedback follows Koch, Morvan, and related reviews of superconducting-qubit engineering[8][12][15].
\end{quote}

The construction of the SPICE-Q framework depends on highly repeatable process stability. In recent years, superconducting qubits have begun to move away from laboratory-style double-angle evaporation and lift-off processes toward CMOS-compatible workflows that are closer to the semiconductor industry. Verjauw et al. demonstrated a manufacturable overlap Josephson-junction route[25]; Van Damme et al. further fabricated transmon qubits on a 300 mm CMOS pilot line and reported cross-wafer statistics, yield, variability, and aging data[29]. These results indicate that the central issue in quantum-chip fabrication is shifting from "whether a single high-coherence device can be made" to "whether large numbers of devices can be repeatedly fabricated in a statistically controllable manner."

Against this background, the "standardized fabrication factory" referred to in this paper is not an already fully mature industry standard, but an engineering abstraction of a quantum-chip manufacturing system. It requires that equipment states, process parameters, material batches, environmental perturbations, cryogenic test results, and layout design variables be recorded in a unified manner and fed back into SPICE-Q/PDK models. In other words, the PDK provides manufacturability constraints, SPICE-Q provides simulation and error-propagation models, and factory data provides the statistical evidence required for continuous calibration.

A standardized manufacturing system can be understood as a combination of five categories of capabilities. The first is the standardization of process and design platforms: through the PDK system, process layers, material parameters, design rules, device models, and statistical corners are unified, abstracting manufacturing capability into a computable design space. The second is consistency of equipment states and calibration workflows: by recording equipment models, key process chambers, maintenance states, calibration curves, and operating histories, statistical comparability of behavior across tools and batches is achieved. The third is the standardization of operating procedures and automated execution: SOPs, automated scheduling, and event logs reduce uncertainty from human operation and transform process execution into a traceable and replayable data stream. The fourth is the standardization of data and communication protocols: unified data formats, sensor interfaces, process logs, and cryogenic-test data structures make the entire factory an observable system, supporting model calibration, anomaly tracing, and closed-loop control. The fifth is consistency control of environmental and material inputs: by recording and controlling temperature, humidity, vibration, electromagnetic environment, particle contamination, and material batches, external disturbances are incorporated into measurable boundary conditions rather than assumed to be completely negligible.

The five aspects above jointly constitute the basic architecture of a semiconductor "yield-controllable system." When SPICE-Q is extended to quantum-chip manufacturing, this system must further introduce cross-domain data associations among cryogenic measurement, microwave loss, quantum coherence, and process statistics, so that the manufacturing system is not merely the physical site that produces chips, but also a data source that continuously updates device models, error models, and yield models.

\subsection{Standardization of Each Process Run}

\begin{quote}
Literature index: Process repeatability, observability of process variables, and manufacturing feedback loops follow Verjauw's CMOS-compatible qubit fabrication route, Van Damme's 300 mm cross-wafer statistics, and the engineering reviews by Krantz/Kjaergaard[7][8][25][29].
\end{quote}

In the standardized factory system on which SPICE-Q relies, the stability of a single process run is a core factor determining the convergence of device-parameter distributions. Because quantum devices, especially JJs and resonators, are highly sensitive to nanometer-scale geometric variations, interface states, and material loss, small inconsistencies in any process step may appear at the system level as frequency drift, coupling mismatch, or broadening of coherence-time distributions. Therefore, "each process run" must be abstracted into a standardized state-transition process that is computable, traceable, and replayable[7][8][25][29].

\textbf{(1) Process as a State-Transition System}

The entire fabrication process can be formalized as a sequence of state transitions:

\[
S_{t+1}=\mathcal{F}(S_t,u_t,\xi_t)
\]

where $S_t$ denotes the current process state, $u_t$ denotes human or mechanical operation inputs, $\xi_t$ denotes environmental perturbations, such as temperature, particles, vibration, electromagnetic background, and material-batch differences, and $\mathcal{F}$ denotes the process-evolution function. The goal of a standardized factory is not to assume that $\xi_t$ completely disappears, but to control it within a range that is measurable, statistical, and feedback-capable, while constraining the batch-to-batch repeatability of $\mathcal{F}$.

\textbf{(2) Standardization of Human Operations}

Human operation is one of the important sources of uncertainty in conventional processes. In the manufacturing system corresponding to SPICE-Q, human behavior needs to be discretized into a set of executable instructions:

\[
u_{human}\rightarrow\{op_1,op_2,\ldots,op_n\}
\]

Each operation should be bound to a timestamp, location coordinate, equipment state, material batch, and operator/automation-system information, thereby realizing a digital mapping of behavior. Human-operation error can be modeled as a random perturbation:

\[
\delta u_{human}\sim\mathcal{N}(0,\sigma^2)
\]

By reducing $\sigma$ through SOPs, automation assistance, training records, and anomaly logs, overall process consistency can be improved.

\textbf{(3) Standardization of Mechanical Operations and Equipment Behavior}

Mechanical operations, namely robotic or automated tool actions, must be deterministically described through the control system, for example:

\[
u_{machine}=\pi(x_t)
\]

where $\pi$ is the control-policy function and $x_t$ is the equipment-state vector. In the SPICE-Q system, each tool is not only an execution unit, but also a modelable dynamic system. Its temperature, pressure, power, chamber state, maintenance history, and sensor drift should all enter the process simulation model as sources of output error.

\textbf{(4) Standardization of Physical Transfer and Trajectory Paths}

During wafer handling and process transfer, the physical path, or trajectory planning, affects particle contamination, vibration perturbations, and ESD risk. Its trajectory optimization can be expressed as:

\[
\min\int_0^T\left(\alpha v(t)^2+\beta a(t)^2+\gamma j(t)^2\right)dt
\]

where $v(t)$ is velocity, $a(t)$ is acceleration, and $j(t)$ is jerk. By limiting kinematic parameters, the influence of micro-vibration and impact on JJs, thin films, package structures, and nanoscale patterns can be reduced.

\textbf{(5) Full-Process Datafication and Observability}

SPICE-Q requires each process step to be mapped into a data event stream:

\[
\mathcal{D}=\{(t_i,S_i,u_i,\xi_i,m_i)\}_{i=1}^N
\]

where $m_i$ denotes an observable process-measurement value or quality-inspection result. This data structure makes the entire factory an observable factory, supporting process replay, error tracing, and inverse estimation of model parameters. At the same time, this data stream can be used to calibrate SPICE-Q model parameters in reverse, realizing DTCO closed-loop optimization.

\textbf{(6) Engineering Significance: From Process Execution to Computable Manufacturing}

Through the standardization described above, a single process run is no longer a collection of empirical operations, but a computable system. The fabrication process is transformed into a state-space model, human and mechanical operations are transformed into control variables, error sources are transformed into statistical perturbation models, and the factory becomes an object that can be simulated, replayed, and calibrated. Therefore, this standardized workflow is the foundational layer for SPICE-Q to realize the closed loop of "manufacturing $\rightarrow$ simulation $\rightarrow$ design."

\subsection{Standardization of Each Operation Step and Tool}

\begin{quote}
Literature index: Standardization of equipment behavior, critical process windows, and the sensitivity of quantum devices to process drift follow Koch, Verjauw, Place, Van Damme, and reviews of superconducting-qubit engineering[7][8][12][19][25][29].
\end{quote}

In a SPICE-Q-driven standardized factory system, equipment is no longer regarded as an isolated execution unit, but is modeled as a "repeatable physical function." During operation, every machine should satisfy a calibratable input-state-output mapping relation, so that consistency across equipment and batches can be evaluated statistically rather than relying only on empirical judgment[25][29].

\textbf{(1) Three-Phase Behavioral Model of Equipment}

The operation of each tool can be divided into a warm-up phase, a process execution phase, and a post-processing phase. The core objective of this three-phase model is to make equipment thermal drift, mechanical drift, chamber memory effects, and sensor drift explicit and to incorporate them into the control system and process logs.

\textbf{(2) Standardized Control of the Warm-Up Phase}

The primary goal of the warm-up phase is to converge the equipment state from the initial unstable state $S_0$ to a stable operating point $S^*$:

\[
S^*=\lim_{t\rightarrow t_w}S(t)
\]

where $t_w$ is the warm-up time. In quantum-device fabrication, for example in thin-film deposition, oxidation, etching, and cryogenic measurement equipment, temperature stability, chamber background, residual atmosphere, and power drift all affect thin-film growth rate, interface roughness, JJ critical-current distributions, and material loss. Therefore, the warm-up phase should not merely be "waiting for the tool to stabilize," but should form a recordable stability criterion.

\textbf{(3) Deterministic Execution Model for the Operation Phase}

The operation phase requires the tool execution process to satisfy a calibratable input-output functional relationship:

\[
y(t)=\mathcal{G}(x(t),\theta_{device},\eta_t)
\]

where $x(t)$ denotes the input process parameters, $\theta_{device}$ denotes intrinsic or calibrated tool parameters, $\eta_t$ denotes residual tool noise and environmental perturbations, and $\mathcal{G}$ denotes the tool response function. In the SPICE-Q framework, $\mathcal{G}$ must be continuously calibrated through historical data, test structures, and cryogenic measurement results so that tool behavior maintains statistical consistency.

\textbf{(4) Post-Operation Phase and Equipment Memory Effects}

The post-operation phase is mainly used to record and eliminate residual states, including thermal hysteresis, charge accumulation, chamber contamination, mechanical stress, and material residues:

\[
S_{post}=S_{final}-\Delta S_{memory}
\]

This phase is particularly important for superconducting quantum chips, because equipment memory effects may enter the final device-parameter distribution through film quality, interface loss, or JJ critical-current drift.

\textbf{(5) Equipment Layout, Vibration Isolation, and ESD Boundaries}

The physical layout of equipment, placement plane, foundation stiffness, vibration isolation, and ESD paths should be regarded as boundary conditions of the equipment model, rather than merely construction requirements. The geometric consistency of the tool placement plane can be constrained by $\Delta h\leq\epsilon_h$, and vibration isolation can be described by

\[
m\ddot{x}+c\dot{x}+kx=F_{ext}(t)
\]

while ESD paths require a unified grounding and charge-discharge network to keep charge accumulation below safe thresholds. Together, these conditions determine whether equipment outputs can be reproduced across batches.

\textbf{(6) Summary of Engineering Significance}

The core objective of equipment standardization is to transform "machine differences" into "parameter differences" and incorporate them into a unified model. Human differences correspond to operation protocols and training records, mechanical differences correspond to equipment calibration and maintenance states, and environmental differences correspond to measurable perturbation terms. Therefore, in the SPICE-Q framework, each tool becomes a physical node that can be modeled, simulated, and calibrated, thereby supporting the statistical repeatability of quantum-chip fabrication.

\subsection{Standardization of Physical Transport}

\begin{quote}
Literature index: The potential effects of transport, packaging, 3D interconnects, and post-process handling on coherence/parasitic coupling mainly follow Rosenberg, Vahidpour, Dunsworth, and Verjauw[16][17][18][25]; the influence of material/interface contamination on coherence follows the work of Place and others[19].
\end{quote}

In the SPICE-Q standardized factory system, the physical transport process, including wafer/material transport, is regarded as a critical hidden process step affecting yield. Existing studies of superconducting quantum devices have shown that material interfaces, surface contamination, package interconnects, and multilayer wiring all affect microwave loss, parasitic coupling, and coherence[16][17][18][19][25]. Therefore, the transport process should not be treated as simple logistics, but should be explicitly modeled as a process boundary condition, with particular attention to recording vibration, particle contamination, humidity variations, ESD events, and personnel contact history.

\textbf{(1) Path Constraints and Spatial Discretization}

By establishing a digital floor routing grid inside the factory, the movement paths of personnel and equipment can be discretized into a graph structure:

\[
\mathcal{G}_{factory}=(V,E)
\]

where $V$ denotes standard workstation nodes and $E$ denotes allowed transport paths. This structure is similar to the routing-constraint graph inside a wafer fab. It can reduce contamination and contact differences caused by random walking and optimize the stability of transport paths.

\textbf{(2) Floor Structure and Vibration-Propagation Control}

Floor structure, equipment supports, and automated material-handling systems affect the propagation function of mechanical vibration:

\[
H(\omega)=\frac{X_{out}(\omega)}{X_{in}(\omega)}
\]

where $H(\omega)$ denotes the vibration transfer function. By increasing floor stiffness, introducing damping layers, and limiting velocity, acceleration, and jerk during transport actions, the impact of mechanical shock on wafers, thin films, package structures, and nanoscale patterns can be reduced.

\textbf{(3) Automated Transport and Trajectory Optimization}

Conventional manual handling can be gradually replaced by controlled automated transport systems. Their dynamics can be simplified as:

\[
m\ddot{x}+c\dot{x}=F(t)
\]

The trajectory-optimization objective function can be written as:

\[
\min\int_0^T\left(\alpha v(t)^2+\beta a(t)^2+\gamma j(t)^2\right)dt
\]

where $j(t)$ is jerk, used to control micro-vibration and impact loads. The emphasis here is not to require a specific electromagnetic or mechanical drive method, but to require transport trajectories, velocity profiles, and anomalous events to be recorded, reproduced, and associated with subsequent test data.

\textbf{(4) Packaging and Environmental Isolation}

Standardized sealed containers or controlled carriers should be used during transport to isolate particle contamination, humidity variations, and electrostatic accumulation. The packaging system can be described by the following boundary conditions:

\[
P_{particle}\rightarrow 0, \quad \Delta RH\leq\epsilon_{RH}
\]

where $RH$ denotes relative humidity. For superconducting quantum-chip manufacturing, moisture, particles, and organic residues may change surface/interface loss. Therefore, transport records need to be associated with subsequent resonator $Q$, qubit $T_1/T_2$, and frequency-drift data.

\textbf{(5) Electrostatic Control and Charge-Path Design}

The transport system must integrate ESD-safe path design so that the charge-discharge process satisfies:

\[
Q(t)\xrightarrow{t\to\infty}0
\]

Conductive flooring, ion neutralizers, unified grounding networks, and carrier-material control can ensure that wafers do not accumulate dangerous potential differences during transport. For nanoscale structures such as Josephson junctions, an ESD event may alter local defects or subsequent yield statistics even if it does not cause visible damage. It should therefore be recorded as a traceable event.

\textbf{(6) Engineering Significance: Transport as Part of the Process}

In the SPICE-Q framework, physical transport is no longer an auxiliary step, but part of the complete process chain. Its core transformation is to turn manual handling into a controlled dynamical system, spatial movement into a trajectory-optimization problem, and empirical operation into computable trajectory and event models. Therefore, standardization of physical transport is an important foundational layer for realizing end-to-end repeatable quantum-chip manufacturing.

\subsection{Standardization of Inspection and Testing}

\begin{quote}
Literature index: Cryogenic testing, $T_1/T_2$, resonant frequency, and statistical distributions of qubit parameters follow Krantz, Kjaergaard, Place, and Verjauw[7][8][19][25]; cross-wafer statistics, yield, variability, and aging data follow the work of Van Damme and others[29].
\end{quote}

In a SPICE-Q-driven standardized factory system, inspection and testing (inspection \textbackslash{}\& metrology) are defined as the key closed-loop node connecting the "manufacturing process" with "model calibration." For superconducting quantum chips, device parameters, such as the critical current $I_c$ of a Josephson junction, resonant frequency $f_r$, quality factor $Q$, coupling strength, and decoherence times $T_1/T_2$, have significant statistical-distribution characteristics. Therefore, the test system must provide cross-batch consistency, high-dimensional parameter observability, and a data structure that can be written back to SPICE-Q/PDK models[7][8][25][29].

\textbf{(1) Standardized Parameter Collection (Metrology Standardization)}

The core objective of parameter collection is to map physical measurement results into a unified data structure:

\[
\mathcal{M}:\mathcal{S}_{device}\rightarrow\mathbb{R}^n
\]

where $\mathcal{S}_{device}$ denotes the physical state of the device, and $\mathbb{R}^n$ denotes the standardized parameter space. Typical parameters include JJ critical current $I_c$, qubit frequency $f_q$, resonant frequency $f_r$, quality factors $Q_i/Q_c$, coupling strength $g$, readout parameters, $T_1/T_2$, frequency drift, and aging metrics. Unified sampling protocols can reduce systematic bias among different test equipment, batches, and analysis scripts, allowing the data to enter a comparable space.

\textbf{(2) Standardized Modeling of the Test Workflow}

The test process can be formalized as an observation function:

\[
y=\mathcal{H}(x)+\epsilon
\]

where $x$ is the true device parameter, $\mathcal{H}$ is the response function of the measurement system, and $\epsilon$ is measurement noise. SPICE-Q requires $\mathcal{H}$ to be calibratable across different equipment, namely:

\[
\mathcal{H}_1\approx\mathcal{H}_2\approx\cdots\approx\mathcal{H}_n
\]

This does not mean that all measurement systems are completely identical; rather, it requires their bias, noise floor, bandwidth, link calibration, and data-processing workflows to be recordable and transformable, thereby ensuring that data across laboratories or production lines can be fused.

\textbf{(3) Cryogenic Measurement Standardization}

For superconducting quantum chips, key performance tests usually need to be performed in a millikelvin cryogenic environment. The error sources of a cryogenic measurement system can be expressed as:

\[
\delta x=\delta x_{thermal}+\delta x_{rf}+\delta x_{vibration}+\delta x_{calibration}
\]

Each term must be suppressed and recorded through engineering measures. RF links must be calibrated to eliminate errors from standing waves, attenuation, phase drift, and gain drift:

\[
S_{21}(\omega)\rightarrow S_{21}^{calibrated}(\omega)
\]

Only after the cryogenic link and room-temperature instrument response have been standardized and recorded are $T_1/T_2$, readout fidelity, and frequency-drift data suitable for inclusion in yield models.

\textbf{(4) Test Automation and Data Consistency}

In the SPICE-Q system, the test workflow should be automated as much as possible, so that each measurement corresponds to a unique traceable data ID:

\[
\mathcal{D}_{test}=\{(id,t,device,config,recipe,result,analysis\ version)\}
\]

This structure supports construction of a process data lake for subsequent DTCO model calibration. The key here is not simply to increase data volume, but to ensure that test recipe, instrument configuration, chip location, process batch, and analysis version can all be tracked simultaneously.

\textbf{(5) Statistical Consistency and Yield Modeling}

Test data are used not only for evaluating single devices, but also for modeling overall yield:

\[
Y=P(f_r\in[f_{min},f_{max}], I_c\in\Omega, T_1>T_{th})
\]

By fitting statistical distributions, one can infer process-bias sources in reverse and thereby optimize the SPICE-Q model-parameter space. Cross-wafer and cross-batch data are particularly important because they reveal local random fluctuations, wafer-level systematic bias, batch drift, and aging behavior[29].

\textbf{(6) Summary of Engineering Significance}

Standardized inspection and testing give SPICE-Q closed-loop capability. Physical measurements are transformed into a unified parameter space, differences among test equipment are transformed into calibratable error models, and discrete experimental data are transformed into continuous statistical models, supporting the continuous updating and convergence of DTCO models. Therefore, a standardized testing system is a key bridge for SPICE-Q to move from a "simulation framework" toward an "industrial-grade quantum manufacturing system."

\subsection{Standardization of the Fabrication Environment}

\begin{quote}
Literature index: Cleanliness, interface contamination, material surface loss, and the constraints imposed by vibration/electromagnetic background on superconducting quantum devices follow Place, Verjauw, Krantz, and Dunsworth[7][18][19][25]; cross-wafer fabrication statistics and CMOS-compatible workflows follow the work of Van Damme and others[29].
\end{quote}

In the large-scale quantum-chip manufacturing system corresponding to the SPICE-Q framework, the "fabrication environment" is a bottom-level constraint that determines device consistency and process yield. Quantum devices, especially Josephson structures, dielectric interfaces, and microwave resonators, are highly sensitive to microparticle contamination, interface residues, electromagnetic background, vibration spectral density, airflow disturbances, and temperature/humidity drift. Therefore, the fabrication environment must be treated as a controlled multiphysics field system[7][19][25].

\textbf{(1) Cleanliness and Particle-Contamination Control (Cleanroom Standardization)}

One of the core metrics of the fabrication environment is particle concentration, whose constraint can be expressed as:

\[
C_{particle}(d>d_0)\leq C_{threshold}
\]

where $d_0$ is the critical particle size. Particle contamination may cause local pattern defects, dielectric-layer defects, increased interface loss, or increased microwave loss. For JJ regions, it may also affect the effective junction area or fringe-field distribution through local pattern deviation. Therefore, the cleanroom not only needs to satisfy ISO class constraints, but also needs to record particle events, cleaning states, and material exposure histories in critical process areas.

\textbf{(2) Airflow Organization and Pressure-Difference Stability (Airflow and Pressure Stability)}

Airflow organization affects particle transport and may also introduce mechanical perturbations through equipment structures, worktables, and carriers. Airflow velocity fluctuations can be expressed as:

\[
v_{air}(t)=v_0+\delta v(t)
\]

Its coupling relation to vibration sources can be approximately written as:

\[
F_{air}(t)\propto\rho v_{air}^2(t)
\]

Therefore, the fabrication environment requires low turbulence, stable laminar flow, and traceable wind-speed/pressure-difference records. For SPICE-Q, airflow is not only a cleanroom management parameter, but can also enter the process model as a boundary condition affecting particle events and vibration spectra.

\textbf{(3) Temperature Field and Humidity Stability}

Although the fabrication environment is different from the cryogenic test environment, cleanroom temperature and humidity still affect lithographic alignment, thin-film deposition rate, material stress, and surface adsorption states:

\[
\Delta T_{fab}\leq\epsilon_T, \quad \Delta RH\leq\epsilon_{RH}
\]

Temperature nonuniformity may lead to material stress gradients:

\[
\sigma\propto E\alpha\Delta T
\]

thereby affecting thin-film morphology, interface state, and subsequent device-parameter distributions. Humidity and surface adsorption may change dielectric loss or residual contamination risk, and therefore should be recorded together with material batch and process waiting time.

\textbf{(4) Electromagnetic Background Control (Electromagnetic Fabrication Shielding)}

Equipment power-supply noise, RF communication, switching power supplies, and ground loops in the fabrication environment form background electromagnetic fields:

\[
E_{env}(t)=\sum_iE_i(\omega_i,t)
\]

These signals do not necessarily directly change the Hamiltonian of fabricated devices, but they may affect sensitive measurements, bias equipment, electron-beam/lithography tool stability, or process-equipment control loops. Therefore, a more accurate statement is that electromagnetic background should be recorded and isolated as an external noise boundary of the fabrication and testing chain. Local shielding, RF isolation, low-noise power-supply zoning, unified grounding, and control of sensitive-equipment areas can be used to reduce systematic bias.

\textbf{(5) Vibration Spectral Control}

Vibration in the fabrication environment comes from equipment, airflow systems, floor conduction, and the external building environment. Its power spectral density can be expressed as:

\[
S_x(\omega)=\langle|X(\omega)|^2\rangle
\]

SPICE-Q requires vibration peaks to be recorded and suppressed in critical frequency bands, preventing them from entering device-parameter distributions through lithographic alignment, deposition uniformity, transport impact, or packaging stress. Here, attention should not be limited to a single low-frequency threshold; statistical constraints on frequency bands, amplitudes, and durations should instead be defined according to specific equipment and process steps.

\textbf{(6) Unified Multiphysics Modeling (Fabrication Environment Model)}

The fabrication environment can be uniformly represented as a perturbation vector:

\[
\xi_{fab}=f(C_{particle},T,RH,E_{env},v_{air},S_x,\text{material batch})
\]

This vector directly enters the SPICE-Q process model for predicting the broadening of device-parameter distributions. More importantly, environmental parameters need to be associated with test structures, cross-wafer statistics, and cryogenic performance data; otherwise, environmental records remain only at the factory-management level and cannot truly serve as a basis for model calibration.

\textbf{(7) Summary of Engineering Significance}

The essence of fabrication-environment standardization is to transform the "factory background" into "computable boundary conditions." Only when environmental perturbations are parameterized can process results be predicted; only when yield fluctuations are modeled can the DTCO loop converge. Therefore, in the SPICE-Q system, the fabrication environment itself constitutes part of the simulation model, rather than an external uncontrollable condition.

\subsection{Data Collection and Standardization}

\begin{quote}
Literature index: Data collection, process-parameter inversion, manufacturing feedback, and yield statistical modeling follow BSIM parameter extraction, Krantz, Morvan, and Verjauw[4][7][15][25].
\end{quote}

In the quantum-device manufacturing system corresponding to SPICE-Q, data collection and standardization constitute one of the foundational links of closed-loop design-technology co-optimization (DTCO). Unlike conventional electronic manufacturing, which mainly relies on discrete data records from sampling inspection, quantum-chip processes require high-density and continuous structured collection of every process step, every device parameter, and every environmental perturbation, in order to support subsequent model calibration and physical-parameter inversion.

This system emphasizes automated, low-contact, and traceable data acquisition paradigms. Specifically, automated robotic arms, vacuum transport systems, and cryogenic-compatible non-contact measurement interfaces are used to reduce contamination, vibration, and ESD uncertainty caused by human operation. This approach is particularly important in superconducting quantum-chip fabrication because slight contamination, vibration, or electrostatic discharge can change the critical-current distribution of Josephson Junctions and thereby affect the overall qubit frequency distribution (Krantz et al., 2019).

At the data-structure level, the SPICE-Q data platform is designed as a unified multilayer data-fusion architecture. Its core objective is to realize the mapping from "physical metrology data" to "simulation-usable parameters." Let the manufacturing observation dataset be $\mathcal{D} = \{x_i, y_i\}_{i=1}^N$, where $x_i$ denotes process inputs, such as exposure dose, film thickness, and annealing temperature, and $y_i$ denotes output physical quantities, such as $J_c$, capacitance $C$, and decoherence time $T_1/T_2$. SPICE-Q learns the mapping function

\[
f_{\theta}: x \rightarrow y
\]

and combines it with Bayesian updating or physics-constrained regression to realize dynamic correction of device-level SPICE model parameters (Koch et al., 2007).

In addition, the data platform is not only responsible for storage; it must also support cross-scale data alignment, namely unified timestamping and topological mapping of process-level data (wafer-level), device-level data, and circuit-level data. This standardized processing enables the DTCO closed loop to operate stably across different design-iteration cycles, thereby realizing a continuous optimization workflow of "manufacturing--measurement--modeling--redesign."

From a systems-engineering perspective, this data architecture is equivalent to constructing a "nervous system" for quantum-chip manufacturing, in which each sensor node corresponds to a traceable physical state variable, while the data platform serves as a global state estimator, realizing real-time modeling of the fabrication process and control of error propagation (Verjauw et al., 2022).

\section{Integrating SPICE-Q with Process Models}

\begin{quote}
Literature index: This chapter integrates SPICE-Q with process parameters, measurement data, and manufacturing control, drawing on the classical logic of BSIM parameter extraction[4], as well as the literature on superconducting-qubit engineering, transmon parameter mapping, frequency allocation, manufacturable JJ processes, and 300 mm cross-wafer statistics[7][8][12][15][25][29]. The focus of this chapter is not to write process models as isolated empirical tables, but to explain how process variables, device parameters, cryogenic measurements, and system-level yield metrics can be organized into a calibratable closed-loop model.
\end{quote}

\subsection{Principles for Selecting SPICE-Q Parameters}

\begin{quote}
Literature index: The "measurability-first" principle for parameter selection follows BSIM parameter extraction[4]; the mapping among $E_J$, $E_C$, $I_c$, $C_\Sigma$, and transmon frequency follows Koch, Krantz, and Kjaergaard[7][8][12]; manufacturing statistics, frequency allocation, and manufacturable process flows follow the work of Morvan, Verjauw, Van Damme, and others[15][25][29].
\end{quote}

Parameter extraction in SPICE-Q is strongly correlated with process variables and measurable quantities. Its essence is to embed "manufacturability constraints" directly into the parameter space of the simulation model. In the conventional SPICE framework, device parameters, such as capacitance, resistance, and transistor threshold voltage, are usually fitted from test structures and process data. In SPICE-Q, the parameters must further be mapped to specific process variables of quantum devices, such as film thickness, edge roughness, Josephson-junction area, oxide-layer uniformity, interface loss, total capacitance, and package parasitics[4][7][12][25].

For superconducting-qubit systems, the core parameters usually include the Josephson energy $E_J$, charging energy $E_C$, resonator/readout-line coupling rate $\kappa$, coupling strength $g$, loss participation ratios, and decoherence channels. Among them,

\[
E_J=\frac{\hbar I_c}{2e}=\frac{\Phi_0 I_c}{2\pi}, \quad E_C=\frac{e^2}{2C_\Sigma}
\]

are determined by the critical current $I_c$ and the total capacitance $C_\Sigma$, respectively, while these two quantities are directly affected by the junction area $A$, critical-current density $J_c$, oxide thickness $t_{ox}$, capacitor-pad geometry, and parasitic capacitance. Therefore, parameter selection in SPICE-Q is no longer an abstract fitting problem, but a physical mapping problem from the "process-variable space $\mathcal{X}$" to the "quantum-Hamiltonian parameter space $\mathcal{H}$."

In practical modeling, the parameter-extraction process can be formalized as the following optimization problem:

\[
\theta^{*}=\arg\min_{\theta}\left\|y_{\text{meas}}-f_{\text{SPICE-Q}}(x_{\text{process}},\theta)\right\|^2+\lambda\mathcal{R}(\theta)
\]

where $f_{\text{SPICE-Q}}$ denotes the hybrid simulation function jointly formed by layout, EM extraction, EPR/circuit quantization, and Hamiltonian solution, and $\mathcal{R}(\theta)$ is a physical-prior regularization term used to constrain parameters within reasonable process ranges[25][27][28].

Furthermore, SPICE-Q emphasizes a "measurability-first" principle for parameter selection; that is, model parameters must be directly or indirectly obtainable through room-temperature test structures, EM extraction, on-chip test chips, cryogenic measurements, or statistical inversion. For example, frequency shifts of a CPW resonator can be used to infer effective dielectric constants or geometric deviations, resonator $Q$ values can reflect loss channels, and two-qubit gate error rates together with frequency-collision statistics can be used to infer coupling strengths and frequency distributions. This approach makes the model no longer depend on purely theoretical assumptions, but instead forms a "measurement-driven closed-loop parameter calibration system."

In addition, in large-scale quantum-chip design, parameter selection must consider statistical distributions rather than single-point nominal values. Process fluctuations are not necessarily always Gaussian; Josephson-junction discreteness, TLS loss, package modes, and local defects may all introduce skewed distributions or tail risks. Therefore, SPICE-Q is better suited to probabilistic parameter models:

\[
\theta\sim p(\theta\mid \mathcal{D}_{process},\mathcal{D}_{test})
\]

where $\mathcal{D}_{process}$ denotes process data, and $\mathcal{D}_{test}$ denotes test-structure and cryogenic-measurement data. This expression is more general than a single $\mathcal{N}(\mu_\theta,\sigma_\theta^2)$ and is also better suited to yield-aware design and statistical robust optimization[15][29].

In summary, parameter selection in SPICE-Q is essentially a cross-scale inverse problem. Its core objective is to realize a consistent mapping of "process $\rightarrow$ device parameters $\rightarrow$ Hamiltonian $\rightarrow$ system metrics," thereby providing a predictable and calibratable engineering foundation for scalable quantum-computing hardware.

\subsection{SPICE-Q Requirements for Process Control}

\begin{quote}
Literature index: Process fluctuations, JJ parameter sensitivity, surface/interface loss, CMOS-compatible manufacturing control, and cross-wafer statistics follow the work of Koch, Krantz, Kjaergaard, Place, Verjauw, Van Damme, and others[7][8][12][19][25][29].
\end{quote}

The reliability of SPICE-Q is mutually coupled to the reliability of production-line processes. This relationship is not a one-way dependency, but a bidirectionally coupled closed-loop system. In this framework, the predictive accuracy of the simulation model depends on process stability, observability, and the quality of test data; the direction of process optimization, in turn, depends on the quantitative characterization by SPICE-Q of system error-propagation pathways[7][8][25][29].

In conventional semiconductor manufacturing, process control mainly focuses on statistical metrics such as critical dimension (CD), doping concentration, film thickness, and interlayer alignment error, with the usual goal of minimizing within-wafer and wafer-to-wafer variability. In superconducting quantum-chip manufacturing, similar errors are mapped through Josephson nonlinearity, total capacitance, material loss, and the package environment into qubit frequency drift, coupling mismatch, and degradation of coherence time. For example, the critical current $I_c$ of a Josephson junction can exhibit strong sensitivity to oxide thickness $t_{ox}$, and a commonly used engineering approximation is:

\[
I_c\propto\exp(-\alpha t_{ox})
\]

where $\alpha$ is a material- and barrier-related parameter. This relationship reminds us that small process fluctuations can lead to systematic shifts in qubit parameters, thereby affecting frequency allocation, gate operations, and yield[12][15][25].

Therefore, one of the core requirements that SPICE-Q imposes on process control is "error-propagation modeling capability"; that is, it must be possible to explicitly model the mapping function from process perturbations to device parameters and then to system metrics:

\[
\Delta y=J_f(x)\Delta x
\]

where $J_f(x)$ denotes the local sensitivity matrix from process space to quantum-circuit performance space. Here, $y$ should not be understood only as the frequency of a single device, but should also include $T_1/T_2$, coupling strength, readout parameters, frequency-collision probability, and gate-error budgets. This structure enables process engineers to move from empirical parameter tuning to model-driven control and to perform interpretable analysis of error sources.

SPICE-Q also requires the process-control system to provide statistical-consistency guarantees. That is, it must control not only mean shifts but also variance, correlations, spatial gradients, and tail distributions. In quantum chips, tail risks in device parameters are often more critical than means, because system-level failure can be dominated by a small number of frequency collisions, low-$T_1$ devices, or anomalous loss channels. Therefore, the process-control objective can be formalized as:

\[
\min\;\mathbb{E}[L(y)] \quad \text{s.t.}\quad \mathrm{Var}(y)\leq\epsilon,\; P(y\in\mathcal{Y}_{fail})\leq p_{max}
\]

where $L(y)$ denotes the performance-loss function of the quantum circuit, and $\mathcal{Y}_{fail}$ denotes failure regions such as frequency collisions, excessively short coherence times, or out-of-range coupling.

From a systems-engineering perspective, SPICE-Q further requires process control to support cross-scale consistency, meaning that statistical models at the wafer level, chip level, device level, and cryogenic-test level must be mutually alignable. This requirement upgrades conventional isolated process-control modules into a unified model-driven process-control system: room-temperature process data, test structures, cryogenic parameters, and simulation residuals should all enter the same versioned model, rather than being scattered across mutually disconnected reports.

Ultimately, the requirements of SPICE-Q for process control can be summarized as four capabilities: high-precision error-propagation modeling, constraints on statistical distributions and tail risks, cross-scale data consistency, and a versionable feedback loop. Together, these capabilities constitute the basic control logic for industrial production of quantum chips.

\subsection{Feedback from SPICE-Q to Process Control}

\begin{quote}
Literature index: Model-measurement residuals, sensitivity analysis, yield-aware design, and frequency-allocation optimization follow BSIM, Koch, Morvan, Verjauw, Van Damme, and related literature[4][12][15][25][29].
\end{quote}

The construction of SPICE-Q produces feedback effects on process control, but these effects should not be understood as automatically solving process problems. Rather, they provide a model-based diagnostic and decision-making mechanism. By unifying physical modeling, parameter extraction, test structures, and cryogenic experimental validation within the same closed-loop framework, SPICE-Q can help process teams determine which process variables are most worth controlling, which model assumptions need updating, and which designs are overly sensitive to the current process window[4][25][29].

First, SPICE-Q identifies hidden error sources in the process system by establishing a "simulation-measurement consistency residual." Let the simulation-predicted output be $y_{\text{sim}}$ and the experimentally measured output be $y_{\text{exp}}$; then the residual can be defined as:

\[
\delta y=y_{\text{exp}}-y_{\text{sim}}
\]

This residual is used not only for model calibration, but also for locating sources of systematic shift, such as lithographic bias, etch nonuniformity, film-deposition drift, changes in JJ oxidation conditions, anomalous dielectric loss, or unmodeled package modes. The key to residual-driven process diagnosis is to distinguish among three cases: model-structure errors, insufficient parameter fitting, and true process drift. Only when these are clearly separated can model updates avoid incorrectly absorbing process anomalies into empirical parameters.

Second, SPICE-Q guides the prioritization of process control through sensitivity analysis. In parameter space, the gradient of a performance metric with respect to a process variable is defined as:

\[
S_i=\frac{\partial y}{\partial x_i}
\]

where $x_i$ denotes the $i$th process variable, such as junction area, oxide thickness, film thickness, etch time, dielectric loss, or package geometry. By computing the sensitivity vector $\mathbf{S}$, one can identify the process factors that most strongly affect qubit frequency drift, resonator $Q$ value, coupling strength, or decoherence time, thereby forming a process strategy of "priority control for key variables."

Furthermore, SPICE-Q can provide a simulation-based digital process sandbox to evaluate the impact of process perturbations on system performance before actual tape-out. Process perturbations can be described by general probabilistic models:

\[
x_i\sim p_i(x_i\mid\mathcal{D}_{process})
\]

rather than being preset as a single Gaussian distribution. Through Monte Carlo sampling, corner simulation, or yield-aware optimization, the probabilities of frequency collisions, coupling mismatch, readout failure, and low-coherence tail events can be predicted, allowing high-risk designs or process paths to be eliminated in advance[12][15].

In addition, a key contribution of SPICE-Q to process control is the establishment of a design-for-manufacturability feedback mechanism. When a design performs well in simulation but fails in manufacturing, SPICE-Q cannot directly locate all bottlenecks, but it can narrow the range of possible causes through residuals, sensitivity, and test-structure data, and feed this information back into the design space. For example, if the frequency distribution exhibits a systematic shift, the JJ model card or target frequency allocation may need to be updated; if resonator $Q$ values show a batch-level decrease, the material/interface loss model may need to be inspected; if coupling strength deviates from prediction, EM extraction, layout bias, or package models need to be revisited.

Finally, from a systems-engineering perspective, SPICE-Q transforms process control from a static constraint system into a dynamic model-update system. Through continuous data feedback and versioned model updates, process-parameter distributions can be progressively corrected:

\[
p_{t+1}(\theta)=\mathcal{U}\big(p_t(\theta),\mathcal{D}_{new}\big)
\]

where $\mathcal{U}$ denotes an update operator based on new data. In practical engineering, such updates should include a version number, sample range, confidence interval, applicable layout range, and regression-test results, so as to avoid overfitting the model to a small number of anomalous batches.

In summary, SPICE-Q is not only a simulation tool, but also a feedback interface for process control. Its core value lies in transforming process uncertainty, through modeling and data loops, into system variables that are quantifiable, diagnosable, optimizable, and governable by version.

\section{Design-Technology Co-Optimization (DTCO)}

\begin{quote}
Literature index: The DTCO section draws on classical EDA/PDK engineering experience[1][3][4][24], together with literature on superconducting-qubit process sensitivity, frequency allocation, manufacturable JJ routes, and 300 mm cross-wafer statistics[7][8][12][15][25][29].
\end{quote}

Design-Technology Co-Optimization (DTCO) plays a central role in the scalable manufacturing of superconducting quantum chips. Its essence is to transform the "design space" and the "process space" from a loosely coupled relationship into a jointly optimized system that can be co-modeled, co-constrained, and co-updated. In classical semiconductors, DTCO relies on PDKs, device models, statistical corners, and manufacturing feedback to constrain the design space. In superconducting quantum chips, the same idea must be further extended to Josephson-junction discreteness, resonator frequency deviations, material/interface loss, package modes, cryogenic measurement, and system-level frequency-collision risks[7][12][15][25][29]. In the SPICE-Q framework, DTCO should not be understood as a correction tool used late in the design flow; rather, it should run through the entire process of device-level modeling, EM parameter extraction, Hamiltonian construction, layout planning, test-structure design, and model feedback.

\subsection{The Need for DTCO in Quantum Manufacturing}

\begin{quote}
Literature index: Within-wafer and run-to-run deviations, JJ process fluctuations, frequency drift, frequency allocation, and statistically robust optimization are discussed with reference to Koch, Krantz, Kjaergaard, Morvan, Verjauw, Van Damme, and related works[7][8][12][15][25][29].
\end{quote}

Current superconducting quantum-chip fabrication processes exhibit pronounced within-wafer and wafer-to-wafer/run-to-run deviations. These deviations mainly arise from nonuniform thin-film deposition thickness, etch-rate drift, variations in the Josephson-junction oxidation process, material-interface states, packaging environments, and differences in cryogenic test chains. In classical semiconductor devices, such process fluctuations are usually handled through device models, statistical corners, and design margins. In superconducting quantum systems, they map more directly onto qubit frequency drift, coupling-strength mismatch, resonator $Q$ variation, and broader distributions of decoherence times[7][8][12][25][29].

In an ideal simulation environment, a quantum-circuit design is usually optimized with respect to a nominal parameter set $\theta_0$, with the goal of maximizing a performance metric $\mathcal{F}(x,\theta_0)$, such as gate fidelity, frequency separation, or readout fidelity. In an actual fabrication process, however, the real system parameters are $\theta=\theta_0+\delta\theta$, where $\delta\theta$ denotes unavoidable process perturbations. The realized performance therefore becomes:

\[
\mathcal{F}_{\text{real}}(x)=\mathcal{F}(x,\theta_0+\delta\theta)
\]

Because $\mathcal{F}$ is usually nonlinear in quantum systems, a small $\delta\theta$ may move the design point away from the optimal region, causing frequency collisions, coupling-resonance mismatch, overlap of readout windows, or increased calibration complexity. Such deviations typically appear as a reduced fraction of usable qubits, degraded gate fidelity, heavier calibration burden, and lower system-level yield.

The essence of this problem is the decoupling between the design space and the manufacturing space. If a conventional design method assumes that process parameters are deterministic, it will underestimate frequency crowding, crosstalk, and tail risks under real manufacturing conditions. The core requirement of DTCO is therefore to establish a unified co-optimization framework in which design variables $x$ and process variables $\theta$ are evaluated in the same mathematical space:

\[
\min_x\;\mathbb{E}_{\theta\sim p(\theta)}[\mathcal{L}(x,\theta)]
\]

This formulation explicitly incorporates process uncertainty into the design objective function, shifting optimization from single-point optimality to statistically robust optimality. If the objective is to improve manufacturable success rate, a yield-aware objective can also be used:

\[
\max_x\;P_{\theta\sim p(\theta)}\left(\mathcal{F}(x,\theta)>\mathcal{F}_{threshold}\right)
\]

This expression is closer to engineering problems such as frequency allocation, readout multiplexing, and cryogenic test pass rate[15][25][29].

At the physical level, key devices in superconducting quantum chips, such as Josephson junctions, CPW resonators, and tunable couplers, are all highly sensitive to parameters. For example, the critical current $I_c$ of a Josephson junction may show exponential sensitivity to the oxide-layer thickness $t_{ox}$:

\[
I_c\propto\exp(-\alpha t_{ox})
\]

This means that nanometer-scale process fluctuations can lead to appreciable frequency shifts. The specific offset depends on the target frequency, $E_J/E_C$, the junction-area distribution, and the process window. In large-scale fixed-frequency transmon arrays, the tails of the frequency distribution and the margin between frequency bands jointly determine the probability of frequency collisions[12][15].

Moreover, in multi-qubit systems, process deviations have a cumulative effect. If the failure probability of the $i$-th unit is $p_i$, the probability that at least one failed unit appears in the system can be written as:

\[
P_{\text{fail}}=1-\prod_i(1-p_i)
\]

This expression shows that scaling amplifies tail risk, although practical systems can reduce the impact through redundancy, screening, frequency reallocation, tunable coupling, and error-correction architectures. Thus, the key goal of DTCO in quantum manufacturing is not merely to minimize average error, but to control tail risk, spatial correlation, and calibratability at the same time.

More broadly, quantum-chip fabrication requires unified cross-scale modeling capability. Device-level errors, such as JJ area deviations and variations in dielectric loss, propagate to the circuit level, such as coupling mismatch and readout offset, and further affect system-level performance, such as gate-error budgets, frequency planning, and error-correction overhead. This cross-scale propagation process can be represented as a composite mapping:

\[
x\rightarrow f_{device}(x,\theta)\rightarrow f_{circuit}\rightarrow f_{system}
\]

The core value of DTCO is to model this chain in a unified way, thereby enabling global optimization rather than local correction.

Therefore, the fundamental need for DTCO in quantum manufacturing can be summarized as follows: explicitly model process uncertainty, establish a cross-scale consistent model for performance propagation, and form a batch-oriented data closed loop among design, manufacturing, measurement, and model updating. The closed loop here is usually not millisecond-level feedback in the sense of real-time control, but an engineering iteration process based on test structures, cross-wafer statistics, cryogenic measurements, and model-version updates.

\subsection{The DTCO Closed Loop}

\begin{quote}
Literature index: The idea of a design-fabrication-measurement-calibration-redesign closed loop follows SPICE/BSIM parameter-extraction flows[1][4], and is combined with literature on cryogenic measurement of quantum devices, frequency planning, manufacturable processes, and cross-wafer statistics[7][8][15][25][29].
\end{quote}

The SPICE-Q framework implements a closed-loop design-technology co-optimization (DTCO) methodology. Its core is an iterative process of "design-simulation-fabrication-measurement-calibration-redesign" that reduces systematic deviations between simulation predictions and real fabrication outcomes. It should be emphasized that this closed loop is not a real-time control loop automatically completed by a single software package. Instead, it is an engineering iteration system jointly composed of PCells, PDKs, SPICE-Q models, test structures, cryogenic metrology, model cards, and version governance[4][25][29].

During the design and simulation stage, the system uses parameterized cells (PCells) to construct the initial topology of the quantum circuit, including core devices such as Josephson junctions, CPW resonators, readout structures, and tunable couplers. At this stage, SPICE-Q not only performs EM extraction, EPR/circuit quantization, and Hamiltonian solution under nominal conditions, but also explicitly introduces a process statistical model $p(\theta)$ to predict the performance distribution of the nominal design in perturbation space:

\[
\mathcal{F}_{sim}(x)=\mathbb{E}_{\theta\sim p(\theta)}[f(x,\theta)]
\]

This process shifts the design stage from deterministic optimization to probabilistically robust optimization, and allows frequency planning, readout multiplexing, coupling matrices, and packaging constraints to be evaluated within the same statistical framework.

During the fabrication and metrology stage, the system should fabricate both the target design and standardized test structures. Test structures are used to separate statistical variations intrinsic to the process platform, such as the JJ critical-current distribution, resonator frequency and $Q$ value, coupling strength, material loss, and package modes. The target chip is used to verify system-level performance. Cryogenic measurements include qubit frequency $\omega_q$, decoherence times $T_1/T_2$, readout parameters, coupling strength $g$, gate errors, and frequency-collision events. These observations form an empirical data set $\mathcal{D}_{exp}$ for subsequent model updating and parameter inversion.

During the model-calibration stage, SPICE-Q feeds experimental data back into the physical model, enabling joint updating of process distributions and device parameters. This process can be abstracted as Bayesian updating, physically constrained regression, or residual minimization:

\[
p_{t+1}(\theta)=\mathcal{U}\big(p_t(\theta),\mathcal{D}_{exp}\big)
\]

Here, $\mathcal{U}$ denotes the model-update operator. In practical engineering, a model update must be accompanied by sample ranges, batch information, confidence intervals, applicable layout ranges, and regression-test results, so as to avoid overfitting or erroneous transfer caused by a small number of anomalous batches.

During the design-optimization stage, SPICE-Q uses the updated model to re-optimize the circuit topology and layout structure, making the design more robust to known process perturbations. For example, by adjusting target frequencies, junction-area distributions, coupling-structure parameters, readout-resonator windows, or by introducing tunable couplers, one can reduce the probability of frequency collisions and improve the overall consistency of the system. At the same time, this stage introduces a yield-aware optimization objective:

\[
\max_x\;P_{\theta\sim p(\theta)}\big(\mathcal{F}(x,\theta)>\mathcal{F}_{threshold}\big)
\]

This objective directly uses manufacturable success rate or usable-chip fraction as the optimization metric, rather than single-point performance under ideal conditions.

In summary, this closed-loop DTCO method unifies design, fabrication, and measurement within a continuous feedback system. SPICE-Q therefore exists not only as a simulation tool, but also as a model interface connecting PDKs, test structures, process statistics, and system-level design decisions. Its core value is to allow manufacturing data to continuously update design models, so that predictions of frequency distribution, coupling strength, coherence, and yield can gradually converge with batch data.

\section{Large-Scale Production Examples and Design Scenarios}

\begin{quote}
Literature index: The examples in this chapter are engineering concepts based on existing literature, mainly drawing on fixed-frequency transmon frequency allocation and yield optimization[12][15], scalable superconducting control architectures[14], 3D integration/flip-chip/TSV/low-loss multilayer interconnect literature[16][17][18], and literature on resonators, cQED readout, and microwave-network modeling[9][10][11][21]. This chapter discusses design scenarios that SPICE-Q can support; it does not claim that these flows have already become complete industrial standards.
\end{quote}

\subsection{Mitigating Frequency Collisions}

\begin{quote}
Literature index: Frequency collisions, JJ fabrication discreteness, and yield optimization in fixed-frequency transmon processors are discussed with reference to Koch, Krantz, Versluis, Morvan, Gambetta, and related works[7][12][14][15][32].
\end{quote}

In large-scale superconducting quantum processors, especially in qubit arrays based on fixed-frequency transmon architectures, controlling the frequency distributions of qubits, readout resonators, and coupling modes is a key issue for avoiding frequency collisions and parasitic coupling. A frequency collision refers to an excessively small frequency separation or undesired hybridization between different qubits, or between a qubit and a resonator/coupling mode, which can induce exchange coupling, leakage, crosstalk, or increased gate errors. This problem is determined not only by the target design frequencies, but also by JJ fabrication discreteness, total-capacitance deviations, package modes, and the coupling graph[12][15].

Within the SPICE-Q framework, process deviations in the Josephson-junction critical-current density $J_c$, junction area $A$, and total capacitance $C_\Sigma$ can be explicitly modeled as random variables and further propagated to the qubit frequency distribution. For a transmon qubit, if $E_J$ and $E_C$ are expressed in energy units, its transition angular frequency approximately satisfies:

\[
\omega_{01}\approx\frac{\sqrt{8E_JE_C}-E_C}{\hbar},
\quad E_J=\frac{\hbar I_c}{2e}, \quad E_C=\frac{e^2}{2C_\Sigma}
\]

where $I_c=J_cA$. Therefore, process-level variations in $J_c$, $A$, or $C_\Sigma$ lead to systematic drift in $\omega_{01}$. The magnitude of this drift depends on the device design point, the process window, and the capacitance structure; it cannot be described by a single fixed deviation.

In a multi-qubit array, the frequency distribution can be modeled as a random process:

\[
\omega_i=\bar{\omega}_i+\delta\omega_i(J_c,A,C_\Sigma,\mathbf{r}_i)
\]

where $\mathbf{r}_i$ denotes the chip position or layout environment. $\delta\omega_i$ may be a Gaussian approximation, or it may come from a within-wafer spatial-correlation model, a batch-drift model, or test-chip statistics. Through Monte Carlo simulation, statistical linearization, or corner analysis, SPICE-Q evaluates the spectral-overlap probability of the array and computes the frequency-collision risk:

\[
P_{collision}=P(|\omega_i-\omega_j|<\Delta_{crit})
\]

where $\Delta_{crit}$ is jointly determined by the coupling strength, anharmonicity, gate-operation bandwidth, and allowable crosstalk threshold.

At the level of design optimization, the DTCO closed loop allows the above statistical results to be fed back into the layout-design stage. Frequency-zoning design can assign different target frequency windows to different regions across the chip, thereby reducing local frequency crowding. Tunable couplers or tunable qubits can partially alleviate frequency mismatch caused by fabrication deviations, but they introduce flux noise and calibration complexity. Layout-parameter optimization can expand the margin between frequency bands by adjusting junction areas, capacitor pads, coupler geometries, and readout-resonator windows. These strategies do not replace one another; instead, they jointly constitute manufacturing-aware frequency planning.

Related studies show that, in superconducting-qubit arrays containing process fluctuations, statistical design and frequency-allocation optimization can reduce the probability of frequency crowding[15][32]. The advantage of SPICE-Q is that it embeds this statistical process into the device-level, EM-level, and Hamiltonian-level simulation chain, thereby enabling manufacturing-aware design.

\subsection{Optimizing 3D Interconnects}

\begin{quote}
Literature index: 3D integration, flip-chip, TSV, and multilayer low-loss interconnects are discussed with reference to Rosenberg, Vahidpour, and Dunsworth[16][17][18]. Microwave parasitic-parameter modeling follows Pozar[21], and scalable surface-code control and packaging/interconnect requirements follow Versluis et al.[14].
\end{quote}

As superconducting quantum processors scale up, conventional two-dimensional planar wiring architectures gradually encounter constraints in readout lines, control lines, ground continuity, and package modes. Therefore, three-dimensional integration technologies, including through-silicon vias (TSVs), flip-chip bonding, airbridges, and multilayer low-loss interconnects, have become important candidate routes for alleviating routing congestion[16][17][18]. However, such 3D interconnect structures introduce parasitic capacitance, inductance, impedance discontinuities, interface loss, and radiation-loss paths in the microwave band, thereby becoming potential sources of crosstalk and decoherence.

In the SPICE-Q modeling framework, 3D interconnects should not be treated as ideal wires, but as coupled multiphysics systems. Their equivalent impedance can be represented by a frequency-dependent network approximation:

\[
Z_{\mathrm{interconnect}}(\omega)=R(\omega)+j\omega L_{\mathrm{eff}}(\omega)+\frac{1}{j\omega C_{\mathrm{eff}}(\omega)}
\]

where both $L_{\mathrm{eff}}$ and $C_{\mathrm{eff}}$ are determined by the three-dimensional geometry, material dielectric constants, grounding boundaries, and packaging environment, and both exhibit significant frequency dependence.

For a TSV structure under an ideal coaxial approximation, its parasitic capacitance can be approximated as:

\[
C_{\mathrm{TSV}}\approx\frac{2\pi\epsilon L}{\ln(b/a)}
\]

where $a$ is the TSV radius, $b$ is the radius of the surrounding ground structure, and $L$ is the via length. This formula is only a rough estimate; real structures must also account for ground-return paths, bump connections, dielectric loss, superconducting boundary conditions, and package-cavity modes.

In flip-chip structures, interface inductance, contact nonuniformity, and microvoid effects introduced by bump bonding may lead to nonuniform coupling paths, thereby forming local mode-leakage channels. These effects are easily oversimplified in conventional lumped-parameter models, but in SPICE-Q they must be captured through full-wave electromagnetic simulation, S-parameter extraction, and equivalent-network mapping[16][21].

The multiphysics co-simulation flow in SPICE-Q usually includes three consecutive steps. First, the EM solver extracts S-parameters, field distributions, package modes, and port impedances. Second, equivalent-circuit extraction maps continuous fields into a frequency-dependent network. Finally, the quantum-system model maps parasitic coupling terms into Hamiltonian corrections:

\[
H=H_0+\sum_{i\neq j}g_{ij}(a_i^\dagger a_j+a_j^\dagger a_i)
\]

where $g_{ij}$ originates from the electromagnetic coupling strength of the interconnect structure or from equivalent-network parameters. Through this hybrid modeling method, the TSV spacing, bump-array symmetry, ground-return paths, dielectric materials, and airbridge placement can be optimized during the design stage to reduce crosstalk, mode localization, and additional loss.

Related studies point out that 3D integration is an important route toward scalable superconducting quantum computing, but its engineering complexity must rely on cross-scale simulation frameworks for systematic optimization[16][17][18]. The role of SPICE-Q here is analogous to SPICE+EM co-simulation in classical IC design, moving quantum interconnects from empirical design toward computable design.

\subsection{Standardized Selection of Resonators}

\begin{quote}
Literature index: Resonator/cQED readout, CPW resonator parameters, EM extraction, and library-based device design are discussed with reference to Blais, Wallraff, Pozar, and Krantz[7][9][10][11][21]. EM-to-Hamiltonian parameter extraction and design-database workflows follow the EPR and SQuADDS literature[26][27].
\end{quote}

In superconducting quantum-chip design, resonators carry core functions such as quantum-state readout, qubit-coupling control, frequency multiplexing, and quantum-information relay. Conventional design flows usually rely on high-precision electromagnetic simulation tools to optimize CPW resonators or lumped-element resonators on a case-by-case basis. This process has relatively high computational cost, and the resulting designs are difficult to reuse if they lack a unified parameter interface. The cQED and superconducting-qubit engineering literature shows that resonator frequency, quality factor, external coupling rate, Purcell limitation, and the readout chain jointly determine readout performance and system scalability[7][9][10][11].

Within the SPICE-Q framework, resonators are no longer treated as isolated geometry-optimization problems, but are incorporated into a standardized resonator-library system. By normalizing and compressing EM simulation results, EPR/parameter-extraction results, and experimental measurement data, this library establishes a mapping from geometry to equivalent-circuit/Hamiltonian parameters:

\[
\{\mathrm{geometry}\}\xrightarrow{\mathrm{EM\ simulation}}\{S(\omega),Z(\omega),p_{mj}\}\xrightarrow{\mathrm{model\ extraction}}\{\omega_r,Q_i,Q_c,\kappa,g,\chi\}
\]

Here, $\omega_r$ denotes the resonant frequency; $Q_i$ and $Q_c$ denote the internal and external quality factors, respectively; $\kappa$ denotes the linewidth or an external-coupling-related rate; $g$ denotes the qubit-resonator coupling strength; and $\chi$ denotes the dispersive readout frequency shift. Through this mapping, complex three-dimensional structures can be abstracted into reusable parameterized cells (PCells) and model cards, enabling a modular reconstruction of the design space[26][27].

For a CPW half-wave resonator, the fundamental resonant frequency can be approximately written as:

\[
f_n\approx\frac{n v_p}{2L_{eff}}, \quad v_p=\frac{c}{\sqrt{\epsilon_{eff}}}
\]

where $v_p$ is the phase velocity of propagation, and $L_{eff}$ is the effective length after accounting for edge capacitance and port-loading corrections. If a quarter-wavelength structure is adopted, the corresponding odd-mode boundary conditions should be used. In the SPICE-Q standardization flow, $L_{eff}$ should not depend on a single structural fit, but should be provided by a statistical model:

\[
L_{eff}=L_0+\delta L(\epsilon_r,t_{metal},\sigma_{fab},\text{boundary})
\]

This expression explicitly introduces fluctuations in dielectric constant, variations in metal thickness, process randomness, and changes in boundary conditions, thereby incorporating manufacturing uncertainty into the design stage.

By establishing a standardized resonator device library, SPICE-Q can condense repeated EM simulations, parameter extraction, and cryogenic test results into reusable models. Designers do not need to scan the geometry space from scratch for every project. Instead, they can select suitable structures within known ranges of frequency, $Q_i/Q_c$, coupling rate, Purcell constraints, and process tolerances. Such a library-based flow can enhance design reusability, improve control over frequency space, and use statistical models to evaluate in advance the risks of frequency drift, mode hybridization, and readout-chain mismatch.

Therefore, incorporating resonators into the SPICE-Q standardized modeling system is equivalent to introducing into quantum-chip design a device-library and parameterized-design paradigm similar to that in classical IC design. It should be noted, however, that the validity of a resonator library depends on the model's applicable range, EM boundary conditions, packaging environment, and cryogenic measurement calibration. If these boundary conditions change, the library model must be revalidated.

\section{Engineering Transition and Large-Scale Quantum Chips}

\begin{quote}
Literature index: The transition from workshop-style quantum-chip design to industrialized, model-driven design is discussed with reference to the history of SPICE/EDA[1][3][24][30], NISQ/FTQC and fault-tolerant system scaling requirements[5][6], reviews of superconducting-qubit engineering and scalable control architectures[7][8][13][14], frequency allocation and manufacturable routes[15][25][29], and the system-engineering perspective on logical qubits[32].
\end{quote}

The industrialization of quantum computing requires abandoning workshop-style design methods. At the current stage of superconducting quantum-chip development, design flows still rely heavily on experience-driven case-by-case optimization, such as tuning the frequency of individual qubits, debugging local coupling structures, calibrating readout chains device by device, and modifying layouts on the basis of limited simulations. This "hand-crafted tuning" approach remains acceptable for small-scale systems with tens of qubits, but when extended to arrays with hundreds or thousands of physical qubits, or even to the large-scale arrays required for fault-tolerant quantum computing, it will quickly fail because of exploding parameter dimensionality, frequency crowding, overloaded test chains, and poor transferability across batches[7][8][14][15].

SPICE-Q, integrated within a robust DTCO (Design-Technology Co-Optimization) framework, provides the necessary model and flow carrier for scaled production of superconducting quantum chips. It should be emphasized that SPICE-Q is not a single solver. Rather, it organizes device physics, electromagnetic extraction, Hamiltonian construction, statistical yield evaluation, and manufacturing feedback into a versioned and calibratable engineering closed loop[26][27][28][29]. Its core idea is to build quantum-device models on the fundamental physics of the Josephson junction and to continuously update model-parameter distributions through cross-level manufacturing and cryogenic test data.

In this framework, the nonlinear current-phase relation of the Josephson junction,

\[
I = I_c \sin(\phi), \quad E_J = \frac{\hbar I_c}{2e}
\]

forms the basic unit for modeling the entire quantum circuit, where $I_c=J_c A$ is determined by process parameters, including the critical-current density $J_c$, junction area $A$, and barrier/interface state[7][12][25]. By embedding this microscopic model together with EM mode participation ratios, total capacitance, and parasitic environments into the SPICE-Q simulation chain, one can predict frequency shifts, coupling mismatch, changes in anharmonicity, and frequency-collision risks caused by fabrication discreteness at the circuit level[12][15][27].

Furthermore, by introducing yield data, test-structure statistics, and cryogenic metrology feedback (yield data \textbackslash{}\& metrology feedback), model parameters are extended from deterministic nominal values into statistical distributions:

\[
\theta_{\mathrm{device}} = \theta_0 + \delta \theta, \quad \delta \theta \sim \mathcal{D}_{\mathrm{fab}}
\]

where $\mathcal{D}_{\mathrm{fab}}$ may include within-wafer spatial correlations, batch drift, and non-Gaussian tails, and need not be limited to a simple Gaussian approximation[15][25][29]. This forms "manufacturing-aware simulation", enabling the design stage to evaluate frequency-distribution overlap, decoherence statistics, and test pass rates without relying entirely on multiple rounds of physical prototyping.

The key significance of this method is that it transforms quantum-chip design from a "single-point optimization problem" into a "statistically robust optimization problem": the design objective is no longer only the optimal fidelity under nominal conditions, but maximizing yield, controlling variance, and preserving system scalability under process uncertainty[15][32]. This is consistent with the design-for-manufacturability philosophy jointly supported by SPICE, PDKs, and statistical corners in the classical semiconductor industry, and it is also a necessary prerequisite for quantum hardware to move from experimental physics systems to engineered products[1][3][24].

\subsection{From Workshop-Style Design to Industrialized Design}

\begin{quote}
Literature index: Industrialized design, device libraries/PCells, yield objectives, and model-driven flows are discussed with reference to CMOS VLSI design, the history of SPICE[1][3][24][30], scalable superconducting quantum circuits[13][14], frequency allocation and statistically robust design[15], SQuADDS/EPR design databases and workflows[26][27], and the manufacturable/300 mm CMOS routes of Verjauw and Van Damme[25][29].
\end{quote}

The transition of quantum-chip design from a workshop-style flow to an industrialized flow is, in essence, a transition from "local experience-driven optimization" to "systematic, model-driven design". In a conventional flow, each quantum processor often needs to be retuned for a specific process batch, including frequency calibration, coupling-strength adjustment, readout-resonator matching, and package-mode debugging. This makes design and manufacturing poorly transferable and also makes it difficult to form a design language shared across teams and production lines[7][8][25].

By introducing parameterized cells (PCell-based quantum components), standardized device libraries, and writable-back model cards, the SPICE-Q framework structures the design space as:

\[
\mathcal{S}_{\mathrm{chip}} = \{ \mathrm{JJ}, \mathrm{resonator}, \mathrm{coupler}, \mathrm{routing} \}
\]

and connects EM extraction, EPR/circuit quantization, Hamiltonian solution, and open-system evaluation through a unified data interface, thereby realizing a reusable mapping from "geometry" to "parameters" to "performance"[26][27][28]. The fixed-frequency Xmon route demonstrated by Barends et al.[13], the scalable surface-code control architecture discussed by Versluis et al.[14], and the frequency-allocation optimization of Morvan et al.[15] all indicate that large-scale array design can no longer rely only on local manual tuning, but must depend on composable device modules and system-level constraints.

The core metrics of industrialized design can be formalized as three classes of objective functions: $\max \mathcal{F}_{\mathrm{fidelity}}$ characterizes the quality of quantum operations under the target gate set and readout scheme; $\max \mathcal{Y}_{\mathrm{yield}}$ characterizes the test pass rate under a given process window and frequency-avoidance rules; and $\min \mathcal{V}_{\mathrm{variance}}$ characterizes the sensitivity of the system to process fluctuations, packaging differences, and variations in the cryogenic chain. These three objectives usually involve trade-offs: excessive pursuit of nominal fidelity may compress frequency margins and reduce yield, whereas an overly conservative design sacrifices area efficiency and increases control complexity[15][32].

Through the DTCO closed loop, these metrics are no longer optimized independently, but are solved through a joint optimization problem:

\[
\theta^* = \arg\max_{\theta} \; \mathbb{E}_{\mathcal{D}_{\mathrm{fab}}} \left[ \mathcal{F}(\theta) \right]
\]

This expression reflects the core idea of industrial-grade quantum-chip design: optimizing expected performance under a distribution of process uncertainty, rather than seeking single-point optimality under ideal conditions. The manufacturable overlap Josephson junction route demonstrated by Verjauw et al.[25] and the cross-wafer statistics reported by Van Damme et al. on a 300 mm CMOS pilot line[29] further show that industrialization is not simply "scaling up a laboratory process"; rather, test structures, statistical corners, model versions, and batch feedback must be incorporated at the design entry point.

Existing studies show that similar model-driven methods substantially improved design efficiency and scaling capability during the industrialization of classical ICs, as in the EDA + PDK + SPICE system[24]. SPICE-Q attempts to replicate this paradigm in the quantum domain: using Josephson physics as the foundation, PCell/PDK as the interface, and statistical yield as the constraint, it aims to move quantum computing from laboratory systems toward verifiable, reusable, and manufacturable engineered products[25][26][29].

\subsection{The Scalability Roadmap of SPICE-Q}

\begin{quote}
Literature index: Device-level, circuit-level, architecture-level, and system-level roadmaps are discussed with reference to Krantz, Kjaergaard, Versluis, Fowler, and Preskill[5][6][7][8][14]. EM-to-Hamiltonian parameter extraction and design-automation toolchains follow EPR, SQuADDS, SQcircuit, and scqubits[26][27][28][33]. Logical-qubit and system-level constraints follow Gambetta[32].
\end{quote}

The scalability roadmap of SPICE-Q can be understood in terms of four progressive layers: the device level, circuit level, architecture level, and system level. This layering is not intended to cut the problem into mutually isolated modules, but to preserve traceable parameter interfaces at each layer, so that upper-level optimization can be propagated back to lower-level geometry and process variables[7][8][27].

At the device level, the core objective is to establish scalable model libraries for Josephson junctions, resonators, couplers, and interconnects, so that individual components have transferable parameterized representations across different process nodes. The key challenge at this layer is to compress material defects, edge effects, oxide-layer variations, interface loss, and packaging parasitics into a measurable and calibratable low-dimensional parameter space, and to align it with test structures and cryogenic statistical data[12][19][25][27]. Existing work such as EPR/pyEPR, SQuADDS, and SQcircuit has already partially realized automated transfer from "layout" to "EM" to "Hamiltonian parameters"[26][27][28].

At the circuit level, SPICE-Q uses a Hamiltonian-circuit hybrid description to write multi-qubit arrays, readout chains, and coupling graphs in a unified form:

\[
H_{\mathrm{total}} = H_{\mathrm{qubit}} + H_{\mathrm{resonator}} + H_{\mathrm{coupling}} + H_{\mathrm{noise}}
\]

This layer focuses on many-body coupling, cumulative crosstalk, frequency crowding, and open-system decoherence. For large-scale arrays, the key is not to perform full-rank solution over all degrees of freedom, but to control computational complexity through sparse coupling graphs, local subsystem decomposition, statistical linearization, and corner analysis, while retaining global constraints that are sensitive to yield[15][27][28][34].

At the architecture level, SPICE-Q must be co-modeled with control-line layout, readout multiplexing, 3D interconnects, package modes, and error-correction code topology. The scalable surface-code control architecture discussed by Versluis et al.[14] shows that chip design can no longer optimize isolated qubits only; it must simultaneously consider control bandwidth, crosstalk budgets, frequency planning, and interconnect congestion. The work of Rosenberg, Vahidpour, Dunsworth, and others on 3D integration and multilayer wiring[16][17][18] also indicates that architecture-level simulation must incorporate interconnect parasitics, mode leakage, and package modes into the same parameter chain.

At the system level, SPICE-Q will further coordinate with error-correction architectures, compilation layers, and resource scheduling, forming different design constraints for NISQ and FTQC[5][6][32]. Its long-term goal is not to immediately replace all existing tools, but to gradually build a "quantum EDA platform" analogous to the classical EDA ecosystem: from the bottom up, it provides PDK/PCell, hybrid simulation, statistical yield evaluation, and model-version governance; from the top down, it receives requirements from the algorithm layer concerning logical-error budgets, connectivity, and calibration complexity[5][6][14][32].

From the roadmap perspective, the development path of SPICE-Q can be abstracted as:

\[
\mathrm{Device} \rightarrow \mathrm{Circuit} \rightarrow \mathrm{Architecture} \rightarrow \mathrm{System}
\]

This progressive structure reflects the transition of quantum-chip design from local physical problems to full-stack system-engineering problems[32]. The current literature already covers much of the physics and modeling in the first two layers in a relatively mature way, whereas the third and fourth layers still depend heavily on specific error-correction codes, control electronics, and production-line data. SPICE-Q is therefore better understood as an "emerging engineering infrastructure" rather than a completed productized platform[7][8][29].

Overall, the scalability of SPICE-Q depends not only on advances in algorithms and computing power, but also on manufacturing-data closed loops, standardization of test structures, and model-version governance, so that simulation, fabrication, cryogenic calibration, and design reuse form a unified engineering ecosystem[25][29].

\section{Summary}

\begin{quote}
Literature index: This summary synthesizes the cited literature on classical SPICE/EDA, MNA, BSIM, and CMOS VLSI[1][2][3][4][24][30]; NISQ/FTQC and surface-code scaling requirements[5][6][32]; the foundations of superconducting qubits, transmons, cQED, and microwave engineering[7][8][9][10][11][12][13][20][21][22][23]; scalable control, frequency allocation, 3D integration, and low-loss interconnects[14][15][16][17][18]; material/interface loss, manufacturable processes, and 300 mm CMOS cross-wafer statistics[19][25][29]; and design and solver toolchains such as EPR, SQuADDS, SQcircuit, scqubits, and QuTiP[26][27][28][33][34].
\end{quote}

This paper proposes, around SPICE-Q, a multiphysics integrated simulation and design framework for superconducting quantum chips. Its core problem is not simply to improve the accuracy of a particular simulator, but to organize device physics, electromagnetic-field extraction, effective Hamiltonians, open-system dynamics, manufacturing statistics, and cryogenic test feedback into an engineering flow that is calibratable, reusable, and governed by versions. Compared with isolated HFSS-style electromagnetic simulation or single-device Hamiltonian solution, SPICE-Q is closer to the SPICE/PDK paradigm in classical EDA: it compresses complex physical processes into transferable model parameters and allows designs to undergo system-level verification before fabrication[1][2][3][24][30].

The paper first reviewed how classical SPICE transformed circuit design from a physical trial-and-error process into a computable and reusable engineering flow through device models, MNA equations, and numerical solvers. Experience from BSIM and CMOS VLSI design further shows that industrial-grade chip design does not rely on resolving the underlying physics from scratch each time; instead, it relies on model cards, process corners, standard-cell libraries, and PDK interfaces to enable collaboration across teams and process nodes[4][24]. By analogy, if quantum chips are to move from laboratory prototypes to scalable processors, they also need a similar "SPICE moment": Josephson junctions, resonators, couplers, routing, packaging, and cryogenic measurement results must be incorporated into a unified parameter chain[7][8][12][25].

The model composition of SPICE-Q was summarized as a two-domain framework comprising physical space and information space. The physical space focuses on microwave propagation, port impedance, S-parameters, parasitic modes, loss participation ratios, and three-dimensional electromagnetic-field distributions. The information space focuses on qubit spectra, anharmonicity, coupling strength, decoherence channels, Lindblad dynamics, and gate errors. The key connection between the two is not a simple file conversion, but a continuous mapping from layout geometry to EM modes, then to EPR/circuit-quantization parameters, effective Hamiltonians, and device performance metrics[9][10][11][21][22][26][27][28]. Therefore, a SPICE-Q netlist must simultaneously express classical ports, microwave networks, quantum nodes, Hamiltonian blocks, and noise channels, so that the same design object can be verified at different abstraction levels.

The device-level model section showed that the Josephson junction is the fundamental nonlinear element of the entire framework. Its critical current, junction area, barrier state, and parasitic capacitance jointly determine the transmon frequency, anharmonicity, and frequency-collision risk[7][8][12][25]. Resonators, couplers, microwave routing, and 3D interconnects extend the device model to the system level: resonators must simultaneously satisfy readout, Purcell-limit, $Q_i/Q_c$, and frequency-multiplexing requirements; couplers must trade off target coupling, residual coupling, tunability, and noise sensitivity; and routing, airbridges, TSVs, and flip-chip structures must be included in budgets for impedance discontinuities, mode leakage, interface loss, and crosstalk[14][16][17][18][21]. These points show that large-scale quantum-chip design cannot optimize a single qubit alone, but must treat multi-qubit frequency planning, readout chains, control lines, and package modes as one system problem[14][15][32].

A standardized manufacturing system is the process foundation that enables SPICE-Q to function. The paper modeled fabrication flows, equipment behavior, physical transport, inspection and testing, manufacturing environment, electrical standardization, and data acquisition as traceable engineering variables rather than treating them as laboratory experiential steps. For superconducting quantum devices, nanometer-scale geometric deviations, thin-film and interface loss, particle contamination, ESD, vibration spectra, airflow, and temperature/humidity drift can all map onto changes in frequency distributions, coherence times, and yield[7][19][25]. The manufacturable Josephson-junction route of Verjauw et al. and the 300 mm CMOS-compatible cross-wafer statistics of Van Damme et al. show that quantum-chip fabrication is shifting from "whether a single high-coherence device can be made" to "whether large numbers of devices can be repeatedly fabricated in a statistically controlled manner"[25][29].

In the sections on combining SPICE-Q with process models and on DTCO, this paper emphasized that the design space and process space must be jointly optimized. Manufacturing deviations should not merely be treated as sources of error after tape-out; rather, they should enter simulation during the design stage through random variables, spatial-correlation models, batch-drift models, and non-Gaussian tail risks. The closed-loop flow can be summarized as "design-simulation-fabrication-measurement-model calibration-redesign", but it is not a real-time control loop automatically completed by a single software package. Instead, it is an engineering iteration system jointly composed of PCells, PDKs, test structures, cryogenic metrology, model cards, and version governance[4][15][25][29]. The goal of such manufacturing-aware design is also not merely to maximize nominal fidelity, but to find a robust balance among frequency avoidance, decoherence budget, coupling strength, readout pass rate, and process yield[15][32].

The large-scale production examples further illustrated application scenarios for SPICE-Q. For fixed-frequency transmon arrays, statistical fluctuations in JJ critical-current density, junction area, and total capacitance translate into frequency distributions and collision probabilities. Therefore, frequency zoning, layout-parameter optimization, and test-structure feedback should be incorporated into the same frequency-planning flow[12][15][32]. For 3D interconnects, TSVs, flip-chip, and multilayer wiring can alleviate two-dimensional routing congestion, but they also introduce parasitic capacitance, inductance, impedance discontinuities, and package modes, which must be jointly evaluated through EM extraction, equivalent circuits, and quantum-system corrections[16][17][18][21]. For resonator standardization, the goal is not simply to reuse geometric shapes, but to condense $\omega_r$, $Q_i$, $Q_c$, $\kappa$, $g$, $\chi$, Purcell constraints, and process tolerances into queryable and calibratable device libraries[9][10][11][26][27].

Finally, this paper positions SPICE-Q as an infrastructure for moving quantum chips from workshop-style design toward industrialized design. Its value does not lie in claiming that a complete commercial quantum EDA platform has already been formed, but in clarifying an engineering path: the lower layer is physically grounded in Josephson physics and microwave electromagnetics; the middle layer uses EPR/circuit quantization, Hamiltonian solution, and open-system simulation as parameter bridges; and the upper layer takes DTCO, statistical yield, control architecture, and error-correction resource constraints as system objectives[5][6][7][8][26][27][28][34]. Within this framework, the scaling of superconducting quantum processors is no longer merely a matter of increasing qubit count. Instead, it requires establishing a predictable, manufacturable, testable, and continuously updatable engineering ecosystem that provides a methodological foundation for larger-scale and fault-tolerantly scalable quantum-computing hardware.

\section{Acknowledgment}

The authors thank Jun Ye for helpful assistance. The authors also acknowledge the use of AI tools for translation assistance and auxiliary text generation.

\section{Reference}

[1] L. W. Nagel, \emph{SPICE2: A Computer Program to Simulate Semiconductor Circuits}, UCB/ERL M520, University of California, Berkeley, 1975. PDF: \texttt{paper02reference/Nagel\_1975\_SPICE2.pdf}.

[2] C.-W. Ho, A. E. Ruehli, and P. A. Brennan, ``The Modified Nodal Approach to Network Analysis,'' \emph{IEEE Transactions on Circuits and Systems}, 22(6), 504-509, 1975. PDF: \texttt{paper02reference/Ho\_Ruehli\_Brennan\_1975\_MNA.pdf}.

[3] A. Vladimirescu, \emph{The SPICE Book}, John Wiley \& Sons, 1994.

[4] B. J. Sheu, D. L. Scharfetter, P. K. Ko, and M.-C. Jeng, ``BSIM: Berkeley Short-Channel IGFET Model for MOS Transistors,'' \emph{IEEE Journal of Solid-State Circuits}, 22(4), 558-566, 1987.

[5] J. Preskill, ``Quantum Computing in the NISQ era and beyond,'' \emph{Quantum}, 2, 79, 2018. arXiv:1801.00862. PDF: \texttt{paper02reference/Preskill\_2018\_NISQ.pdf}.

[6] A. G. Fowler, M. Mariantoni, J. M. Martinis, and A. N. Cleland, ``Surface codes: Towards practical large-scale quantum computation,'' \emph{Physical Review A}, 86, 032324, 2012. arXiv:1208.0928. PDF: \texttt{paper02reference/Fowler\_2012\_Surface\_Codes.pdf}.

[7] P. Krantz, M. Kjaergaard, F. Yan, T. P. Orlando, S. Gustavsson, and W. D. Oliver, ``A Quantum Engineer's Guide to Superconducting Qubits,'' \emph{Applied Physics Reviews}, 6, 021318, 2019. arXiv:1904.06560. PDF: \texttt{paper02reference/Krantz\_2019\_Quantum\_Engineers\_Guide.pdf}.

[8] M. Kjaergaard, M. E. Schwartz, J. Braumuller, P. Krantz, J. I.-J. Wang, S. Gustavsson, and W. D. Oliver, ``Superconducting Qubits: Current State of Play,'' \emph{Annual Review of Condensed Matter Physics}, 11, 369-395, 2020. arXiv:1905.13641. PDF: \texttt{paper02reference/Kjaergaard\_2020\_Superconducting\_Qubits\_State\_of\_Play.pdf}.

[9] A. Blais, A. L. Grimsmo, S. M. Girvin, and A. Wallraff, ``Circuit Quantum Electrodynamics,'' \emph{Reviews of Modern Physics}, 93, 025005, 2021. arXiv:2005.12667. PDF: \texttt{paper02reference/Blais\_2021\_Circuit\_QED.pdf}.

[10] A. Blais, R.-S. Huang, A. Wallraff, S. M. Girvin, and R. J. Schoelkopf, ``Cavity quantum electrodynamics for superconducting electrical circuits: An architecture for quantum computation,'' \emph{Physical Review A}, 69, 062320, 2004. arXiv:cond-mat/0402216. PDF: \texttt{paper02reference/Blais\_2004\_Cavity\_QED\_Circuits.pdf}.

[11] A. Wallraff et al., ``Strong coupling of a single photon to a superconducting qubit using circuit quantum electrodynamics,'' \emph{Nature}, 431, 162-167, 2004. arXiv:cond-mat/0407325. PDF: \texttt{paper02reference/Wallraff\_2004\_Strong\_Coupling\_cQED.pdf}.

[12] J. Koch et al., ``Charge-insensitive qubit design derived from the Cooper pair box,'' \emph{Physical Review A}, 76, 042319, 2007. arXiv:cond-mat/0703002. PDF: \texttt{paper02reference/Koch\_2007\_Transmon.pdf}.

[13] R. Barends et al., ``Coherent Josephson qubit suitable for scalable quantum integrated circuits,'' \emph{Physical Review Letters}, 111, 080502, 2013. arXiv:1304.2322. PDF: \texttt{paper02reference/Barends\_2013\_Coherent\_Josephson\_Qubit\_Xmon.pdf}.

[14] R. Versluis et al., ``Scalable Quantum Circuit and Control for a Superconducting Surface Code,'' \emph{Physical Review Applied}, 8, 034021, 2017. arXiv:1612.08208. PDF: \texttt{paper02reference/Versluis\_2017\_Scalable\_Surface\_Code\_Control.pdf}.

[15] A. Morvan et al., ``Optimizing frequency allocation for fixed-frequency superconducting quantum processors,'' \emph{Physical Review Letters}, 129, 240504, 2022. arXiv:2112.01634. PDF: \texttt{paper02reference/Morvan\_2021\_Frequency\_Allocation.pdf}.

[16] D. Rosenberg et al., ``3D integrated superconducting qubits,'' \emph{npj Quantum Information}, 3, 42, 2017. arXiv:1706.04116. PDF: \texttt{paper02reference/Rosenberg\_2017\_3D\_Integrated\_Qubits.pdf}.

[17] M. Vahidpour et al., ``Superconducting Through-Silicon Vias for Quantum Integrated Circuits,'' arXiv:1708.02226, 2017. PDF: \texttt{paper02reference/Vahidpour\_2017\_Superconducting\_TSV.pdf}.

[18] A. Dunsworth et al., ``A method for building low loss multi-layer wiring for superconducting microwave devices,'' \emph{Applied Physics Letters}, 112, 063502, 2018. PDF: \texttt{paper02reference/Dunsworth\_2018\_Low\_Loss\_Multilayer\_Wiring.pdf}.

[19] A. P. M. Place et al., ``New material platform for superconducting transmon qubits with coherence times exceeding 0.3 milliseconds,'' \emph{Nature Communications}, 12, 1779, 2021. arXiv:2003.00024. PDF: \texttt{paper02reference/Place\_2021\_Tantalum\_Transmon.pdf}.

[20] G. Wendin, ``Quantum information processing with superconducting circuits: a review,'' \emph{Reports on Progress in Physics}, 80, 106001, 2017. arXiv:1610.02208. PDF: \texttt{paper02reference/Wendin\_2017\_Superconducting\_Circuits\_Review.pdf}.

[21] D. M. Pozar, \emph{Microwave Engineering}, 4th ed., Wiley, 2011.

[22] J.-M. Jin, \emph{The Finite Element Method in Electromagnetics}, 3rd ed., Wiley, 2014.

[23] D. M. Pozar and related microwave network theory references are used for S-parameter, CPW, and distributed microwave-line notation in the SPICE-Q physical-space sections.

[24] N. H. E. Weste and D. Harris, \emph{CMOS VLSI Design: A Circuits and Systems Perspective}, 4th ed., Addison-Wesley, 2010.

[25] J. Verjauw et al., ``Path toward manufacturable superconducting qubits with relaxation times exceeding 0.1 ms,'' \emph{npj Quantum Information}, 8, 93, 2022. arXiv:2202.10303. PDF: \texttt{paper02reference/Verjauw\_2022\_Manufacturable\_Superconducting\_Qubits.pdf}.

[26] S. Shanto et al., ``SQuADDS: A validated design database and simulation workflow for superconducting qubit design,'' \emph{Quantum}, 8, 1465, 2024. arXiv:2312.13483. PDF: \texttt{paper02reference/Shanto\_2024\_SQuADDS.pdf}.

[27] Z. K. Minev et al., ``Energy-participation quantization of Josephson circuits,'' \emph{npj Quantum Information}, 7, 131, 2021. arXiv:2010.00620. PDF: \texttt{paper02reference/Minev\_2021\_EPR\_Quantization.pdf}.

[28] T. Rajabzadeh et al., ``Analysis of arbitrary superconducting quantum circuits accompanied by a Python package: SQcircuit,'' \emph{Quantum}, 7, 1118, 2023. arXiv:2206.08319. PDF: \texttt{paper02reference/Rajabzadeh\_2023\_SQcircuit.pdf}.

[29] J. Van Damme et al., ``Advanced CMOS manufacturing of superconducting qubits on 300 mm wafers,'' \emph{Nature}, 634, 74-79, 2024. arXiv:2403.01312. PDF: \texttt{paper02reference/VanDamme\_2024\_CMOS\_300mm\_Qubits.pdf}.

[30] L. W. Nagel and D. O. Pederson, \emph{SPICE (Simulation Program with Integrated Circuit Emphasis)}, UCB/ERL M382, University of California, Berkeley, 1973. PDF: \texttt{paper02reference/Nagel\_Pederson\_1973\_SPICE.pdf}.

[31] R. P. Feynman, ``Simulating Physics with Computers,'' \emph{International Journal of Theoretical Physics}, 21, 467-488, 1982. PDF: \texttt{paper02reference/Feynman\_1982\_Simulating\_Physics\_With\_Computers.pdf}.

[32] J. M. Gambetta, J. M. Chow, and M. Steffen, ``Building logical qubits in a superconducting quantum computing system,'' \emph{npj Quantum Information}, 3, 2, 2017. arXiv:1510.04375. PDF: \texttt{paper02reference/Gambetta\_2017\_Building\_Logical\_Qubits.pdf}.

[33] P. Groszkowski and J. Koch, ``scqubits: A Python package for superconducting qubits,'' \emph{Quantum}, 5, 583, 2021. arXiv:2107.08552. DOI: https://doi.org/10.22331/q-2021-11-17-583.

[34] J. R. Johansson, P. D. Nation, and F. Nori, ``QuTiP 2: A Python framework for the dynamics of open quantum systems,'' \emph{Computer Physics Communications}, 184(4), 1234-1240, 2013. DOI: https://doi.org/10.1016/j.cpc.2012.11.019.

\section{Appendix A}

\begin{quote}
Literature index: In the appendix, the entries on HFSS, Hamiltonian Simulation, and Traditional SPICE correspond respectively to the literature on electromagnetic finite elements and microwave engineering [21][22], reviews of cQED and superconducting qubits [7][8][9][10][11], literature on transmons and device models [12][13], the EPR/SQuADDS/SQcircuit/scqubits/QuTiP toolchain [26][27][28][33][34], and the classical SPICE, MNA, and device-model literature [1][2][3][4][30].
\end{quote}

This appendix supplements the discussion of the three existing classes of simulation methods repeatedly mentioned in this paper: HFSS-type three-dimensional electromagnetic simulation, Hamiltonian Simulation-type quantum-dynamical simulation, and Traditional SPICE-type classical circuit simulation. They cover different layers in the quantum-chip design flow. HFSS is better suited to solving electromagnetic modes and port parameters from geometric and material boundary conditions; Hamiltonian Simulation is better suited to describing quantum degrees of freedom, energy-level structures, coupling, and open-system dynamics; and Traditional SPICE provides the mature paradigm of device models, netlists, MNA equations, and transient/small-signal analysis in the classical circuit industry [1][2][3][21][22].

These methods are not substitutes for one another. For superconducting quantum chips, HFSS-type tools can output resonant frequencies, impedances, S-parameters, field distributions, and energy participation ratios; tools such as EPR, SQcircuit, and scqubits can further map these results into effective Hamiltonian parameters; and QuTiP or similar open-system solvers can be used to analyze driving, decoherence, and measurement processes [26][27][28][33][34]. The significance of SPICE-Q is that it organizes the inputs, outputs, model boundaries, and cryogenic test feedback of these distributed tools into a traceable data flow, rather than compressing them into a single solver.

\subsection{HFSS}

The core task of HFSS-type three-dimensional electromagnetic simulation tools is to solve Maxwell's equations under specified geometric structures, material parameters, port boundaries, and packaging environments, thereby obtaining microwave field distributions, resonant modes, impedances, S-parameters, radiation losses, and parasitic coupling paths [21][22]. In superconducting quantum-chip design, such simulations are commonly used to analyze CPW resonators, readout cavities, couplers, airbridges, TSVs, flip-chip interconnects, and package cavity modes, because the performance of these structures depends not only on two-dimensional layout dimensions but also on three-dimensional boundary conditions, ground continuity, dielectric loss, and port definitions [16][17][18][21].

Mathematically, HFSS-type tools solve electromagnetic boundary-value problems in the frequency-domain or eigenmode form, for example:

\[
\nabla \times \mu^{-1}\nabla \times \mathbf{E}-\omega^2\epsilon\mathbf{E}=0
\]

Here, $\epsilon$ and $\mu$ denote the permittivity and permeability, respectively, while the boundary conditions are jointly determined by metal layers, dielectric layers, ports, packaging, and radiation boundaries. From the solution, one can further extract parameters such as $S_{ij}(\omega)$, $Z(\omega)$, eigenfrequencies, quality factors, field-energy distributions, and energy participation ratios. These parameters are not directly equivalent to qubit frequencies or gate fidelities, but they are key inputs for subsequent EPR/circuit quantization and Hamiltonian construction [9][21][27].

Therefore, the role of HFSS in the SPICE-Q flow is that of a "physical-space solver." It is responsible for transforming layout geometry and the packaging environment into quantifiable electromagnetic parameters, but it cannot by itself complete quantum dynamics, decoherence, or yield prediction. If designers rely only on HFSS results without entering the Hamiltonian, noise-model, and cryogenic-calibration stages, they can only see whether the electromagnetic structure is reasonable; they cannot determine whether multi-qubit frequency collisions, readout backaction, Purcell limitations, or system-level crosstalk satisfy the requirements of a quantum processor [7][8][9][15].

\subsection{Hamiltonian Simulation}

Hamiltonian Simulation focuses on quantum degrees of freedom and their time evolution. For superconducting quantum circuits, electromagnetic simulations or equivalent circuit models usually need to be further converted into effective Hamiltonians before one can calculate energy-level structures, anharmonicities, coupling strengths, matrix elements, drive responses, and decoherence channels [7][8][9][12]. A typical transmon model can be written as:

\[
H=4E_C(\hat n-n_g)^2-E_J\cos\hat\phi
\]

Here, $E_C$ is determined by the total capacitance, and $E_J$ is related to the critical current of the Josephson junction. For multi-qubit systems, resonators, couplers, drive terms, and noise terms must also be included, for example $H_{\mathrm{qubit}}+H_{\mathrm{resonator}}+H_{\mathrm{coupling}}+H_{\mathrm{noise}}$. Such models can be constructed and solved through workflows such as EPR/pyEPR, black-box quantization, SQcircuit, or scqubits [26][27][28][33].

The outputs of Hamiltonian Simulation usually include eigenfrequencies, anharmonicities, coupling strengths, dispersive frequency shifts, transition matrix elements, leakage probabilities, and open-system dynamical results. For cases involving driving and dissipation, it is also necessary to use the Lindblad master equation or a master-equation solver, such as QuTiP-type tools, to describe behaviors including $T_1/T_2$, readout, leakage, and gate errors [34]. Its basic form can be written as:

\[
\frac{d\rho}{dt}=-\frac{i}{\hbar}[H,\rho]+\sum_k\mathcal{L}_k(\rho)
\]

The role of this class of simulation in the SPICE-Q flow is that of an "information-space solver." It can convert electromagnetic and device parameters into quantum performance metrics, but its accuracy depends on whether the input parameters come from reliable EM extraction, process statistics, and cryogenic test calibration. For large-scale chips, the full Hilbert space expands rapidly; therefore, practical engineering often requires local subsystem decomposition, effective models, sparse coupling graphs, statistical corners, and Monte Carlo yield analysis, rather than full-dimensional exact solution of all quantum degrees of freedom [5][6][15][32].

\subsection{Traditional SPICE Simulation Model}

Traditional SPICE simulation models consist of device models, netlist representations, matrix-based circuit-equation solvers, time-domain transient analysis, and frequency-domain small-signal/noise analysis [1][2][3]. Their core idea is not to solve each device directly from microscopic semiconductor physics, but to abstract objects such as transistors, resistors, capacitors, inductors, and voltage or current sources into composable model cards, and to form sparse algebraic/differential equation systems through methods such as MNA (Modified Nodal Analysis). A typical circuit equation can be summarized as:

\[
F(x,\dot{x},t;\theta_{\mathrm{device}})=0
\]

Here, $x$ denotes node voltages and branch variables, while $\theta_{\mathrm{device}}$ denotes device-model parameters. Device models such as BSIM further illustrate that industrial-grade simulation relies on model cards that can be calibrated through process testing and parameter extraction, rather than resolving the underlying material physics anew for every design [4].

Traditional SPICE offers three main sources of inspiration for this paper. First, models and solvers are separated: device models can be updated with process nodes, while netlists, matrix solving, and simulation interfaces remain relatively stable. Second, design and manufacturing are connected through model cards and process corners, enabling designers to evaluate nominal performance, corner conditions, and noise sensitivity before tape-out. Third, standardized netlists and toolchains reduce the cost of reuse across teams, processes, and projects, thereby supporting the formation of classical EDA and the fabless industry [1][3][24][30].

However, Traditional SPICE cannot directly cover all the requirements of superconducting quantum chips. Quantum chips must also describe three-dimensional microwave fields, Josephson nonlinearity, cQED coupling, Hamiltonian quantization, open-system noise, cryogenic test feedback, and process-statistical yield [7][8][9][12]. Therefore, what this paper draws from Traditional SPICE is its engineering paradigm, rather than applying the Traditional SPICE equations unchanged to quantum chips. On top of the ideas of traditional device models, netlists, and solvers, SPICE-Q needs to add EM parameter extraction, EPR/circuit quantization, Hamiltonian construction, noise model cards, and a manufacturing-feedback closed loop [26][27][28][34].

From the perspective of the appendix, HFSS, Hamiltonian Simulation, and Traditional SPICE correspond respectively to the physical space, information space, and engineering organization paradigm in the SPICE-Q flow. Only when the three are combined can they form the complete engineering closed loop described in this paper: "process/PDK constraints $\rightarrow$ layout $\rightarrow$ EM parameters $\rightarrow$ Hamiltonian $\rightarrow$ noise/yield $\rightarrow$ model update."

\end{document}